\documentclass[twocolumn]{aastex631}
\usepackage{amsmath}

\begin{document}

\title{Formation of Complex Organic Molecules in Prestellar Cores: The Role of Non-Diffusive Grain Chemistry}

\author[0000-0001-5618-4660]{Katerina Borshcheva}
\affiliation{Research Laboratory for Astrochemistry, Ural Federal University \\
Mira st. 19 Yekaterinburg, Russia}
\affiliation{Institute of Astronomy of the Russian Academy of Sciences \\ Pyatnitskaya str. 48 Moscow, Russia}


\author[0000-0003-2434-2219]{Gleb Fedoseev}
\affiliation{Xinjiang Astronomical Observatory, Chinese Academy of Sciences \\ Urumqi 830011, China}
\affiliation{Xinjiang Key Laboratory of Radio Astrophysics \\ Urumqi 830011, China}
\affiliation{Research Laboratory for Astrochemistry, Ural Federal University \\
Mira st. 19 Yekaterinburg, Russia}

\author[0000-0001-6004-875X]{Anna F. Punanova}
\affiliation{Onsala Space Observatory \\
Observatoriev\"agen 90, R\aa\"o, 43992 Onsala, Sweden}

\author[0000-0003-1481-7911]{Paola Caselli}
\affiliation{Max-Planck-Institute for Extraterrestrial Physics \\
Giesenbachstrasse 1 86748 Garching, Germany}

\author[0000-0003-4493-8714]{Izaskun Jim\'enez-Serra}
\affiliation{Centro de Astrobiologia (CSIC-INTA) \\ 
Torrejon de Ardoz Madrid, Spain}

\author[0000-0003-1684-3355]{Anton I. Vasyunin}
\affiliation{Research Laboratory for Astrochemistry, Ural Federal University \\
Mira st. 19 Yekaterinburg, Russia}\email{anton.vasyunin@gmail.com}



\begin{abstract}
We present the results of astrochemical modeling of complex organic molecules (COMs) in the ice and gas of the prestellar core L1544 with the recently updated MONACO rate equations-based model. The model includes, in particular, non-diffusive processes, new laboratory verified chemical routes for acetaldehyde and methane ice formation and variation of H and $\rm H_2$ desorption energies depending on the surface coverage by $\rm H_2$ molecules. For the first time, we simultaneously reproduce the abundances of several oxygen-bearing COMs in the gas phase, the approximate location of the peak of methanol emission, as well as the abundance of methanol in the icy mantles of L1544. Radical-radical reactions on grains surface between species such as $\rm CH_3$, $\rm CH_3O$ and $\rm HCO$ efficiently proceed non-diffusively. COMs are delivered to the gas phase via chemical desorption amplified by the loops of H-addition/abstraction surface reactions. However, gas-phase chemical reactions as well provide a noticeable input to the formation of COMs in the gas, but not to the COMs solid-state abundances. This particularly applies for CH$_3$CHO and CH$_3$OCH$_3$. The simulated abundances of COMs in the ice are in the range 1\%--2\%  (for methyl formate ice) or $\sim$~0.1\% (for CH$_3$CHO and CH$_3$OCH$_3$) with respect to the abundance of H$_2$O ice. We stress a similarity between the simulated abundances of icy COMs in L1544 and the abundances of COMs in the gas phase of hot cores/corinos. We compare our non-diffusive model with the diffusive model and provide constraints for the species' diffusion-to-desorption energy ratios.
\end{abstract}

\keywords{Astrochemistry (75) --- Interstellar Dust Processes (838) --- Star Formation (1569) --- Molecule Formation (2076)}


\section{Introduction} \label{sec:intro}


In astrochemistry, complex organic molecules (COMs) are usually defined as carbon-bearing species containing six or more atoms~\citep[see e.g.][]{HerbstVanDishoeck09}. COMs are detected at all stages of star formation. Their detection in cold dark gas of 
the earliest stages of low-mass star formation
was unexpected. Probably, the first detection of a nearly saturated complex organic molecule toward a cold dark cloud was given by~\citet[][]{Marcelino_ea2007}, who reported the discovery of propylene, $\rm CH_2CHCH_3$, toward TMC-1. The detection of O-bearing COMs in cold dense cores took place in 2012~\citep[][]{Bacmann_ea12, Cernicharo_ea12}. During the last decade, COMs were found in many cold cores that appear to be on different stages of chemical and dynamical evolution along the star formation process, defined by central number densities, a density profile and deuterium fraction~\citep[][]{Crapsi_ea2005, Keto_Caselli2008}. The most dynamically evolved starless cores, on the verge of star formation, are named prestellar cores. COMs were detected in L1521E~\citep[][]{Nagy_ea19, Scibelli_ea21}, L1517B~\citep[][]{Megias_ea23} and L1498~\citep[][]{Jimenez-Serra21}, which can be considered starless cores, as well as in L1544~\citep[][]{Jimenez-Serra16, Caselli_ea22}, which is a prestellar core. An attempt to establish an evolutionary sequence of starless cores has been recently made by~\citet[][]{Megias_ea23}. Different evolutionary statuses possibly explain observed differences in abundances of COMs across the cores. Thus, currently, it is possible to assert that COMs are ubiquitous at the earliest stages of the evolution of low-mass star-forming regions.

Although COMs have been detected towards many starless and prestellar cores, their distribution within the cores remains unknown, with the exception of methanol. In starless and prestellar cores, methanol is most abundant in a shell around the dust emission peak \citep[e.g.][]{Tafalla_ea2002, Bizzocchi14}. This area is often referred to as methanol emission peak, or ``methanol peak''. In turn, the dust emission peak, or ``dust peak'' corresponds to the part of the core with the highest gas density. According to the model by \cite{KetoCaselli10}, in L1544 gas density at the dust peak exceeds $10^7$~cm$^{-3}$ with $ A_{\rm V} > 50$, while the methanol peak is characterized by the moderate density of $10^5$~cm$^{-3}$ and $ A_{\rm V} \approx 5$. The two peaks are separated by $\approx$~4000~au from each other~\citep[][]{Bizzocchi14}. The observations by \citet[][]{Jimenez-Serra16} revealed presence of COMs towards both peaks in L1544 with abundances higher by a factor of $\approx$~2--10 towards the methanol peak than towards the dust peak.

During the last decade, several scenarios explaining the presence of COMs in cold dense molecular gas were proposed based on the possible role of cosmic rays, Eley-Rideal kinetics, and surface carbon insertion reactions~\citep[e.g.,][]{Ruaud_ea15, Shingledecker_ea18, Bergner_ea17}. Essentially, the majority of proposed scenarios can be divided into two classes based on whether gas-phase chemistry or chemistry on interstellar grains is responsible for the formation of COMs. The first proposed scenario \citep[][]{VasyuninHerbst13} relied on gas-phase formation of COMs from precursors formed on grains (methanol). The model by \citet[][]{VasyuninHerbst13} and its extended version \citep[][]{Vasyunin17} was partially successful in explaining the observed abundances of COMs in L1689b and B1-b, as well as abundances and spatial distribution of COMs in L1544. Nevertheless, the scenario faced several difficulties. First, methanol appeared to be overproduced in the gas phase. Second, when applied to other starless cores that are presumably less evolved, the model failed to reproduce observed abundances of some COMs \citep[e.g.][]{Scibelli_ea21}. Finally, the model by \citet[][]{Vasyunin17} was not able to produce appreciable amounts of CO$_{2}$ in the ice under the physical conditions of the static model of L1544 taken from \citet[][]{KetoCaselli10}. In that model of L1544, dust temperature is below 10~K in the inner part of the core and reaches about 15~K at its outer edge. According to~\citet[][]{Oeberg11} and \citet[][]{Boogert_ea2015}, CO$_{2}$ is one of the major ice constituents with an abundance of more than 20\% with respect to solid water, although it is not completely clear whether or not such high abundance holds in the inner parts of cold dense cores. Similar abundance ratios of $\rm CO_2$ and water ices (10--20\%) are also found in dense molecular clouds according to the results of the Ice Age program~\cite[][]{McClure_ea23, Dartois_ea24}. 

The model by \citet[][]{VasyuninHerbst13} and \citet[][]{Vasyunin17} strongly relied on the parametrization of reactive desorption (RD) proposed in \citet[][]{Minissale16}. This parametrization is based on a pioneering series of experiments aimed at quantifying this important process ~\citep[][]{Minissale_Dulieu_14, Cazaux_ea16}. Reactive desorption is a process of ejection of product(s) of an exothermic surface reaction to the gas phase during the relaxation of energy released in the reaction event. The quantification of this process is difficult, as its efficiency depends on many factors including the properties of the underlying surface, binding energies, and complexity of product species, etc. The formula for the efficiency of RD (i.e., for the probability of product species of a reaction to desorb) proposed in \citet[][]{Minissale16} predicts a wide range of efficiencies for products of various surface reactions, including large (5\% or more) in some cases. For example, H$_2$CO is likely overproduced in the gas phase in \citet[][]{Vasyunin17} due to high RD efficiency of CH$_3$ and H$_2$CO itself. High abundances of gaseous O$_2$ and H$_2$O are related to the extremely high RD efficiencies ($>$60\%) for the surface reaction O~+~O~$\rightarrow$~O$_2$. Several other species such as H$_2$S have surface formation channels with high RD efficiencies ($>10$\%).

Later studies reported on average lower and more similar efficiencies for reactive desorption~\citep[][]{chuang_ea18, Oba_ea18, furuya_ea22, Santos_ea23}. Estimated values of RD efficiency vary in the range of 1--3\% with significant uncertainties. Also, new parametrization of RD was proposed in~\citet[][]{FredonCuppen18} in addition to the earlier parametrizations by \citet[][]{Garrod07} and \citet[][]{Minissale16}. Thus, while extending our knowledge on RD and confirming that it is a ubiquitous phenomenon, the aforementioned studies also proved that quantitative estimation of RD efficiency still may contain significant uncertainties. Therefore, experimental data on RD should be taken with caution when including it into astrochemical models. As a result, it is reasonable to consider more conservative parametrizations of RD,~\citep[e.g.,][]{Garrod07}. 

While it was shown that at least under certain assumptions, COMs in the cold gas can be formed in gas-phase chemical reactions, the role of chemistry on interstellar grains to synthesize COMs at $\sim$~10~K was assessed very differently in previous studies. While in some works \citep[][]{Balucani15, Skouteris_ea18} this role was found secondary, newer studies suggest grain surface chemistry plays a pivotal role in detriment of gas phase chemistry~\citep[][]{JinGarrod20}. The novel view on grain surface chemistry as a key source of complex organic molecules at low temperatures ($\sim$~10~K) is based on experimental findings by \citet[][]{Fedoseev_ea15}, \citet[][]{Butscher_ea2016} and \citet[][]{Ioppolo_ea21}. In those works, it was shown that complex organic molecules containing multiple carbon atoms, such as glycol aldehyde, ethylene glycol and even the simplest amino acid, glycine, are efficiently formed in the laboratory experiments under physical conditions similar to those in cold dark clouds. Such results cannot be explained within the traditional paradigm of grain surface chemistry, that is based on the assumption that chemical reactions on grains proceed solely via the diffusive Langmuir-Hinshelwood mechanism. At 10~K, only the lightest species, that are atomic and molecular hydrogen, shall be sufficiently mobile on the surface to induce efficient diffusive chemical reactions, see~\citet[][]{HasegawaHerbst92}. As a result, by means of diffusive chemistry at 10~K and below only hydrogenation reactions can occur. Radical-radical reactions that are required to form complex organic molecules efficiently proceed on grains at $T$$\sim$~30--40~K~\citep[][]{GarrodHerbst06}, and are not efficient at 10~K because radicals are immobile at this temperature. Similarly, reactive atoms heavier than hydrogen shall be poorly mobile as well. 

One of the possible ways to explain the results of \citet[][]{Fedoseev_ea15} and \citet[][]{Ioppolo_ea21} is to assume that radical-radical chemical reactions leading to the formation of complex organic species are still efficient at 10~K. This is due to statistical probability for the otherwise immobile radicals to appear next to each other as the products of processes that are efficient at 10~K. Such processes include accretion from the gas phase and production in efficient diffusive surface reactions, photo- and cosmic ray-induced reactions. Thus, one can assume that chemical reactions on the surface between immobile species appearing next to each other can occur beyond the diffusive Langmuir-Hinshelwood mechanism, i.e., non-diffusively. As a result, complex organic molecules as well as other species such as CO$_2$ that cannot be formed at 10~K in the chemical models utilizing the traditional diffusive paradigm can be produced via non-diffusive chemical reactions on the grain surface.

Basic mathematical formulation of non-diffusive surface chemistry suitable for inclusion into rate equations-based astrochemical models has been proposed by~\citet[][]{JinGarrod20}. They studied the role of non-diffusive chemistry on the formation of complex organic molecules in the L1544 prestellar core~\citep[][]{JinGarrod20}, as well as the role of non-diffusive chemistry in the formation of icy mantles of interstellar grains in the Cha-MMS1 Class~0 protostar~\citep[][]{Jin_ea22}, and in the formation of COMs in hot cores~\citep[][]{Garrod_ea22}. The new models generally successfully reproduce the observed values of abundances of COMs in the gas phase, and typical composition of interstellar ices. However, in the case of the prestellar core L1544, the model by \citet[][]{JinGarrod20} has difficulty in reproducing the radial profiles of methanol and other complex organic species measured by~\citet[][]{Jimenez-Serra16}. Importantly, \citet[][]{JinGarrod20} also conclude that gas-phase chemistry does not contribute significantly to the formation of COMs in L1544. This is in contrast to the earlier work by \citet[][]{Vasyunin17} who claimed that gas-phase chemical reactions play an important role in the formation of COMs toward the same prestellar core.

The goal of this study is to investigate the impact of non-diffusive chemical processes in icy mantles of interstellar grains on the formation of complex organic molecules in gas and ice under the conditions typical of the earliest stages of low-mass star formation. For that, we modify the MONACO code by adding the description of non-diffusive chemical processes to it. Then, we apply the updated code to the 1D static physical model of L1544. The physical model of L1544 is the same as the one utilized in our previous study of COMs chemistry~\citep[][]{Vasyunin17}. This choice enables us to compare new results to those in~\citet[][]{Vasyunin17}, and thus reconsider the roles of gas-phase and grain-surface chemical processes in the formation of COMs in prestellar cores. With this goal, we also updated the utilized chemical network with some most recent laboratory and theoretical results.

The paper is organized as follows. In Section~\ref{sec:model}, modifications introduced to the MONACO code and the physical model of L1544 are described. In Section~\ref{sec:results}, results are presented obtained with the model that includes enabled non-diffusive chemical processes and treatment of reactive desorption described in~\cite{Garrod07}. Consequently, these results are compared with the results of the model with non-diffusive chemical processes but with RD treatment following~\cite{Minissale16}. In Section~\ref{sec:discussion}, the results of this study are discussed. Finally, in Section~\ref{sec:conclusions}, the summary of this study is given.

\section{Model} \label{sec:model}

\subsection{A Three-phase Code with Non-diffusive Chemistry} \label{subsec:3p-code}
Our astrochemical model is based on the three-phase code MONACO described in \cite{Vasyunin17}, which treats the gas phase, outer layers of icy mantles of interstellar grains, and the bulk ice as distinct phases interacting with each other. The code utilizes chemical rate equations to govern chemistry in each of the three phases. Several important updates are introduced to the original model. Those include treatment of non-diffusive chemistry on the ice surface and in the icy bulk, dependence of H and H$_{2}$ binding energies on the H$_{2}$ surface coverage, and reaction/diffusion competition for barrier-mediated chemical reactions in the solid phase. As in similar models, gas-phase rate equations are given with the expression:
\begin{multline}\label{eq:chem_gas}
\frac{dn_i^{\rm gas}}{dt} = \sum_{j,l}k_{jl}^{\rm gas}n_j^{\rm gas}n_l^{\rm gas} - n_i^{\rm gas}\sum_jk_{ij}^{\rm gas}n_j^{\rm gas} + \\ 
\sum_{j}k_{\mathrm{ext},j}^{\rm gas}n_j^{\rm gas} - k_{\mathrm{ext},i}^{\rm gas}n_i^{\rm gas} - \\ k_{\mathrm{ads},i}n_i^{\rm gas} + R_{\mathrm{des},i},
\end{multline}
where $n_{i(j,l)}^{\rm gas}$ is the gas-phase abundance ($\mathrm{cm}^{-3}$) of the species $i(j,l)$, $k_{ij}$ and $k_{jl}$ are the rate coefficients for gas-phase two-particle reactions, $k_{\mathrm{ext},i}$ and $k_{\mathrm{ext},j}$ are the rate coefficients for the reactions caused by external factors (those reactions are cosmic-ray ionization, photoionization, cosmic-ray-induced photoreactions), $k_{\mathrm{ads},i}$ is the adsorption rate coefficient for the $i$th species, and $R_{\mathrm{des},i}$ is its desorption rate. 

Desorption processes included in the model are thermal evaporation, cosmic ray-induced desorption \citep{HasegawaHerbst93}, desorption by cosmic ray-induced UV-photons \citep{PrasadTarafdar83}, photodesorption, and chemical (or reactive) desorption.  Photodesorption yield per incident photon of $10^{-5}$ is assumed for all species except CO \citep{Bertin16,Cruz-Diaz16}. For carbon monoxide, a different yield equal to $10^{-2}$ is adopted \citep{Fayolle11}. More detailed discussion on the implication of photodesorption yields on modeling results can be found in \cite{Punanova22}.

Chemistry on ice surface and within the ice bulk is governed by the following equations:
\begin{equation}\label{eq:total_surf}
\frac{dn_i^{\rm sur}}{dt} = \left(\frac{dn_i^{\rm sur}}{dt}\right)^{\rm chem} - \left(\frac{dn_i^{\rm sur}}{dt}\right)^{\rm tran} - R_i^{\rm diff,s2b},
\end{equation}
\begin{equation}\label{eq:total_bulk}
\frac{dn_i^{\rm bulk}}{dt} = \left(\frac{dn_i^{\rm bulk}}{dt}\right)^{\rm chem} + \left(\frac{dn_i^{\rm sur}}{dt}\right)^{\rm tran} - R_i^{\rm diff,b2s}.
\end{equation}
Here, $n_i^{\rm sur}$ is the abundance of $i$th species on the surface, and $n_i^{\rm bulk}$ is the abundance of $i$th species in the bulk. The first term in the right side of the equation (\ref{eq:total_surf}) describes the evolution of the abundance of species $i$ due to chemical reactions on the surface, accretion and desorption. In the equation (\ref{eq:total_bulk}), the first term includes chemical reactions in the bulk ice. The second and third terms in equations (\ref{eq:total_surf}) and (\ref{eq:total_bulk}) describe the transport of chemical species between the surface layers of icy mantles and the bulk ice. They are explained below.

In this work, we consider both diffusive and non-diffusive chemical reactions on the grain/ice surface. The diffusive mechanism of surface reactions implies that a reaction occurs when two reactants encounter each other on the surface as a result of a two-dimensional random walk. In other words, at least one reactant must be mobile on the surface under the physical conditions of interest. At grain temperatures below 10~K, typical of prestellar cores, only atomic and molecular hydrogen are believed to be highly mobile due to their low binding energies to the surface or, possibly, due to quantum tunneling through the potential barriers between binding sites on the surface. However, it is also possible that two reactants appear on the surface in close proximity to each other, already in a position to react. The reactants may appear as products of a prior chemical reaction, or accrete from the gas. In this case, the mobility of reactants is not necessary for a reaction to occur, and reactions between heavier species such as radicals can proceed at low temperatures. 

Diffusive chemical reactions in the first term of equations (\ref{eq:total_surf}) and (\ref{eq:total_bulk}) can be described similarly as gas-phase reactions in Eq.~(\ref{eq:chem_gas}) or using the modified rate equations approach~\citep[][]{Garrod08, Garrod_ea09}. Non-diffusive chemical reactions on the surface and in the bulk are introduced following the approach by \cite{JinGarrod20}. The expression for the rate of a non-diffusive chemical reaction is as follows:
\begin{multline}\label{eq:nondiff}
R_{AB} = f_{\rm act}(AB)R_{\rm comp}(A) \frac{N_B}{N_S} + \\ + f_{\rm act}(AB)R_{\rm comp}(B) \frac{N_A}{N_S},
\end{multline}
where $f_{\rm act}(AB)$ is a reaction efficiency for barrier-mediated reactions calculated taking into account reaction-diffusion competition (see below), $N_A$ and $N_B$ are the average numbers of atoms or molecules of the reactants on a single grain, $N_S$ is the binding sites number on the surface of an average grain, and $R_{\rm comp}(i)$ is the so-called ``completion rate'' for the $i$th species. In the case of the regular diffusive mechanism, the ``completion rate'' would be equal to $k_{\rm hop}(i) N_i$. This would account for the events of the appearance of the reactant $i$ as the result of hopping. In the case of the non-diffusive mechanism implementation, the completion rate is equal to:
\begin{equation}\label{eq:comprate}
    R_{\rm comp} (A,B) = \frac{1}{1/R_{\rm app}(A,B) + t_{AB}}
\end{equation}
Here, $R_{\rm app}{(A,B)}$ is an ``appearance rate'' for species A or B, i.e., the sum of rates of all processes that ``deliver'' reactants A or B on the surface~\citep[][]{JinGarrod20}. Those processes could be accretion from the gas phase, photoreactions in the ice, surface diffusive reactions, etc. The $t_{AB}$ is a timescale against any possible event to occur to species A and B when they are already in a position to react. 

To the MONACO code, we introduced all types of non-diffusive chemical processes described in~\cite{JinGarrod20} except ``3-body excited reactions''. Also, in this study we prefer to name ``three-body reactions'' as ``sequential reactions'', as in chemistry, termolecular reactions are often referred to as ``three-body'', and because in our point of view term ``sequential'' better reflects the essence of occurring non-diffusive reactivity. The individual impact of all considered non-diffusive mechanisms is discussed in Section~\ref{sec:discussion}.

\citet[][]{JinGarrod20} allow three rounds of non-diffusive sequential reactions. They note that the impact of non-diffusive processes on species' abundances diminishes after the second round. In our simulations, the influence of non-diffusive reactions is negligible after the third round, thus we have also chosen to perform three rounds. 

The reaction efficiency $f_{\rm act}(AB)$ for reactions with activation barriers is calculated taking into account so-called ``reaction-diffusion competition''. The existence of an activation barrier effectively means that at any single encounter of reactants, the probability of a reaction event to occur is below unity. When two reactants of a barrier-mediated surface reaction stay next to each other, two possibilities compete: the possibility to react and the possibility to ``diffuse away'' from each other. The resulting probability of a reaction between species A and B to occur thus is
\begin{equation}\label{eq:f_act}
    f_{\rm act}(AB) = \frac{\nu_{AB}\kappa_{AB}}{\nu_{AB}\kappa_{AB}+k_{\rm hop}(A)+k_{\rm hop}(B)},
\end{equation}
where $\nu_{AB}$ is the highest of the vibrational frequencies of reactants A and B, $\kappa_{AB}$ is the probability for the reaction to occur, $k_{\rm hop}(A)$ and $k_{\rm hop}(B)$ are the thermal hopping rates for species A and B, correspondingly~\citep[][]{HasegawaHerbst92}.

The second term in equations (\ref{eq:total_surf}) and (\ref{eq:total_bulk}) is responsible for redefining the surface when existing surface material is ``buried'' to the bulk by newly accreted species, or, on the contrary, ``excavated'' from the bulk when surface material is desorbed, or its amount is changed due to surface chemical reactions. Expressions (4) and (5) in \cite{Vasyunin17} describe the second term completely. Note that the parameter $\alpha_{\rm tran}$ in their Expression (4) is defined as in \cite{GarrodPauly11}. The third term in equations (\ref{eq:total_surf}) and (\ref{eq:total_bulk}) describes thermal diffusion of material between the surface layers and the bulk ice. We utilized the basic expression for bulk-surface diffusion given in \cite{Garrod13}, although a more elaborated approach exists \citep[][]{Garrod_ea17}. However, at the very least, for low dust temperatures typical of prestellar cores, the rates of such diffusion shall be slow and will not affect modeling results significantly.

In contrast to \cite{Vasyunin17}, in the current model we allow H$_{2}$ accretion on grains. Indeed, this is the most abundant molecule in molecular clouds with the abundance 10$^{4}$ times higher than the undepleted abundance of the second most abundant gas-phase molecule, carbon monoxide (CO). This ensures the huge accretion rate of H$_{2}$. Since dust temperature in dark regions of prestellar cores is likely to be lower than 10~K \citep[see e.g.][]{Crapsi_ea2007}, one may expect the significant abundance of H$_{2}$ on grain surfaces despite the low binding energy of this species. Moreover, the simulations may lead to a nonphysical result of almost complete freeze-out of H$_{2}$ from the gas phase, and the building up of giant icy mantles that consist mainly of solid H$_{2}$.

To avoid a non-physical scenario of the complete freezout of H$_2$ at low temperatures, in our new model binding energies of H and $\rm H_2$ are adjusted according to the surface composition --- namely, to the surface fraction covered by $\rm H_2$, following the approach proposed in \cite{GarrodPauly11}. Following \cite{Cuppen09}, they note that binding on the mixed surface composed of $\rm H_2$ and $\rm H_2O$ is weaker than on that composed of pure $\rm H_2O$ ice. Since \cite{Cuppen09} also estimated the binding energy to $\rm H_2$ surface as about 10 times weaker than to CO surface, \cite{GarrodPauly11} suggested an expression for the time-dependent ``effective desorption energy'',
\begin{equation}\label{eq:Edes_reducing}
E_{\rm des,eff} = E_{\rm des} \left(1 - \theta(\rm H_2)\right) + 0.1E_{\rm des}\theta(\rm H_2).
\end{equation}
Here $\theta(\rm H_2)$ is the fraction of grain surface covered by molecular hydrogen. Therefore, the binding energy of a species becomes in general a time-dependent value. In \cite{GarrodPauly11}, binding energies of all surface species are corrected using the expression~(\ref{eq:Edes_reducing}). However, in our model, we apply it only for atomic and molecular hydrogen. The correction is not applied to heavier species because it is possible that they can penetrate the surface $\rm H_2$ layer and bind to heavier species below it \citep[see][]{Fuchs_ea2009}.

Diffusive chemistry in the bulk ice is also treated similarly as in \cite{Vasyunin17} with the exception for chemical reactions with atomic and molecular hydrogen. In \cite{Vasyunin17}, following \cite{Garrod13}, it was assumed that all diffusive chemical reactions that occur on the grain surface also proceed in the bulk ice via the swapping of species. The rate of swapping was calculated similarly to the rate of thermal diffusion but assuming that swapping energies $E_{\rm swap}$ of species in the bulk are twice as high as their surface diffusion energies $E_{\rm diff}$. This reflects the assumption that binding energies of species in the bulk are twice higher than for the surface species because bulk species on average have twice more neighbors. We made an exception for atomic and molecular hydrogen: it is assumed that swapping energies of H and H$_{2}$ in the bulk are only by a factor of 1.5 higher than their surface diffusion energies. Atoms and molecules of hydrogen are geometrically smaller than other species in the ice. This may result in higher mobility of H and H$_2$ in the ice bulk in comparison to other species. In the models presented in this work, the bulk mobility of species other than H and H$_2$ has a limited impact on the chemical evolution of ice and gas.

The probability for H atoms to stick to grain surface upon collision is calculated according to \cite{Hollenbach_McKee1979} and depends on dust temperature. The sticking probabilities for other species are taken as unity \citep[see e.g.][]{Jones_Williams1985}. Other suggested values and the impact of the sticking probabilities on the modeling results are discussed in Section~\ref{subsec:sticking}.

Chemical processes on ice surface and in the bulk ice induced by cosmic rays and UV photons are treated as in the gas phase. Cross-sections of photoprocesses on the surface are equal to that of the corresponding processes in the gas phase. Note that photodissociation rates in the ice can be smaller than corresponding rates in the gas phase~\citep[see e.g.][]{Oeberg16, Kalvans18, Paardekooper_ea2016}. Radiolysis of interstellar ices and their laboratory analogs is studied extensively in recent years~\citep{Pilling_ea10,Fedoseev_ea18,Shingledecker_ea18,Shingledecker_ea19,Shingledecker_ea20,Ivlev_ea23}. The results of those works suggest that the impact of cosmic rays on the chemistry of interstellar ices is more complicated than it is assumed in our model. On the other hand, \cite{Fedoseev_ea18} investigated the formation of $\rm OCN^-$ by cosmic ray processing of $\rm N_2$- and $\rm NH_3$-containing interstellar ice analogs. The results suggest that in dense dark clouds significant processing of $\rm N_2$-containing ices by cosmic rays requires timescales of $\sim$~$10^{6}$--$10^{7}$~years. At the same time, dynamical and chemical timescales of evolution in prestellar cores are approximately one--two orders of magnitude shorter. Thus, we believe that the aforementioned simplistic approach to cosmic ray processing of ices ``as in the gas phase'' is sufficient for this work.

\subsection{Physical Model of L1544 and Initial Conditions for Chemistry} \label{subsec:physmod}
\begin{figure}
\begin{center}
\includegraphics[width=1.0\linewidth]{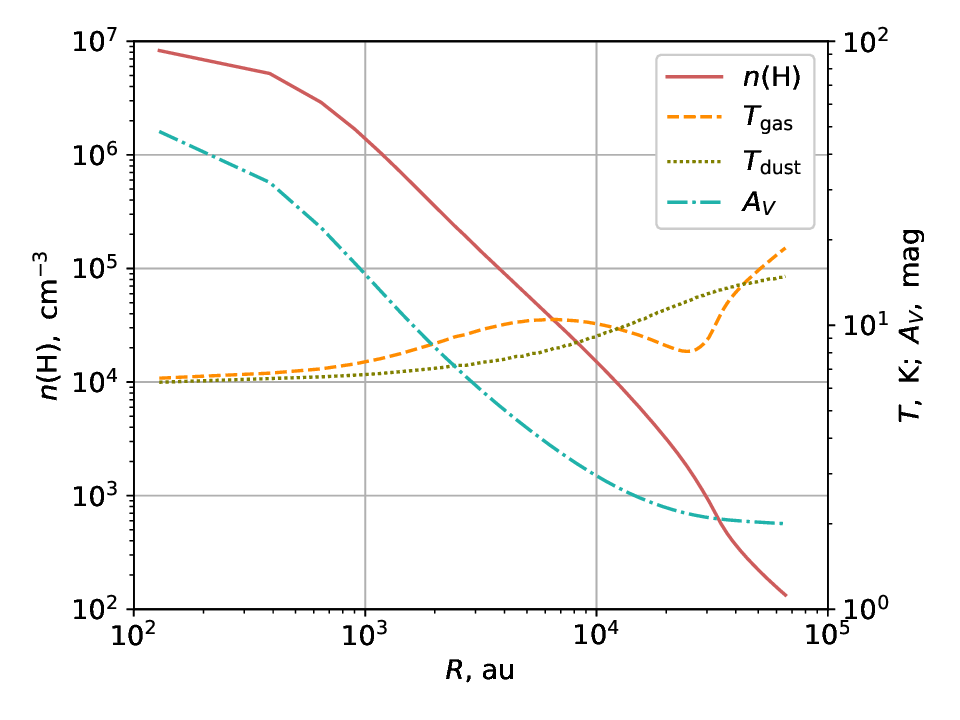}
\end{center}
\caption{The radial profile of physical conditions in L1544 from \cite{KetoCaselli10}. The dust temperature in the interior of the core is below 10~K.} \label{fig:phys_profile}
\end{figure}

The physical model of L1544 prestellar core and initial conditions for chemistry are similar to those used in \cite{Vasyunin17}. In particular, we use a static 1D radial profile of gas density, dust and gas temperatures in L1544 with central densities of $\sim$~10$^{7}$~cm$^{-3}$ first obtained in \citet{KetoCaselli10} and confirmed recently by~\citet{Caselli_ea2019, Caselli_ea22} (see Fig.~\ref{fig:phys_profile}). The visual extinction at the edge of the core in our modeling equals 2~mag to simulate the fact that L1544 is embedded in a molecular cloud~\citep[][]{Redaelli_ea22}. There are 128 radial points in our profile. In every point, chemical evolution is calculated independently with our 0D MONACO model. Initial abundances for the calculation of the prestellar core chemistry are the same in every radial point and taken as final chemical composition of a ``translucent cloud'' after 10$^6$~years of evolution. The ``translucent cloud'' is characterized by gas density of 10$^3$~cm$^{-3}$ and visual extinction Av~=~2.0~mag. During the 10$^6$~years of evolution, gas and dust temperatures linearly drop from 15~K to 10~K. The initial abundances for the translucent cloud phase are atomic ``low metals'' values listed in Table~1 in~\citet[][]{WakelamHerbst08}. Hydrogen is assumed to initially reside in molecular form. 

Grain surface chemistry in cold dark environments is mainly controlled by the mobility of species. In classic gas-grain astrochemical models, only diffusive grain-surface chemistry is considered~\citep[e.g.,][]{HasegawaHerbst92}, so the diffusion rates of species are the natural control parameters. In the new grain-chemistry models that include non-diffusive chemical processes, the diffusion rates may seem to be less crucial. However, they still are, as diffusion rates render the relative importance of diffusive and non-diffusive chemical reactions that occur simultaneously. In this work, we assume that all species diffuse on the surface and in the bulk only via thermal hopping. There is still a debate on whether quantum tunneling plays an important role in the diffusion of atomic and molecular hydrogen at low temperatures~\citep[e.g.,][]{Watanabe_ea10, Hama_ea12, HamaWatanabe13, Cuppen_ea17}. However, it is likely that the diffusion of H atoms on amorphous water ice is much slower than predicted by the simple model of tunneling through rectangular potential barriers~\citep[][]{HasegawaHerbst92} even if quantum tunneling through diffusion barriers indeed contributes to its rate along with thermal hopping~\citep[][]{Kuwahata_ea15, Senevirathne_ea17}. Thus, we assume only thermal hopping for H and H$_2$ mobility. The quantum tunneling through reaction activation barriers is enabled, and the barriers are assumed to be rectangular with the width of 1~\AA. Several reactions included in the CO hydrogenation chain have different barrier widths (see Appendix~\ref{subsec:network_updates}).

Rates of thermal hopping in rate equations-based models are typically controlled by the diffusion-to-desorption energy ratio(s), $E_{\rm diff}/E_{\rm des}$. In principle, there is no fundamental physical argument that  $E_{\rm diff}/E_{\rm des}$ shall be the same for all species on surface~\citep[][]{Fredon_ea21}. Taking this into account, and the results of \citet[][]{Minissale_ea16b} and \citet[][]{KarssemeijerCuppen14} we adopted $E_{\rm diff}/E_{\rm des} = 0.5$ for atomic species and $E_{\rm diff}/E_{\rm des} = 0.3$ for molecular species. Note that~\citet[][]{Furuya_ea2022_EbEd_not_clear} reported a wide range of diffusion-to-desorption energy ratios (0.2--0.7) depending on species. In Section \ref{subsec:EbEd_grid} we discuss how the change in these parameters modifies the results of the simulations.

In this work, we compare two models with different parametrizations of reactive desorption. The first is based on Rice-Ramsperger-Kessel (RRK) theory and introduced in an astrochemical context by~\citet[][]{Garrod_ea06} and \citet[][]{Garrod07} (hereafter, GRD model). The value of the parameter $a$ in Expression (2) in~\citet[][]{Garrod07} for the probability of RD is taken equal to 0.01. The second parametrization of RD considered in this study is based on experimental findings by~\citet[][]{Minissale_ea16b} (hereafter, MRD). It was utilized in the study by~\citet[][]{Vasyunin17} and was recently updated in~\citet[][]{Riedel_ea23}. In this work, we utilize it with the previously used value of effective mass of surface element $M$ equal to 100 a.m.u. Note that even where the treatment of RD efficiency follows \cite{Garrod07}, it is allowed only from the non-water fraction of the ice surface, as discussed in \cite{Vasyunin17}. The key parameters of the utilized models are summarized in Table~\ref{model_parameters}. The major updates to the chemical network are described in Appendix~\ref{subsec:network_updates}.

\begin{deluxetable}{l|l}
\tablecaption{Summary of parameters of considered chemical models.}\label{model_parameters}
\tablehead{
\colhead{Parameter} & \colhead{Value}
}
\startdata
 Cosmic ray ionization rate (s$^{-1}$)        & $1.3(-17)$ \\
 Photodesorption yield for CO (mol./photon)   & $1.0(-2)^{a}$ \\
 Photodes. yield, other species (mol./photon) & $1.0(-5)^{b,c}$\\
 Grain size (cm)                              & $1.0(-5)$ \\
 Dust-to-gas mass ratio                       & $1.0(-2)$ \\
 Surface site density (cm$^{-2}$)             & $1.5(+15)$ \\
 Grain density (g/cm$^3$)                     & $3.0$      \\
 $E_{\rm diff}/E_{\rm des}$, atomic species   & $0.5$      \\
 $E_{\rm diff}/E_{\rm des}$, \# atoms $>$ 1   & $0.3$      \\
 $E_{\rm swap}/E_{\rm diff}$, H and $\rm H_2$ & $1.5$      \\
 $E_{\rm swap}/E_{\rm diff}$, other species   & $2.0$      \\
 Number of ''surface'' monolayers             & $4^{d}$ \\
 Number of rounds for non-diffusive reactions        & $3^{e}$ \\
 Tunneling through diffusion barriers         & Off$^{e}$\\
 Hopping/reaction activation competition      & On$^{f}$ \\
 Tunneling through reaction barriers          & On         \\
 Reaction bar. width, \AA\ (exceptions: see Table~\ref{tab:CH3OH_chain}) & 1.0
\enddata
\tablecomments{$^{a}$~\cite{Fayolle11}; $^{b}$~\cite{Bertin16}; $^{c}$~\cite{Cruz-Diaz16}; $^{d}$~\cite{VasyuninHerbst13}; $^{e}$~\cite{JinGarrod20}; $^{f}$~\cite{GarrodPauly11}; $x(y)$ means $x \cdot 10^{y}$.}
\end{deluxetable}

\section{Results} \label{sec:results}
\subsection{CO Depletion Factor in Models} \label{subsec:CO_depletion}
Comparison of modeling results with the observational data requires the establishment of some formal approach. Below, we describe the approach utilized in this work. First, it is applied to the comparison of the results of the GRD model with the observed abundances of COMs in L1544. Then, in Section~\ref{subsec:compare_to_Minissale}, it is utilized to test the results of the MRD model against the same observations. 

\begin{figure}
\begin{center}
\includegraphics[width=1.0\linewidth]{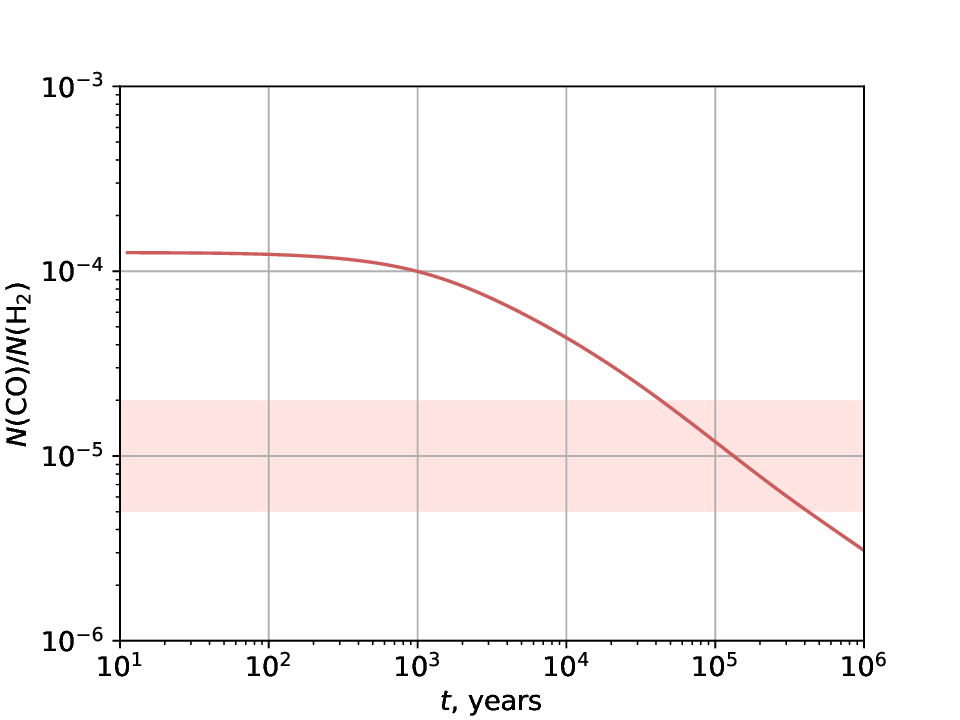}
\end{center}
\caption{Temporal evolution of gas-phase CO-to-H$_{2}$ column density ratio in our simulations. Colored area denotes the factor two uncertainty for the observed CO depletion factor taken from \cite{Caselli99}.} \label{fig:COdepl}
\end{figure}

As in~\cite{Vasyunin17}, to determine the range of evolutionary time where modeled and observed abundances can be compared, we calculated the temporal evolution of CO depletion factor towards the dust continuum peak (which is considered to be the core center) in the model, and compared it to the observed value obtained by \cite{Caselli99}. According to their data, CO experiences significant depletion in the central parts of the prestellar core L1544 --- they estimate the observed CO depletion factor as $\approx 10$ at the position of the dust peak. In Figure \ref{fig:COdepl}, the temporal evolution of CO-to-$\rm H_2$ column density ratio in our GRD model is presented; from here on, every gas column density is convoluted with a Gaussian function representing the IRAM radio telescope beam with FWHM = $26''$ (which is the beam size in the \cite{Jimenez-Serra16} observations). The colored area denotes the factor two uncertainty for the observed depletion factor of CO. In our modeling, the observed CO depletion factor is reached at $\approx 1.2 \cdot 10^5$ years of simulation time. However, the uncertainty margin of CO depletion covers the time range starting at $\approx 4.0 \cdot 10^4$~years and up to $\approx 4.0 \cdot 10^5$ years.

\subsection{Definition for the Agreement of Modeling and Observational Results} \label{subsec:agreement}
In our spherical model of the physical structure of L1544, the dust peak is located in the center of the core, and the location of the methanol peak corresponds to the radial distance of 4000~au from the core center. Modeling results are compared to the observed values of abundances of species towards the dust peak and methanol peak reported in~\citet[][]{Jimenez-Serra16}.

\begin{figure*} [ht!]
  \includegraphics[width=0.5\linewidth]{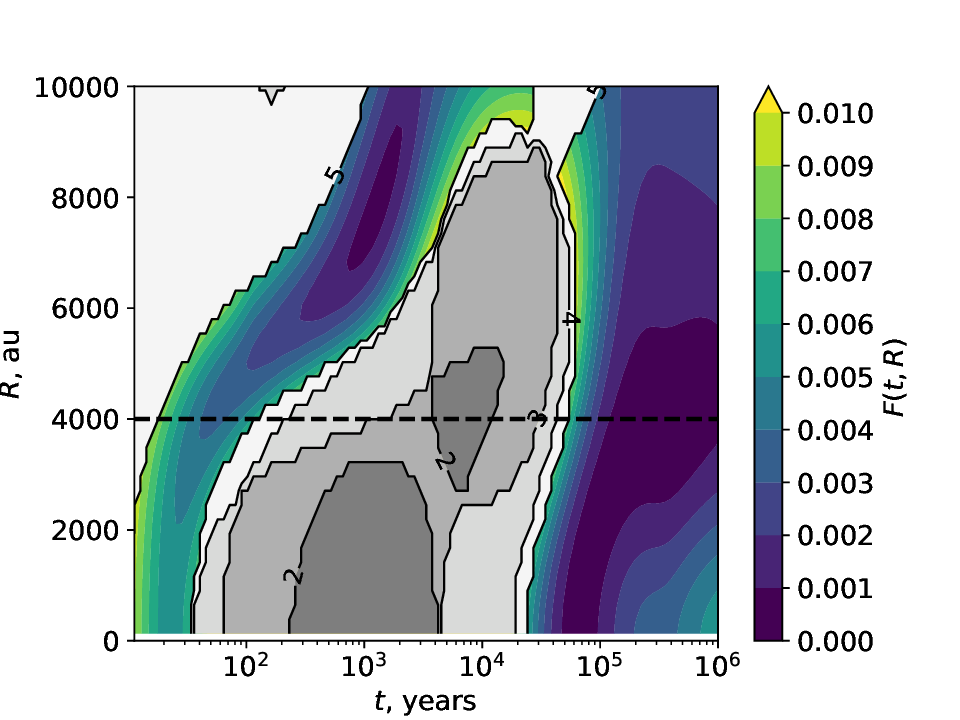}
  \includegraphics[width=0.5\linewidth]{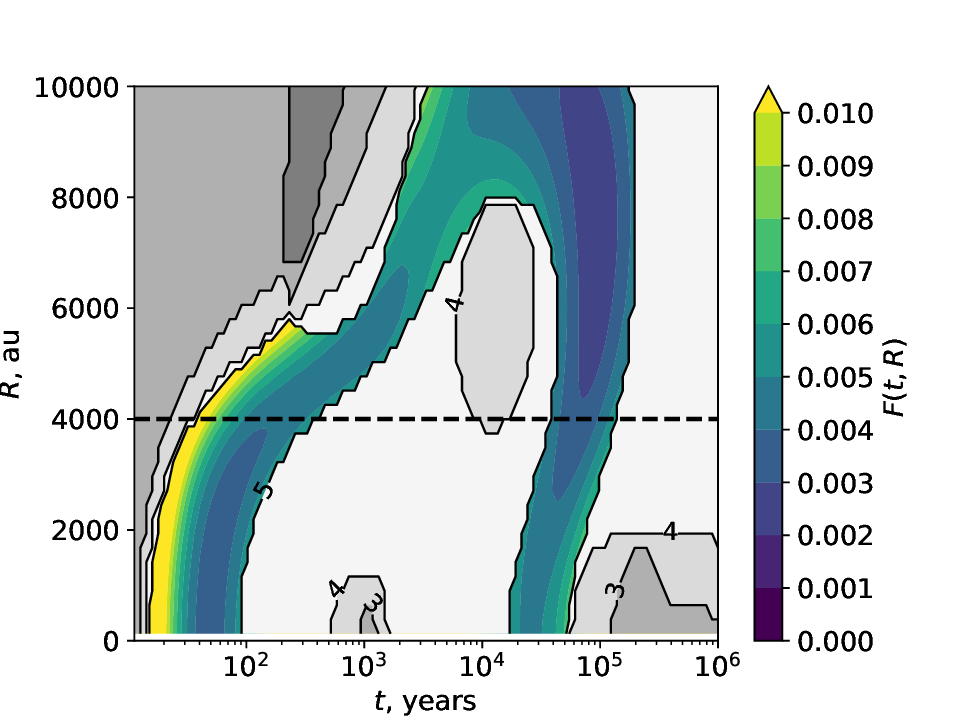}
\caption{Agreement maps for the dust peak (left) and for the methanol peak (right) for the GRD model. Vertical axis represents the radial distance from the core center. Horizontal axis represents the evolutionary time in the model. The position of the dust peak corresponds to 0~au on vertical axis, the position of the methanol peak corresponds to 4000~au and is marked with a dashed horizontal line. Grey scale is for the areas where the abundances of $\leq 5$ of studied species ($\rm CH_3OH$, $\rm CH_3O$, $\rm CH_3CHO$, $\rm HCOOCH_3$, $\rm CH_3OCH_3$, $\rm NH_2CHO$) are simultaneously in agreement with the observational data; numbers indicate the number of species that are simultaneously in agreement with the observations inside a contour line. The area with all the studied species are simultaneously in agreement with the observations is colored. The higher the $F(t,R)$ value defined in Expression~(\ref{eq:F(r,t)}), the worse the agreement with the observations; 0.000 means complete agreement. (We do not show similar agreement maps for the MRD model since there is no $(R,t)$ domain for which simultaneous agreement for all the studied species is attained.)} \label{fig:agreement_maps}
\end{figure*}

To compare our modeling results with the observations, we created agreement maps in the phase space $(t, R)$, where $t$ is a time moment in simulations, and $R$ is a radial point in a spherically symmetric model of L1544. In the left panel of Figure~\ref{fig:agreement_maps}, we show a comparison of the abundances obtained by our model with the abundances observed toward the dust peak in L1544, while in the right panel, we present a comparison of modeled abundances to the abundances observed toward the low-density, $\rm CH_3OH$-rich shell detected by \citet{Bizzocchi14} at~$\approx 4000$ au from the continuum peak (the ``methanol peak''). For the comparison, we use the gas species $\rm CH_3OH$, $\rm CH_3O$, $\rm CH_3CHO$, $\rm HCOOCH_3$, $\rm CH_3OCH_3$, and $\rm NH_2CHO$. We are primarily interested in those phase areas where the modeled abundances of all the species used for the comparison are \textit{in agreement} with observed values. By \textit{agreement} we mean the situation when the modeled abundance at a certain moment of time differs no more than by an order of magnitude from the observed value. If an upper limit is established for a species from observations, we consider its modeled abundance to be \textit{in agreement} with an upper limit if the modeled abundance is less than or exceeds by no more than an order of magnitude the observed upper limit. Upper limit means non-detection of a species, and exceeding the limit by more than an order of magnitude may be astrochemically interpreted as poor agreement with observations. Those areas of the parameter space where the modeled abundances of all six species are \textit{in agreement} with observations are filled with color.
The areas where five or fewer species have abundances \textit{in agreement} with the observational results are filled with grey scale (the darker the filling, the fewer species have abundances {\it in agreement} with the observational values). The colors inside the best fit area denote the values of the function $F(t, R)$, which resembles the function in equation (17) from \cite{Vasyunin17}:
\begin{equation}\label{eq:F(r,t)}
F(t,R) = \sum\limits_{i = 1}^{6}\left(\frac{\log\chi_{\rm obs}({\rm X}_i) - \log\chi_{\rm mod}^{(t,R)}({\rm X}_i)}{\log\chi_{\rm obs}({\rm X}_i) + \log\chi_{\rm mod}^{(t,R)}({\rm X}_i)}\right)^2,
\end{equation}
where $\chi_{\rm obs}({\rm X}_i)$ is the observed abundance of the species ${\rm X}_i$ (at the dust peak or at the methanol peak, correspondingly) and $\chi_{\rm mod}^{(t,R)}({\rm X}_i)$ is the modeled abundance of the species ${\rm X}_i$ at the point $(t,R)$ derived from the column densities. The modeled abundances, in turn, are defined as $\chi_{\rm mod}^{(t,R)}({\rm X}_i) = N_{\rm mod}^{(t,R)}({\rm X}_i)/N_{\rm mod}^{\rm (t,R)}({\rm H_2})$; here $N_{\rm mod}^{(t,R)}({\rm X}_i)$ is the modeled column density of the species ${\rm X}_i$ convoluted over the $26''$ Gaussian beam, and $N_{\rm mod}^{\rm (t,R)}({\rm H_2})$ is the $\rm H_2$ smoothed column density.

In the map for the dust peak (left panel of Figure~\ref{fig:agreement_maps}), the minimum value of the function $F(t,R)$ --- which indicates the best agreement with the observations --- is reached at the distance of 3700~au from the core center at $3.9 \cdot 10^5$ years of the simulation time for the GRD model. This time point is located on the border of the time interval allowed by the CO depletion factor. The lowest value of $F(t,R)$ for the methanol peak is reached at 7600~au and $10^5$ years of the simulation time, which fits the CO depletion timescale much better (see Figure~\ref{fig:COdepl}). At $10^5$ years, all the species also demonstrate modeling abundances {\it in agreement} with the observed values both at the dust peak and at the observational methanol peak positions. Thus, in our modeling, we consider $10^5$ years as the time of the best agreement with the observational data. Note that our best agreement time is similar to the best fit time from \cite{Vasyunin17}, which is equal to $1.6 \cdot 10^5$ years.

\subsection{Results of the GRD Model}
\label{subsec:results_Garrod}
\begin{figure*} [p!]
  \includegraphics[width=0.5\linewidth]{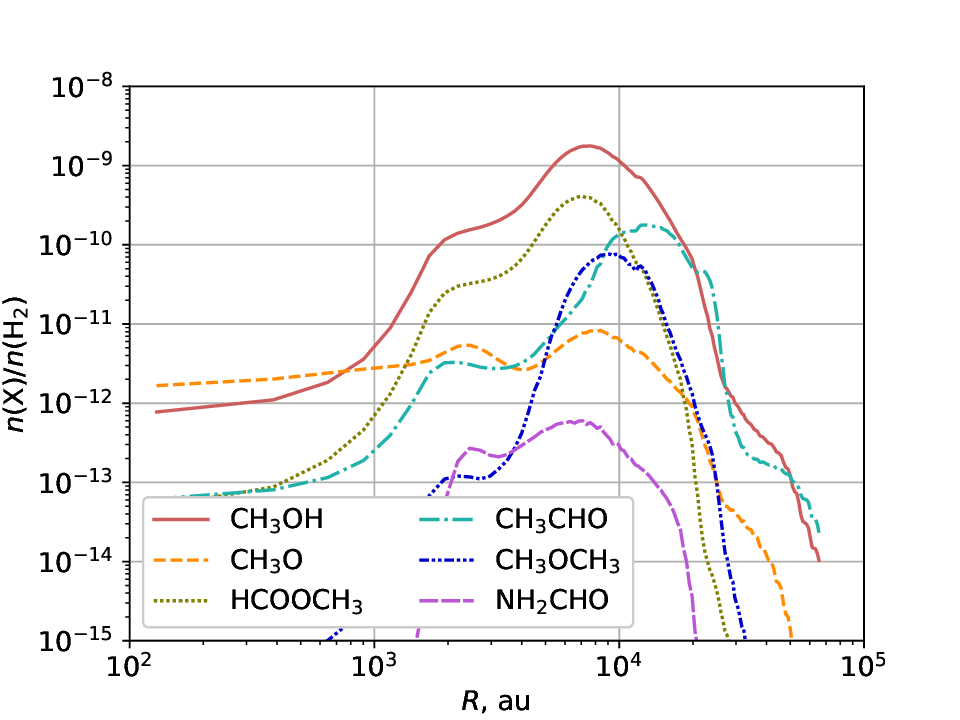}
  \includegraphics[width=0.5\linewidth]{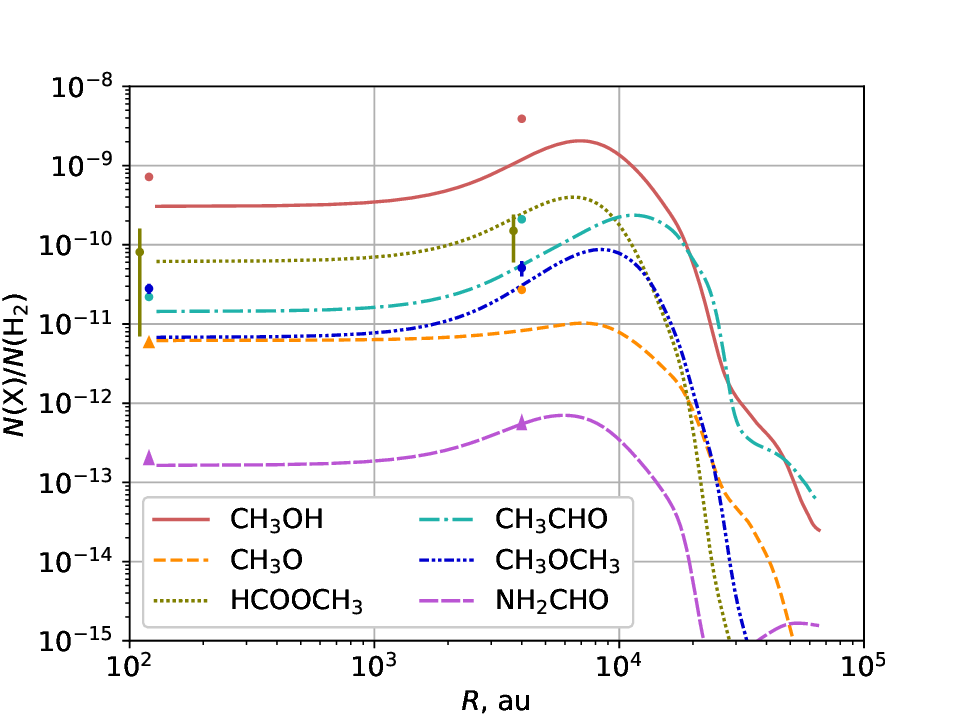} \\
  \includegraphics[width=0.5\linewidth]{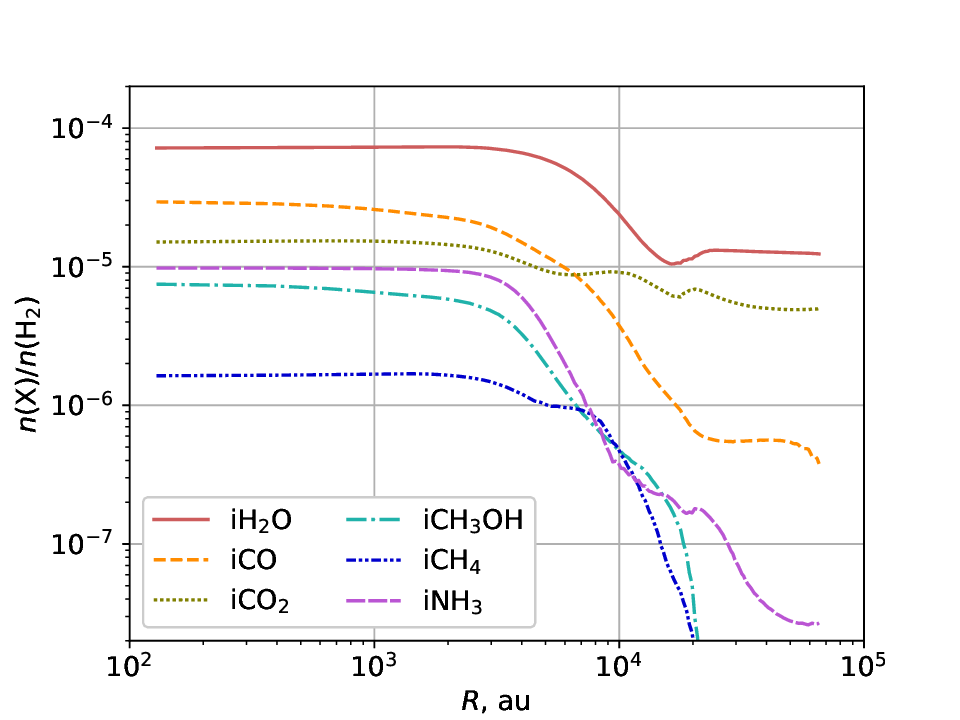}
  \includegraphics[width=0.5\linewidth]{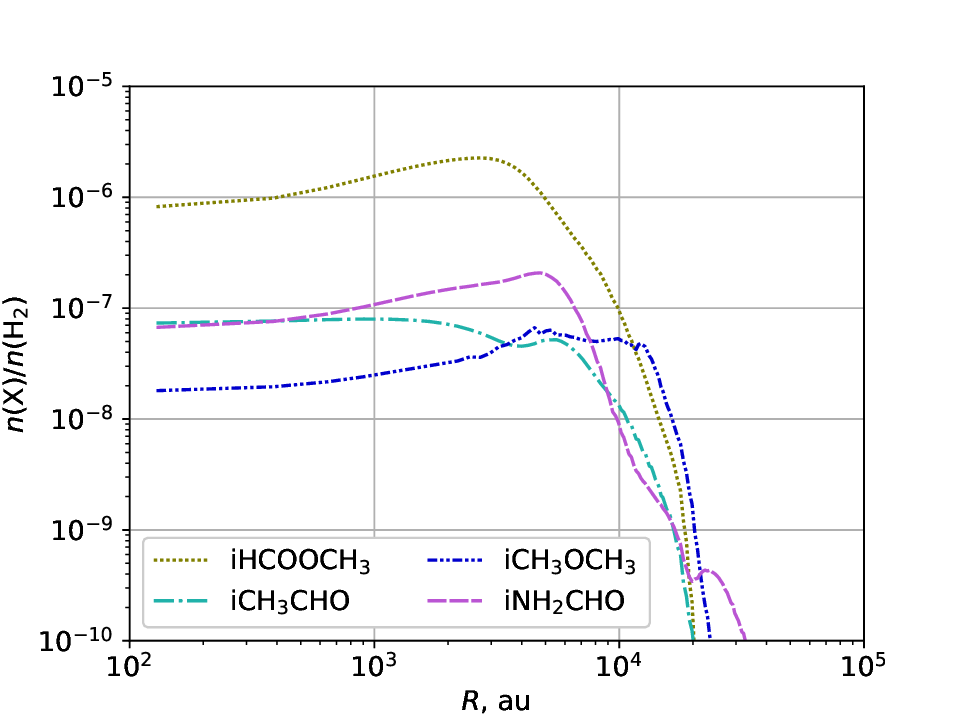} \\
  \includegraphics[width=0.5\linewidth]{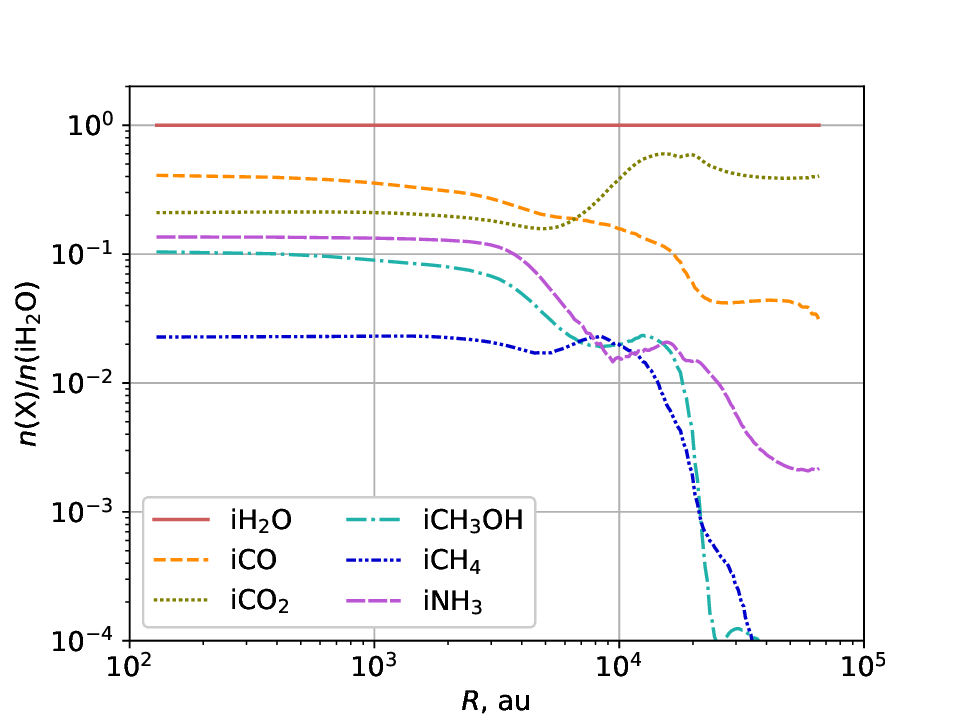}
  \includegraphics[width=0.5\linewidth]{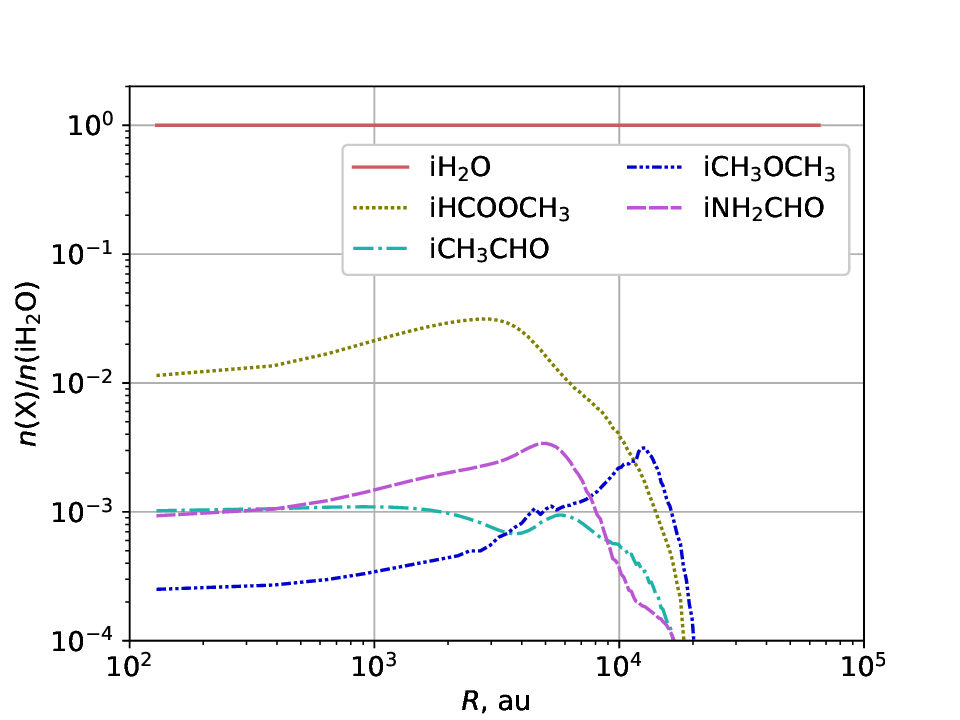}
\caption{Radial profiles obtained with the Model~GRD at the time of the best fit with observational results (10$^{5}$ years). Top: profiles of modeled abundances (top left) and abundances derived as column density ratios (top right) of complex organic species. Abundances derived as column densities are smoothed over the 26$''$ Gaussian beam. Middle: profiles of abundances of major ice constituents (middle left) and selected complex organic molecules in the ice (middle right) w.r.t. H$_2$. Bottom: same as in the middle, but w.r.t. solid H$_2$O. Colored dots in the top right panel denote observational values by \cite{Chacon-Tanarro_ea2019_CH2DOH} for $\rm CH_3OH$ and by \cite{Jimenez-Serra16} for other species, vertical lines are for error bars, arrows are for observational upper limits.} \label{fig:COM_bestfit}
\end{figure*}

Let us now consider modeling results in detail. Figure~\ref{fig:COM_bestfit} presents the modeled fractional abundances of the studied gaseous species (top left) and the abundances derived as the ratios of the smoothed modeled column densities (top right). In Table~\ref{tab:gasCOMs}, modeled abundances for the time of the best agreement are compared to the observed values. In contrast to \cite{Vasyunin17}, our model does not overproduce methanol. At the methanol peak, the model of \cite{Vasyunin17} gives the $\chi({\rm CH_3OH})$ of $2.7 \cdot 10^{-8}$, while our model gives $1.2 \cdot 10^{-9}$, which is closer to $3.9 \cdot 10^{-9}$ reported by \cite{Chacon-Tanarro_ea2019_CH2DOH} based on the IRAM 30m single-dish observations. It should be noted that $\chi({\rm CH_3OH}) = 8.0 \cdot 10^{-9}$ used in \cite{Jimenez-Serra21} was obtained with a $\approx 5 ''$ beam of NOEMA \citep{Punanova_ea2018}. Thus, such results should not be directly compared with the COMs abundances obtained with the 26$''$ beam.

The COMs-to-methanol gas-phase abundance ratios are significantly higher in the present model in comparison to those in~\citet[][]{Vasyunin17}. At 4000~au, the modeled abundances of such COMs as $\rm CH_3OCH_3$, $\rm CH_3CHO$ and $\rm HCOOCH_3$ are 3\%, 4\% and 21\% of the modeled methanol abundance, correspondingly (see Table~\ref{tab:gasCOMs_to_methanol}). In \cite{Vasyunin17}, the gap between modeled abundances of COMs and methanol abundance is generally wider: $\rm HCOOCH_3$ and $\rm CH_3OCH_3$ amount 0.2\% and 0.1\% of the modeled methanol abundance, correspondingly, and $\rm CH_3CHO$ amounts 2\% of $\rm CH_3OH$ abundance. \citet[][]{Scibelli_ea21} report the COMs-to-methanol ratios in prestellar cores derived from observations to be in a range from a few and up to $\approx$~10 percent (see their Figure~12). Our new model fits this range better than the previous one. The modeled radial distances of COMs peak abundances from the core center vary with species. While in the model the HCOOCH$_3$ and NH$_2$CHO peaks coincide with the methanol peak, peaks of CH$_3$CHO and CH$_3$OCH$_3$ are somewhat shifted from the methanol center. Such behavior is related to the formation routes of COMs. Some of the routes are not directly related to methanol (see below).

\begin{deluxetable}{l|rr}
\tablecaption{Observed abundances of COMs in the gas phase $\chi_{\rm obs}({\rm X}) = N_{\rm obs}({\rm X})/N_{\rm obs}({\rm H_2})$ in comparison to modeled abundances of COMs $\chi_{\rm mod}({\rm X}) = N_{\rm mod}({\rm X})/N_{\rm mod}({\rm H_2})$ derived from smoothed column densities. \label{tab:gasCOMs}}
\tablehead{
\colhead{Species}  &  \colhead{$\chi_{\rm obs}({\rm X})$, cm$^{-3}$} & \colhead{$\chi_{\rm mod}({\rm X})$, cm$^{-3}$}
}
\startdata
                & \multicolumn{2}{c}{Dust Peak}                          \\
$\rm CH_3OH$    & $(7.2 \pm 0.8) \cdot 10^{-10}$  & $3.1 \cdot 10^{-10}$ \\
$\rm CH_3O$     & $\leq (5.1-6.7) \cdot 10^{-12}$ & $6.2 \cdot 10^{-12}$ \\
$\rm CH_3CHO$   & $2.2 \cdot 10^{-11}$            & $1.3 \cdot 10^{-11}$ \\
$\rm CH_3OCH_3$ & $(2.8 \pm 0.4) \cdot 10^{-11}$  & $6.8 \cdot 10^{-12}$ \\
$\rm HCOOCH_3$  & $(8.1 \pm 7.4) \cdot 10^{-11}$  & $6.2 \cdot 10^{-11}$ \\
$\rm NH_2CHO$   & $\leq (2.4-3.1) \cdot 10^{-13}$ & $1.6 \cdot 10^{-13}$ \\
\hline
                &  \multicolumn{2}{c}{$\rm CH_3OH$ Peak}                 \\
$\rm CH_3OH$    & $(3.9 \pm 0.4) \cdot 10^{-9}$   & $1.2 \cdot 10^{-9}$  \\
$\rm CH_3O$     & $2.7 \cdot 10^{-11}$            & $8.3 \cdot 10^{-12}$ \\
$\rm CH_3CHO$   & $2.1 \cdot 10^{-10}$            & $5.0 \cdot 10^{-11}$ \\
$\rm CH_3OCH_3$ & $(5.1 \pm 1.1) \cdot 10^{-11}$  & $3.1 \cdot 10^{-11}$ \\
$\rm HCOOCH_3$  & $(1.5 \pm 0.9) \cdot 10^{-10}$  & $2.5 \cdot 10^{-10}$ \\
$\rm NH_2CHO$   & $\leq (6.7-8.7) \cdot 10^{-13}$ & $5.4 \cdot 10^{-13}$ \\
\enddata
\tablecomments{$N_{\rm obs}({\rm X})$ for $\rm CH_3OH$ are obtained from \cite{Chacon-Tanarro_ea2019_CH2DOH}; $N_{\rm obs}({\rm X})$ for the rest species and $N_{\rm obs}({\rm H_2})$ towards the dust peak ($5.4 \cdot10^{22}$~cm$^{-2}$) and the methanol peak ($1.5 \cdot 10^{22}$~cm$^{-2}$) are taken from \cite{Jimenez-Serra16}.}
\end{deluxetable}

\begin{deluxetable}{l|rr|rr}
\tablecaption{Observed and modeled abundances ratios $\chi({\rm X}) / \chi({\rm CH_3OH})$ of selected COMs to $\rm CH_3OH$, \% 
\label{tab:gasCOMs_to_methanol}}
\tablehead{
\colhead{Species}  & \multicolumn{2}{c}{Dust Peak}
& \multicolumn{2}{c}{$\rm CH_3OH$ Peak}
}
\startdata
                &  Observ. & Model. & Observ. & Model. \\
$\rm CH_3CHO$   & $3.1 \pm 0.3$  & $4.2$ & $5.4 \pm 0.6$  & $4.2$ \\
$\rm CH_3OCH_3$ & $3.9 \pm 0.6$  & $2.2$ & $1.3 \pm 0.3$  & $2.6$ \\
$\rm HCOOCH_3$  & $11.3 \pm 10.3$  & $20.0$ & $3.8 \pm 2.4$  & $20.8$ \\
\hline
\enddata
\tablecomments{The observational abundances for $\rm CH_3OH$ are taken from \cite{Chacon-Tanarro_ea2019_CH2DOH}, the observational abundances for the rest species are from \cite{Jimenez-Serra16}.}
\end{deluxetable}

The composition of icy mantles of interstellar grains for the prestellar core L1544 in the current model differs from that in the model by \cite{Vasyunin17}, and may be considered as more reasonable. In Figure~\ref{fig:COM_bestfit}, the modeled fractional abundances $n({\rm X})/n({\rm H_2})$ of key ice components (middle left panel) and of ice COMs (middle right panel) are presented, as well as their fractional abundances with respect to the abundance of the water ice (bottom panels). Table~\ref{tab:ice_ab} summarizes the ice fractional abundances of selected species, their percentage to the $\rm H_2O$ ice abundance and calculated column densities $N({\rm X})$. Here and below, prefix ``g'' denotes species in the surface layers of icy mantles. Prefix ``b'' denotes species in the ice bulk. Finally, prefix ``i'' is used when the total amount of ice species in the ice is considered (``i'' = ``g''+``b''). The CO ice is very abundant at the center of the core, reaching the fractional abundance of 40\% w.r.t. water ice. At the location of the methanol peak, the fraction of CO ice is approximately twice smaller, and further drops to the edge of the core. In contrast to the model by ~\citet[][]{Vasyunin17}, $\rm CO_2$ is now abundant in the ice, reaching the fraction of 21\% w.r.t. water ice at the location of the dust peak and 16\% at the location of the methanol peak. Solid methanol is less abundant in the current model than in the model presented in~\cite{Vasyunin17}. Its abundance is the highest in the center of the core (about 10\% of solid water), and is decreasing towards the core edge, reaching 5\% of solid water at the methanol peak. Nevertheless, methanol has larger abundance in our model than the average abundances reported in similar ISM objects \citep[3\% of water ice toward low-mass protostars, 4\% toward high-mass protostars and cloud cores according to][]{Oeberg11}. However, \cite{Goto21} reported the abundance of methanol ice equal to 11\% of the water ice, consistent with our modeling results. Note however that the position of the background star (Source~3) used for the estimation of methanol ice abundance in~\citet[][]{Goto21} does not coincide neither with the dust peak nor with the methanol peak. The abundance of ammonia ice is approx. 30\% larger than that of methanol ice, and follows the same radial trend. The abundance of solid methane is approximately similar at the locations of dust and methanol peaks: $\approx$2\% of water ice.

\begin{deluxetable}{l|rrr}
\tablecaption{Modeled abundances of the most important ice components and ice COMs in L1544 ($n(\rm{X})$), their ratios to ice $\rm H_2O$ abundance in percentage ($n({\rm X})/n(\rm{iH_2O})$, \%) and calculated column densities $N({\rm X})$. Ratios of column densities are not the same as ratios of abundances since column densities are integrated along the line of sight while the abundances belong to a certain spatial position in the core. \label{tab:ice_ab}}
\tablehead{
\colhead{Species} &   \colhead{$n(\rm{X})$, cm$^{-3}$}   & \colhead{$n({\rm X})/n(\rm{iH_2O})$, \%} & \colhead{$N({\rm X})$, cm$^{-2}$}
}
\startdata
                  &   \multicolumn{3}{c}{Dust Peak} \\
$\rm iH_2O$       &   $7.2 \cdot 10^{-5}$     & 100.00 & $1.2 \cdot 10^{19}$\\
$\rm iCO$         &   $2.9 \cdot 10^{-5}$     & 40.86  & $4.5 \cdot 10^{18}$\\
$\rm iCO_2$       &   $1.5 \cdot 10^{-5}$     & 21.07  & $2.5 \cdot 10^{18}$\\
$\rm iCH_4$       &   $1.6 \cdot 10^{-6}$     & 2.27   & $2.7 \cdot 10^{17}$\\
$\rm iNH_3$       &   $9.6 \cdot 10^{-6}$     & 13.44  & $1.6 \cdot 10^{18}$\\
$\rm iCH_3OH$     &   $7.5 \cdot 10^{-6}$     & 10.43  & $1.1 \cdot 10^{18}$\\
$\rm iCH_3CHO$    &   $7.3 \cdot 10^{-8}$     & 0.10   & $1.2 \cdot 10^{16}$\\
$\rm iCH_3OCH_3$  &   $1.8 \cdot 10^{-8}$     & 0.02   & $4.0 \cdot 10^{15}$\\
$\rm iHCOOCH_3$   &   $8.3 \cdot 10^{-7}$     & 1.15   & $2.0 \cdot 10^{17}$\\
$\rm iNH_2CHO$    &   $6.6 \cdot 10^{-8}$     & 0.09   & $1.5 \cdot 10^{16}$\\
\hline
                  &   \multicolumn{3}{c}{$\rm CH_3OH$ Peak} \\
$\rm iH_2O$       &   $6.6 \cdot 10^{-5}$     & 100.00 & $7.5 \cdot 10^{17}$\\
$\rm iCO$         &   $1.5 \cdot 10^{-5}$     & 22.66  & $1.5 \cdot 10^{17}$\\
$\rm iCO_2$       &   $1.1 \cdot 10^{-5}$     & 16.51  & $1.5 \cdot 10^{17}$\\
$\rm iCH_4$       &   $1.2 \cdot 10^{-6}$     & 1.81   & $1.3 \cdot 10^{16}$\\
$\rm iNH_3$       &   $5.9 \cdot 10^{-6}$     & 8.99   & $4.5 \cdot 10^{16}$\\
$\rm iCH_3OH$     &   $3.3 \cdot 10^{-6}$     & 4.95   & $2.7 \cdot 10^{16}$\\
$\rm iCH_3CHO$    &   $4.5 \cdot 10^{-8}$     & 0.07   & $5.4 \cdot 10^{14}$\\
$\rm iCH_3OCH_3$  &   $5.4 \cdot 10^{-8}$     & 0.08   & $8.0 \cdot 10^{14}$\\
$\rm iHCOOCH_3$   &   $1.7 \cdot 10^{-6}$     & 2.53   & $1.3 \cdot 10^{16}$\\
$\rm iNH_2CHO$    &   $1.9 \cdot 10^{-7}$     & 0.29   & $2.0 \cdot 10^{15}$\\
\enddata
\tablecomments{The prefix ``i'' denotes the total amount of a species in icy grain mantles, which is the sum of surface and bulk ice abundances. The calculated $\rm H_2$ column density is $8.5 \cdot 10^{22}$ for the dust peak and $7.9 \cdot 10^{21}$ for the methanol peak. }
\end{deluxetable}

Such ice COMs as acetaldehyde, dimethyl ether and formamide have abundances in the range of $10^{-8}-10^{-7}$ at the dust peak (0.1\% or less w.r.t. water ice). Methyl formate is the most abundant modeled COM in the ice. At the dust peak, its abundance amounts to $8.3 \cdot 10^{-7}$ (1\% w.r.t. water ice) and is comparable to methane ice abundance (which is $1.6 \cdot 10^{-6}$). Solid-phase acetaldehyde demonstrates some decrease towards the methanol peak, while the three other ice COMs have peaks towards the core edge.

In contrast to \cite{Vasyunin17}, where all the ice components have fractional abundances below $10^{-7}$ at the edge of the core, non-negligible ice thickness is found now. In our simulations, the fractional abundance roughly corresponding to one surface monolayer is about $2 \cdot 10^{-6}$. At the core edge, water and carbon dioxide ices show the modeling abundances of $1.2 \cdot 10^{-5}$ and $5.1 \cdot 10^{-6}$, correspondingly. This difference in ice thickness is likely due to different Phase 1 of simulation: now we adopt a ``translucent cloud'' with the hydrogen number density of $10^3\ \rm cm^{-3}$ and the temperature linearly decreasing from 15 K to 10 K, while previously it was a ``diffuse cloud'' with a smaller gas number density of $10^2\ \rm cm^{-3}$ and constant temperature of 20~K (the visual extinction is the same for both cases and equals 2 mag).

\begin{figure} [ht!]
  \includegraphics[width=1.0\linewidth]{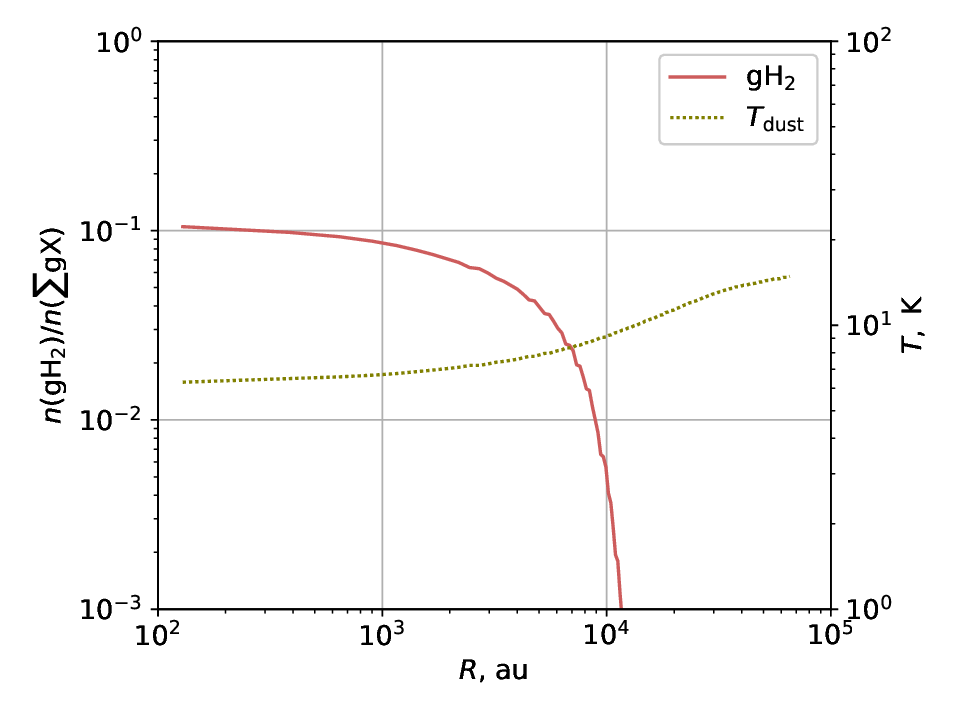}
\caption{Fraction of molecular hydrogen on ice surface vs. radius. The value gets balanced almost instantaneously, and exhibits no appreciable changes throughout the simulation.} \label{fig:surf_h2_fraction}
\end{figure}

Molecular hydrogen amounts $\approx$~10\% of total surface coverage at the central area of the core, with a decrease to the core edge (Figure~\ref{fig:surf_h2_fraction}). This results in a noticeable change of H and H$_2$ binding energies, which, in turn, affects both surface chemistry and reactive desorption rates. The high surface abundance of H$_2$ also enables chemical reactions with molecular hydrogen. As it is discussed below, such reactions are important for the formation of solid water and methane.

The time profiles for the COMs abundances obtained with our GRD model at times relevant for the astrochemical modeling of starless cores are given in Appendix~\ref{sec:time_dependence}.

\subsection{Chemistry at the Methanol Peak with the GRD model}
\label{subsec:chemistry_Garrod}
In this section, the underlying chemical processes at the position of the methanol peak as observed at 4000 au are discussed. All the numbers are given for the moment of the best agreement between the observed values and modeling results obtained with the GRD model.

It should be noted that diffusive reactions on the grain surface efficient at $T_{\rm dust}$~$\sim$~10~K, such as H-addition, also proceed efficiently in a non-diffusive mode. Thus, non-diffusive mechanisms effectively accelerate the rates of diffusive reactions at low temperatures. As an example, surface H-addition reactions forming COMs and methanol become about 20\% faster due to the introduction of non-diffusive mechanisms. We now describe the main formation routes observed in our model for the following molecular species: $\rm CH_3OH,\ HCOOCH_3,\ CH_3CHO,\ CH_3OCH_3,\ NH_2CHO$ and the major ice components $\rm H_2O,\ CO_2,\ CH_4,\ NH_3$.

$CH_3OH$. Methanol in the model is mainly produced on grains as a result of hydrogenation of CO molecules via the H-atom addition diffusive reactions presented in Table~\ref{tab:CH3OH_chain}, Appendix~\ref{subsec:network_updates}. This chain of reactions is also the main source of intermediate radicals, HCO, CH$_2$OH and CH$_3$O, that play a key role in the formation of complex organic molecules via radical-radical chemical reactions that proceed non-diffusively (see below). The hydrogen abstraction reactions that are now included in the CO hydrogenation sequence increase surface production of those radicals. The reactions also increase the abundance of methanol in the gas phase by a factor of two in comparison to the test model run with hydrogen abstraction reactions switched off, because multiple acts of formation and destruction of CH$_3$OH ultimately increase the probability of RD for methanol molecules. This is also true for its precursor species. The non-diffusive reaction $\rm gCH_3O + gCH_3O \rightarrow gCH_3OH +  gH_2CO$ accounts only for 4\% of surface methanol production.

One of the intermediates in the methanol formation chain, formaldehyde, can also be formed in our model via the reaction $\rm gC + gH_2O \rightarrow gH_2CO$ \citep[][]{Molpeceres21}. However, its role was found to be negligible in producing formaldehyde ice, except for the outer edge of the core. The reaction also does not impact gas-phase formaldehyde abundance. Note however, that under the conditions of the translucent cloud where more atomic carbon is available, the non-diffusive version of this reaction produces most of $\rm gH_2CO$ at the early times of translucent cloud evolution. To the end of the translucent cloud phase in our simulations, the abundance of $\rm H_2CO$ ice amounts to $\approx 10^{-7}$, which is about an order of magnitude less than one ice monolayer.

Gas-phase processes account for 10\% of methanol production. Among the major gas-phase routes there is dissociative recombination of protonated or ionized COMs such as HCOOCH$_3$ and CH$_3$OCH$_3$. 

In L1544, the gas-phase methanol peak lies about 4000 au from the continuum “dust peak,” where CO is most heavily depleted. The major source of methanol in the gas phase according to our model is reactive desorption of solid state methanol during its formation on surface of grain ice mantles. There, CH$_3$OH is formed via successive hydrogenation of CO molecules accreting from the gas. This hydrogenation must be completed while the reactants remain in the surface layer of ice; once they are buried in the bulk ice, reactive desorption can no longer occur. Thus, methanol ejected to the gas phase is formed in reactions involving CO molecules that have only recently accreted onto the ice surface. The ``methanol peak'' therefore marks the radius where the accretion rate of CO is high and atomic hydrogen on surface is sufficiently abundant, not the region containing the largest reservoir of CO buried in the ice bulk.

$HCOOCH_3$. Almost all methyl formate, both in the gas phase and in the icy mantles of grains, is surface-formed. It is ejected to gas via RD. In our model, $\rm HCOOCH_3$ is a product of the surface reaction $\rm gHCO + gCH_3O \rightarrow gHCOOCH_3$ that proceeds non-diffusively. Its RD efficiency is 0.02\%. The reactive desorption rate is calculated according to \cite{Garrod07} and multiplied by the non-$\rm H_2O$ surface fraction, which equals 65\% at $\approx 10^5$~years and consists of 31\% gCO and 15\% $\rm gCO_2$ at the methanol peak position. The RD efficiency of the reaction is effectively multiplied by a factor of $\approx$~3 due to the hydrogen abstraction/addition loop $\rm gHCOOCH_3 + \rm H \rightarrow gCH_3OCO + \rm H_2;\ \rm gCH_3OCO + \rm H\rightarrow \rm HCOOCH_3$.

Although the gas-phase formation routes of methyl formate proposed in previous studies are also present in this work, their contribution is minor in comparison to the surface route. The reaction $\rm CH_3OCH_2 + O \rightarrow HCOOCH_3 + H$, which was incorporated into astrochemical modeling by \cite{Balucani15}, contributes 0.4\% of total $\rm HCOOCH_3$ production rate. In \cite{Vasyunin17}, an important route of protonated methyl formate production is the reaction $\rm (CH_3OH)H^+ + HCOOH \rightarrow (HCOOCH_3)H^+ + H_2O$. In this work, the contribution of this reaction to the total formation rate of methyl formate is 0.1\%.  

$CH_3OCH_3$. Dimethyl ether in the model is mainly produced on grains via two channels. First is the reaction $\rm gCH_3 + gCH_3O \rightarrow gCH_3OCH_3$ that proceeds non-diffusively. The probability of RD for dimethyl ether formed in this reaction is 0.03\%. However, this value is further increased by a factor of $\approx$~6 via the hydrogen abstractions/addition loop: $\rm gCH_3OCH_3 + \rm gH \rightarrow \rm gCH_3OCH_2 + \rm gH_2;\ \rm gH + \rm gCH_3OCH_2 \rightarrow \rm gCH_3OCH_3$. The second channel is the non-diffusive reaction $\rm gCH_2 + \rm gCH_3O \rightarrow \rm gCH_3OCH_2$ with subsequent addition of a hydrogen atom. Thus, the second channel intervenes with the aforementioned hydrogen abstraction/addition loop by producing additional gCH$_3$OCH$_2$ radical. Therefore, the reaction $\rm gH + \rm gCH_3OCH_2 \rightarrow \rm gCH_3OCH_3$ is not merely a part of a loop, but also a reaction responsible for the partial production of $\rm gCH_3OCH_3$. Loop-amplified RD of surface-formed 
dimethyl ether is the major source of CH$_3$OCH$_3$ in the gas phase, that accounts for 99\% of its ``production rate'' in the gas.

Gas-phase processes account for less than 1\% of dimethyl ether production. The most efficient gas-phase formation routes of dimethyl ether are the dissociative recombination of $\rm (CH_3OCH_3)H^+$, which is in turn produced in the radiative association of CH$_3$OH and CH$_{3}^{+}$ ion, and neutral-neutral reaction $\rm CH_3O + CH_3 \rightarrow CH_3OCH_3$. The role of those processes is suppressed in comparison to previous works. The formation of protonated dimethyl ether is less efficient than in, e.g.,~\citet[][]{Vasyunin17} because the abundance of gas-phase methanol in this work is significantly lower in comparison to that previous study. For the reaction $\rm CH_3O + CH_3 \rightarrow CH_3OCH_3$, in this work we adopt the ``phase-space'' reaction rate recently calculated in a detailed study by~\citet[][]{Tennis21}. This rate is an order of magnitude smaller than that used previously in \citet[][]{Balucani15} and \citet[][]{Vasyunin17}. Therefore, the role of this reaction in formation of dimethyl ether is correspondingly reduced.

$CH_3CHO$. Gas-phase acetaldehyde is mainly produced in surface processes too. The major productive reaction is $\rm gCH_3 + gHCO \rightarrow gCH_3CHO$ that proceeds non-diffusively. The probability of RD for acetaldehyde formed in this reaction is 0.1\%. The loop of H-addition/abstraction reactions $\rm gH + gCH_3CO \rightarrow gCH_3CHO$ and $\rm gH + gCH_3CHO \rightarrow gH_2 + gCH_3CO$ effectively increases the dust-to-gas acetaldehyde transfer rate by 10 times. Another route of $\rm gCH_3CO$ and following $\rm gCH_3CHO$ formation (approx. 25\% of total production) is the chain $\rm gCO \rightarrow gCCO \rightarrow gHC_2O \rightarrow gCH_2CO \rightarrow gCH_3CO \rightarrow gCH_3CHO$ experimentally verified by \cite{Fedoseev22} (and a similar chain for the bulk of grain mantles). The reactions of H-addition proceed diffusively with the acceleration by similar non-diffusive processes. The probability of RD for CH$_3$CHO formed in the last reaction of this chain is 0.1\%. CCO is produced in the non-diffusive reaction $\rm gC + gCO \rightarrow gCCO$. In the bulk, the non-diffusive reaction $\rm bO + bC_2 \rightarrow bCCO$ is also efficient; however, its rate is 4 times lower than the rate of the reaction $\rm bC + bCO \rightarrow bCCO$. At the methanol peak position, $\rm bCH_2CO$ is efficiently produced via the photodissociation of ethanol ice by cosmic-ray photons.

In total, 98\% of the delivery of acetaldehyde to the gas phase is due to RD of acetaldehyde formed on the surface in the two reactions mentioned above. Gas-phase processes account for only about 2\% of acetaldehyde production. The fastest gas-phase reaction with acetaldehyde as a product is the dissociative recombination of protonated acetaldehyde $\rm (CH_3CHO)H^+$. Protonated acetaldehyde is formed in the reaction $\rm H_3O^+ + C_2H_2 \rightarrow (CH_3CHO)H^+$. The reaction $\rm O + C_2H_5 \rightarrow CH_3CHO + H$ also producing acetaldehyde is about 10 times slower than the dissociative recombination of $\rm (CH_3CHO)H^+$. 

$NH_2CHO$. Gas-phase route of formamide formation $\rm NH_2 + H_2CO \rightarrow NH_2CHO + H$ proposed in the OSU database~\citep[][]{Garrod_ea08, Barone_ea15} was shown later to be inefficient~\citep[][]{SongKaestner16, Douglas_ea22}. On the other hand, the efficiency of the surface formation route of formamide through the reaction $\rm gNH_2 + gHCO \rightarrow gNH_2CHO$ has recently been confirmed experimentally~\citep[][]{Fedoseev_ea2016, Chuang_ea22}. Therefore, in this study, this surface reaction is the only route to produce formamide. At low temperatures, it proceeds non-diffusively. The probability of RD for the NH$_2$CHO formed in this reaction is 0.02\%. In our model, this is the only channel of delivery of formamide to the gas phase in the cold environment of the prestellar core L1544.

According to the performed simulations, methyl formate is the most abundant COM in the ice. The non-diffusive reaction $\rm gHCO + gCH_3O \rightarrow gHCOOCH_3$ and its bulk analog provide the major fraction of $\rm HCOOCH_3$ ice. The combined rate of these reactions is 9 times faster than the total rate of the non-diffusive reaction $\rm gCH_3 + gHCO \rightarrow gCH_3CHO$ and its bulk analog (both producing $\rm CH_3CHO$ ice) and 20 times faster than the total rate of the non-diffusive reaction $\rm gCH_3 + gCH_3O \rightarrow gCH_3OCH_3$ and its bulk analog (both producing $\rm CH_3OCH_3$ ice). Although all the reactions are barrierless, the abundances of the reactants differ significantly --- $2.5 \cdot 10^{-7}$ for $\rm gHCO$ ($4.8 \cdot 10^{-6}$ for $\rm bHCO$), $7.9 \cdot 10^{-8}$ for $\rm gCH_3O$ ($2.2 \cdot 10^{-6}$ for $\rm bCH_3O$), and only $10^{-14}$ for $\rm gCH_3$ ($7.7 \cdot 10^{-7}$ for $\rm bCH_3$) at $\approx 10^5$~years. As it was mentioned before, the ice radicals HCO and $\rm CH_3O$ are efficiently produced in the methanol formation chain. The $\rm CH_3$ ice is mainly produced via the methoxy radical photodissociation reactions (including reactions with cosmic-ray-induced photons), along with the $\rm CH_3OH$ ice photodissociation by cosmic ray-induced photons and the diffusive reaction $\rm gH_2 + gCH_2 \rightarrow gCH_3 + gH$. The contribution of the reaction $\rm gH_2 + gCH \rightarrow gCH_3$ is minor. The production rates for $\rm CH_3$ ice are significantly lower than those for the $\rm HCO$ and $\rm CH_3O$ ice radicals.

\begin{figure*}
\begin{center}
\includegraphics[width=0.9\linewidth]{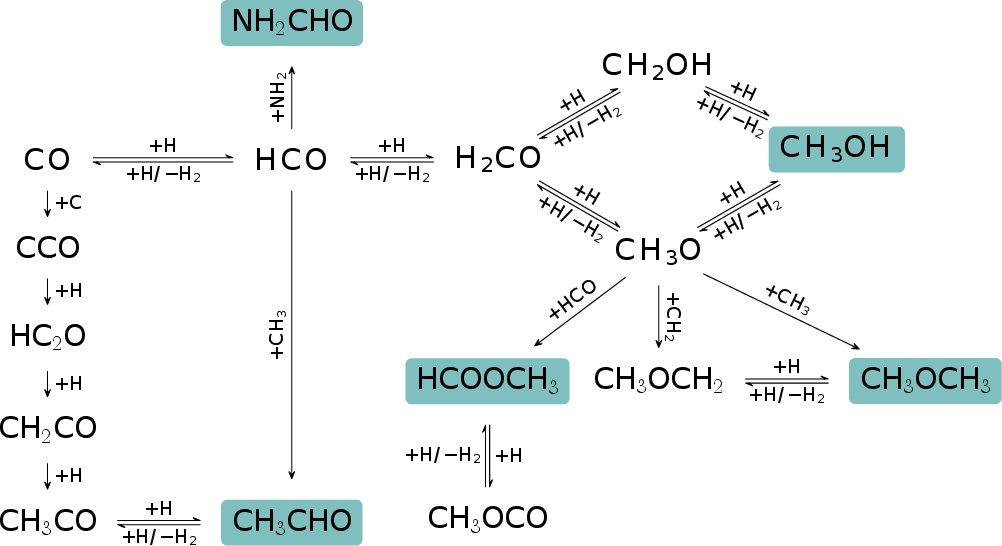}
\end{center}
\caption{The major COMs formation routes on grain surface. The COMs studied in our simulations have teal background. The species involved in the methanol formation chain are highlighted in bold. The chemical desorption in the COMs surface formation reactions is the main source of the gaseous COMs. The expressions $+\rm H$ and $+{\rm H}$/$-{\rm H_2}$ together denote a pair of reactions, H-atom addition and H-atom abstraction.} \label{fig:reactions_scheme}
\end{figure*}

The main grain surface paths of the COMs formation are presented in Figure~\ref{fig:reactions_scheme}. In the gas, COMs appear due to reactive desorption during the final step of their formation.

{\it Chemistry of main ice components.} In this work, accretion of H$_2$ molecules on interstellar grains is considered. Given the large abundance of H$_2$, it is not surprising that chemical reactions with it play an important role in the formation of some ice species. The most abundant ice is solid water, and the main $\rm gH_2O$ formation path is the diffusive reaction $\rm gH_2 + gOH \rightarrow gH_2O + gH$, which accounts for 84\% of surface water production at the position of the observational gas-phase methanol peak \citep{Oba_ea2012}. The non-diffusive H-atom abstraction reactions by OH radical from $\rm gH_2CO$, $\rm gHCO$, $\rm gCH_3O$ and $\rm gCH_2OH$ account for 15\% of $\rm gH_2O$ production in total. As for CO ice, like in many similar models, CO predominantly freezes out from gas.

$\rm CO_2$ ice is produced on grains entirely due to non-diffusive processes. The main source of $\rm gCO_2$ is the non-diffusive reaction $\rm gOH + gCO \rightarrow gCO_2 + gH$ (89\% of its total production rate). In our model, this reaction is a simplified representation for the two-step process $\rm CO + OH\rightarrow HOCO$,
$\rm HOCO + H\rightarrow CO_2 + H_2$~\citep[see][]{Qasim_ea2019}. This channel remains significant even at larger radii, and despite photodissociation, the carbon dioxide ice maintains a non-negligible abundance even at the core edge. Another channel is the non-diffusive reaction $\rm gO + gHCO \rightarrow gCO_2 + gH$ (10\%). The bulk analogs of the aforementioned reactions have rates about 5 times slower than the surface rates.

The source of $\rm gNH_3$ is the diffusive reaction $\rm gH_2 + gNH_2 \rightarrow gNH_3 + gH$ (99\% of its production rate). This reaction has an activation barrier of $\approx$~3000~K~\citep[][]{LiGuo14}. However, it is efficient due to a large abundance of H$_2$ on the surface and quantum tunneling of H$_2$ through the activation barrier. Although barrierless, the surface reaction $\rm gH + gNH_2 \rightarrow gNH_3$ appears to be inefficient because of a low abundance of atomic hydrogen on the surface in comparison to the surface abundance of H$_2$.

More than half of $\rm CH_4$ is delivered to grain surfaces by freeze-out from gas (65\% of its formation rate) where it is formed in ion-molecule reactions. The main surface reaction producing $\rm gCH_4$ is the diffusive reaction $\rm gH_2 + gCH_3 \rightarrow gCH_4 + gH$ (34\% of $\rm gCH_4$ formation rate). $\rm gCH_3$ is produced in a reaction $\rm gCH_2 + gH_2 \rightarrow gCH_3 + gH$. In turn, the main formation route of $\rm gCH_2$ is the reaction between hydrogen molecules and atomic carbon: $\rm gC + gH_2 \rightarrow gCH_2$. Bulk diffusive and non-diffusive reactions $\rm bH + bCH_3 \rightarrow bCH_4$ have rates similar to the rate of the aforementioned surface reaction producing $\rm gCH_4$.

\begin{figure*} [p!]
  \includegraphics[width=0.5\linewidth]{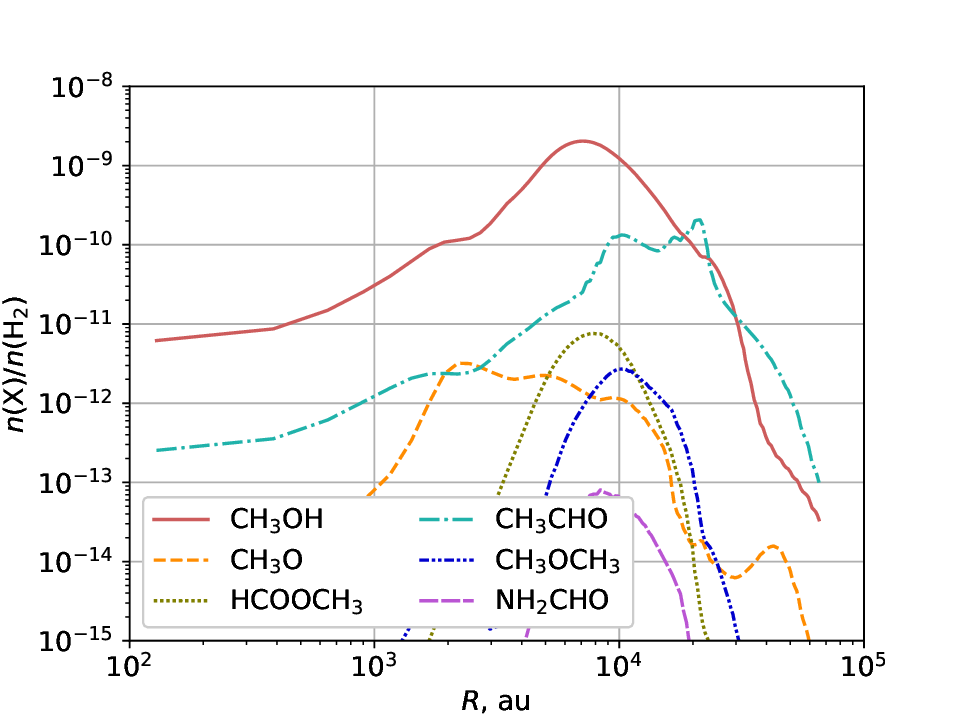}
  \includegraphics[width=0.5\linewidth]{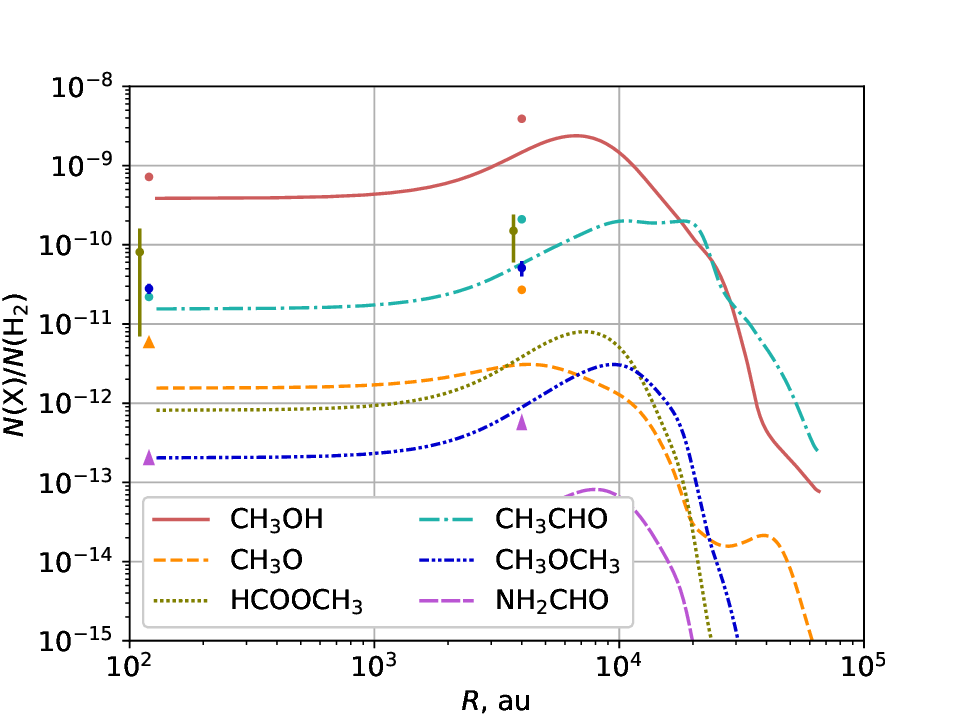} \\
  \includegraphics[width=0.5\linewidth]{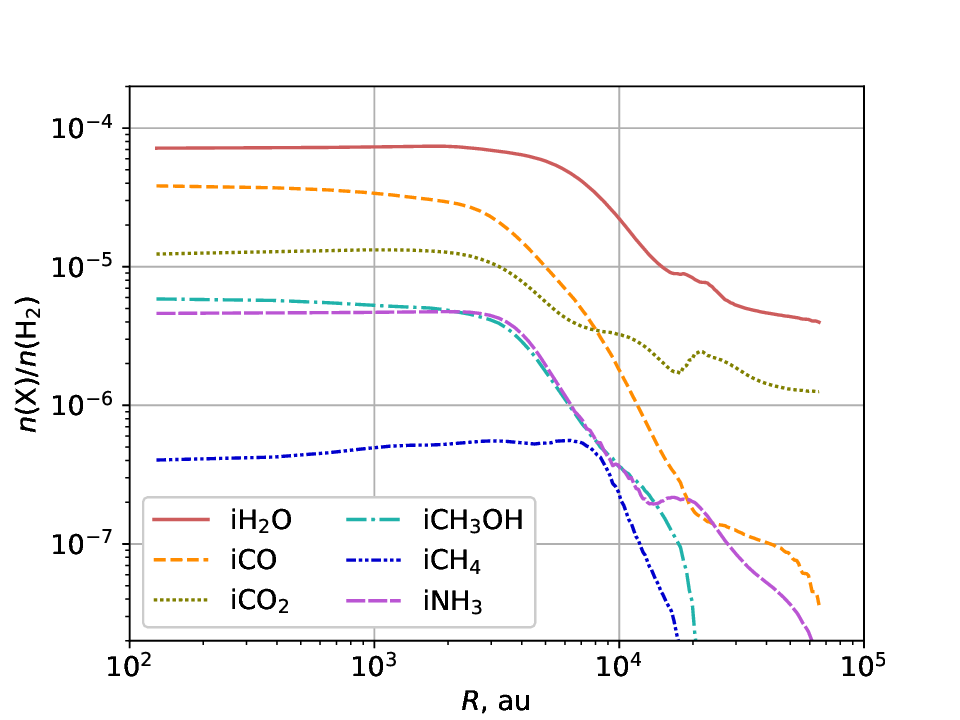}
  \includegraphics[width=0.5\linewidth]{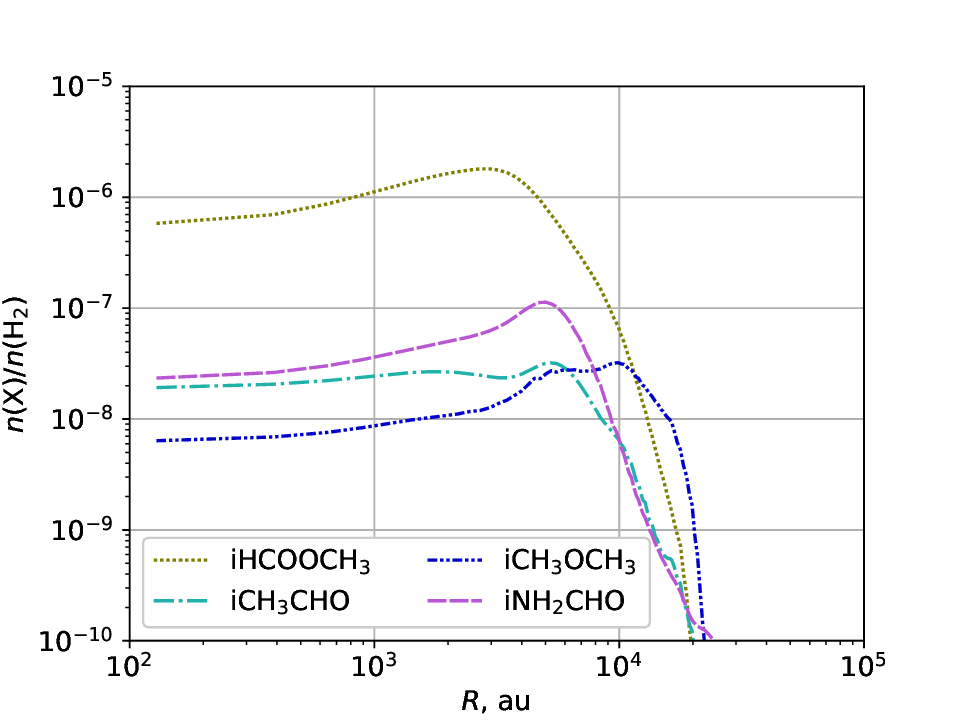} \\
  \includegraphics[width=0.5\linewidth]{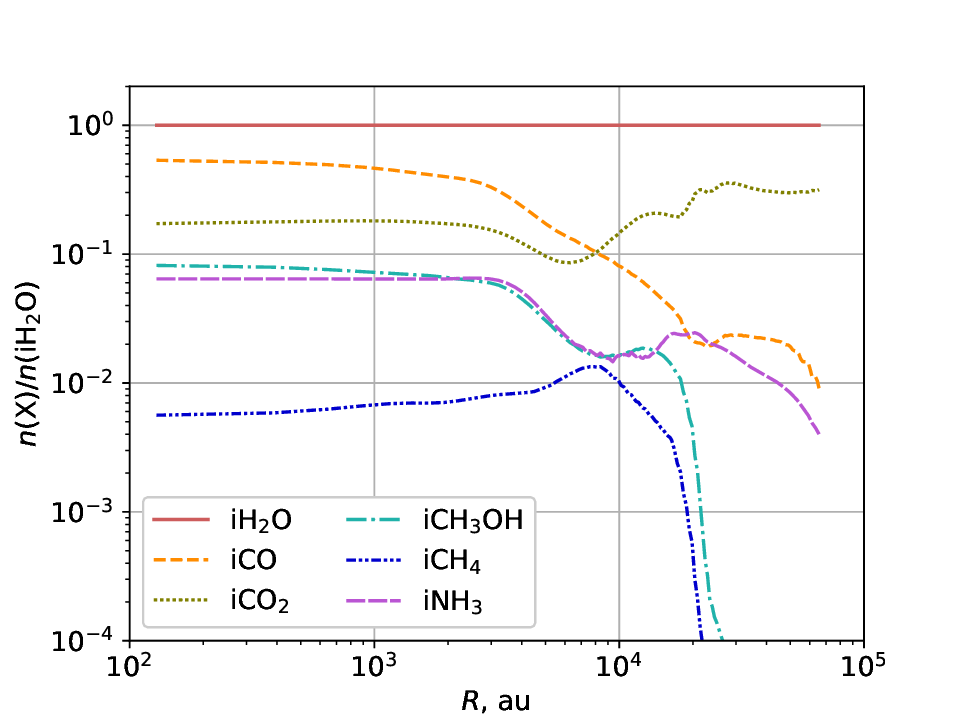}
  \includegraphics[width=0.5\linewidth]{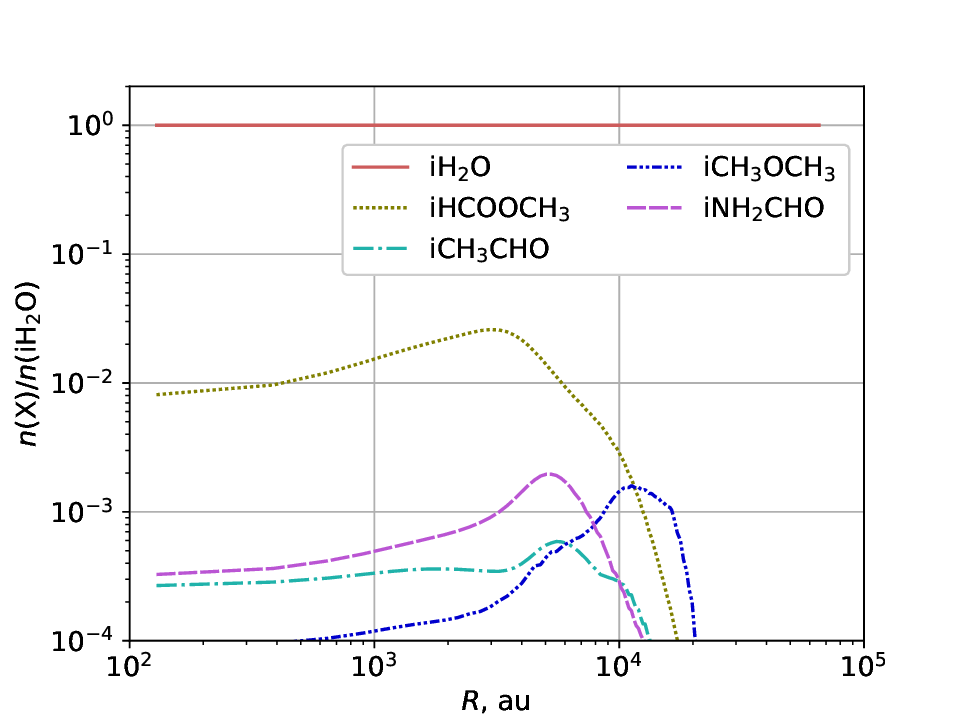} \\
\caption{Radial profiles for the MRD model with reactive desorption treatment following \cite{Minissale16} at 10$^{5}$ years. Top: profiles of modeled abundances (top left) and abundances derived as column density ratios (top right) of complex organic species. Abundances derived as column densities are smoothed over the 26$''$ Gaussian beam. Middle: profiles of abundances of major ice constituents (middle left) and selected complex organic molecules in the ice (middle right) w.r.t. H$_2$. Bottom: same as in the middle, but w.r.t. H$_2$O. Colored dots in the top right panel denote observational values by \cite{Chacon-Tanarro_ea2019_CH2DOH} for $\rm CH_3OH$ and by \cite{Jimenez-Serra16} for other species, vertical lines are for error bars, arrows are for observational upper limits.
} \label{fig:COM_Minissale}
\end{figure*}

\subsection{Comparison to the Results and Chemistry with RD by \cite{Minissale16}}
\label{subsec:compare_to_Minissale}
In the previous section, we presented the results of the GRD model that employs the parametrization of reactive desorption based on the RRKM theory and discussed in e.g.~\citet[][]{Garrod07}. On the other hand, in the previous work where both the observed abundances of COMs as well as the location of the ``methanol peak'' in L1544 were simultaneously reproduced~\citep{Vasyunin17}, RD was parametrized differently. They employed an original parametrization of the results of laboratory experiments proposed in~\cite{Minissale16}. Generally, two mentioned parametrizations provide very different efficiencies of RD for the same reacting systems. Given the large uncertainty on the efficiency of RD for many reacting systems, we made a dedicated attempt to investigate the dependence of gas-phase COMs abundances and ice composition on the chosen parametrization of RD. For that, we constructed a model similar to the above-described GRD model but with the parametrization of RD following~\citet[][]{Minissale_ea16b}. Below, it will be referred to as MRD model.

We ran the MRD model for the same conditions as GRD model described above. We explored different values of $E_{\rm diff}/E_{\rm des}$ for atomic and molecular species in both models (see Section~\ref{subsec:EbEd_grid} below for details). In our MRD model, no agreement for all species simultaneously has been found. Thus, for comparison with the GRD model, we have chosen the same $E_{\rm diff}/E_{\rm des}$ as in our main GRD model to the MRD model. The results were compared for the moment of time when the best agreement of the regular GRD model is achieved, i.e.,~$10^5$~years. 

The results of the MRD model are presented in Figure~\ref{fig:COM_Minissale}. The abundances of gaseous methyl formate and dimethyl ether are lower than the observed values by about one--two orders of magnitude (Fig.~\ref{fig:COM_Minissale}, top panels). Efficiencies of chemical desorption calculated according to \citet[][]{Minissale_ea16b} for the reactions $\rm gCH_3 + gCH_3O \rightarrow gCH_3OCH_3$ and $\rm gHCO + gCH_3O \rightarrow gHCOOCH_3$ are very low. Those are of the order of $10^{-7}$ and $10^{-14}$ correspondingly, with respect to corresponding surface reaction rates. H-addition surface reactions which produce these COMs in loops demonstrate similar RD efficiencies. Thus, in the case of MRD, surface processes cannot supply enough quantities of $\rm HCOOCH_3$ and $\rm CH_3OCH_3$ to the gas phase to match observations. The abundance of acetaldehyde (CH$_3$CHO) is similar in the GRD and MRD models, because efficiencies of RD for surface formation routes of acetaldehyde are similar in both the GRD and MRD models. As for formamide, the efficiency of chemical desorption for the reaction $\rm gNH_2 + gHCO \rightarrow gNH_2CHO$ is negligible in the MRD model. Thus, in the absence of efficient gas-phase production paths in both models, gaseous formamide exhibits abundances lower than $10^{-13}$ in the MRD model. The RD efficiencies for both the GRD and MRD models are given in Table~\ref{tab:RD_efficiency}. The shown calculated RD values are multiplied by the non-$\rm H_2O$ surface fraction, which is about 65\% in the GRD model and 70\% in the MRD model at the methanol peak position.

\begin{deluxetable}{l|r|r}\label{tab:RD_efficiency}
\tablecaption{Reactive desorption efficiency (in \% of the products amounts) for the most important COMs forming reactions in the GRD and MRD models.} 
\tablehead{
\colhead{Reaction} & \colhead{GRD, \%} & \colhead{MRD, \%}
}
\startdata
 $\rm gCH_2OH + gH \rightarrow gCH_3OH$          & 0.05 & 0.03  \\
 $\rm gCH_3O + gH \rightarrow gCH_3OH$           & 0.06 & 0.04  \\
 \hline
 $\rm gHCO + gCH_3O \rightarrow gHCOOCH_3$       & 0.02 & $\sim 10^{-12}$  \\
 $\rm gCH_3OCO + gH \rightarrow HCOOCH_3$        & 0.02 & $\sim 10^{-12}$  \\
 \hline
 $\rm gCH_3 + gCH_3O \rightarrow gCH_3OCH_3$     & 0.03 & $\sim 10^{-5}$  \\
 $\rm gCH_3OCH_2 + gH \rightarrow gCH_3OCH_3$    & 0.04 & $\sim 10^{-4}$  \\
 \hline
 $\rm gCH_3 + gHCO \rightarrow gCH_3CHO$         & 0.1  & 0.05  \\
 $\rm gCH_3CO + gH \rightarrow gCH_3CHO$         & 0.1  & 0.07  \\
 \hline
 $\rm gNH_2 + gHCO \rightarrow gNH_2CHO$         & 0.02 & $\sim 10^{-6}$  \\
\enddata
\tablecomments{In both GRD and MRD cases, the calculated RD values are already multiplied by the non-$\rm H_2O$ surface fraction.}
\end{deluxetable}

In contrast to the gas phase, the ice composition does not experience dramatic variations with the change of the chemical desorption treatment. At the central parts of the core, all ice species exhibit changes in abundances within an order of magnitude compared to the ones obtained with GRD (Fig.~\ref{fig:COM_Minissale}, middle and bottom panels). The most noticeable changes concern ices such as $\rm NH_3$, $\rm CH_4$, $\rm CH_3CHO$, $\rm CH_3OCH_3$ and $\rm NH_2CHO$, whose abundances become lower by about a factor of 2 in comparison to the Model~GRD. In particular, the chemical desorption for $\rm gNH_2$ formation via the reaction $\rm gH + gNH \rightarrow gNH_2$ is about 20 times more efficient in the Model~MRD than in Model~GRD. $\rm gNH_2$ is a precursor of both $\rm gNH_3$ and $\rm gNH_2CHO$. Thus, in the Model~MRD a larger fraction of gNH$_2$ is lost to the gas causing lower abundances of ammonia and formamide in the ice. The precursor of $\rm gCH_4$, $\rm CH_3OCH_3$ and $\rm gCH_3CHO$ --- $\rm gCH_3$ --- is formed more efficiently in the case of the Model~GRD. Its major production path is the reaction $\rm gH_2 + gCH_2 \rightarrow gCH_3 + gH$, and the abundance of $\rm gCH_2$ is greater in the case of the Model~GRD because the chemical desorption in its formation path $\rm gH_2 + gC \rightarrow gCH_2$ is about 40 times less efficient for the Model~GRD than for the Model~MRD. Thus, unless the rates of chemical desorption are quite low compared to the rates of original surface reactions, the changes in these rates may moderately affect the abundances of ice constituents too.

\section{Discussion} \label{sec:discussion}
In this work, we reproduced simultaneously the abundances of complex organic molecules observed in the L1544 prestellar core as well as the approximate location of the methanol peak with respect to the dust peak of the core. The modeled distance between the $\rm CH_3OH$ peak and the dust peak is somewhat larger than the observed ($\approx$~6000~au vs. $\approx$~4000~au). This difference may arise from the fact that we are assuming spherical symmetry while, in reality, L1544 is not spherically symmetric.

Moreover, our model presented in this study provides the composition of ices in L1544 that can be considered as more reasonable than in previous studies~\citep[e.g.,][]{Vasyunin17}: the fraction of methanol ice is close to the values obtained observationally by~\citet[][]{Goto21} for L1544, and the fraction of CO$_2$ ice is similar to that obtained by JWST for other dense dark clouds~\citep[see][]{McClure_ea23}. The modeled abundances of COMs in the ices are similar to the abundances of COMs in the gas phase of hot cores/corinos~\citep[see e.g.][]{Bottinelli_ea04}. The modeled abundances of icy COMs obtained in this study are compared to the corresponding gas-phase abundances in hot cores/corinos in Table~\ref{tab:ice_vs_hotcores}. It can be seen that the abundances of icy COMs in most cases are equal to or higher than the abundances of organic molecules in the gas phase of hot cores/corinos. It may imply that COMs observed in the gas phase of later, more advanced stages of star formation, represented by hot cores/corinos, may already be formed during the earliest stages of protostellar development, in prestellar cores. Chemistry that proceeds in cold ices via non-diffusive mechanisms therefore appears to exhibit similar efficiency in the production of COMs as diffusive radical-radical chemistry that occurs in warm ($T>30$~K) ices during the development of hot cores/corinos as proposed in~\citet[][]{Garrod_ea06}. The smaller abundance of organic molecules in the gas phase of hot cores compared to the ice of the cold core is consistent with the rapid destruction of COMs in gas-phase reactions upon sublimation. Untangling warm-up-driven and cold formation routes of COMs would require isotopic analysis as proposed in~\citet[][]{Jorgensen_ea18}. This is out of scope of this work.

\begin{deluxetable*}{lrrrrl}
\tablecaption{Comparison of the modeled abundances of ice COMs in L1544, $n(\rm{X})$, and the observational abundances of these COMs in the gas-phase of hot cores/corinos. \label{tab:ice_vs_hotcores}}
\tablehead{
\colhead{} & \colhead{$\rm CH_3OH$} &   \colhead{$\rm CH_3CHO$}
           & \colhead{$\rm CH_3OCH_3$} & \colhead{$\rm HCOOCH_3$} & \colhead{Reference}
}
\startdata
\multicolumn{6}{c}{\textbf{ice COMs in L1544}} \\
Dust Peak         & $7.5 \cdot 10^{-6}$ & $7.3 \cdot 10^{-8}$ & $1.8 \cdot 10^{-8}$ & $8.3 \cdot 10^{-7}$ & This study, modeled\\
$\rm CH_3OH$ Peak & $3.3 \cdot 10^{-6}$ & $4.5 \cdot 10^{-8}$ & $5.4 \cdot 10^{-8}$ & $1.7 \cdot 10^{-6}$ & This study, modeled\\
\hline
\multicolumn{6}{c}{\textbf{gas-phase COMs in hot cores/corinos}} \\
IRAS 4A    
                 & $\leq 7.0 \cdot 10^{-9}$ & 
                 & $\leq 2.8 \cdot 10^{-8}$ 
                 & $7.0 \cdot 10^{-8}$ 
                 & \cite{Bottinelli_ea04} \\ 
IRAS 4A2        & & $1.1-7.4 \cdot 10^{-9}$ 
                 & $1.0 \cdot 10^{-8}$
                 & $1.1 \cdot 10^{-8}$
                 & \cite{Lopez-Sepulcre2017} \\
IRAS 16293       & $3.0 \cdot 10^{-7}$ & $5.1\cdot10^{-8}$ 
                  & $2.4 \cdot 10^{-7}$
                  & $4.0 \cdot 10^{-7}$
                  & \cite{Cazaux2003} \\
OMC-1             & $1.0 \cdot 10^{-7}$ & 
                  & $8.0 \cdot 10^{-9}$
                  & $2.0 \cdot 10^{-8}$
                  & \cite{Sutton1995} \\
                  & $1.0 \cdot 10^{-6}$ 
                  & $6.0 \cdot 10^{-10}$
                  & 
                  & $9.0 \cdot 10^{-8}$
                  & \cite{Ikeda2001} \\
VLA 3             &  $6.1 \cdot 10^{-8}$ && 
                 & $3.3 \cdot 10^{-9}$ 
                 & \cite{Gieser_ea2019} \\ 
SVS13-A          & $3.0 \cdot 10^{-8}$ 
                 & $4.0 \cdot 10^{-9}$
                 & $5.0 \cdot 10^{-8}$
                 & $4.0 \cdot 10^{-8}$ 
                 & \cite{Bianchi_ea2019} \\
IRAS4B           & $7.0 \cdot 10^{-7}$ 
                 &
                 & $< 1.2 \cdot 10^{-6}$
                 & $1.1 \cdot 10^{-6}$ 
                 & \cite{Bottinelli_ea2007} \\
IRAS2A           & $3.0 \cdot 10^{-7}$ 
                 &
                 & $3.0 \cdot 10^{-8}$
                 & $< 6.7 \cdot 10^{-7}$ 
                 & \cite{Bottinelli_ea2007} \\
HOPS 373SW       & $9.0 \cdot 10^{-8}$ 
                 & $2.2 \cdot 10^{-9}$
                 & 
                 & $1.7 \cdot 10^{-8}$ 
                 & \cite{Lee_ea2023} \\
\enddata
\end{deluxetable*}

In \cite{JinGarrod20}, several non-diffusive mechanisms of surface reactivity are considered. Those are Eley--Rideal reactions, photodissociation-induced reactions, 3-body reactions (we prefer to name them as ``sequential reactions''), and 3-body excited formation in addition to 3-body reactions. In our GRD and MRD models, we employ all these types of non-diffusive processes, except for 3-body excited reactions. We have also explored the individual impact of each of the aforementioned mechanisms. At low temperatures of prestellar cores, the role of Eley--Rideal reactions is found to be negligible. The impact of photodissociation-induced non-diffusive reactions on chemistry is significant. However, the model that includes only this type of non-diffusive reactivity is not capable of reproducing the observed quantities of $\rm HCOOCH_3$ in the gas phase: the modeled abundance of $\rm HCOOCH_3$ is $\approx$~1 order of magnitude lower than the observational values. In the ice, the abundance of methyl formate is $>$~3 orders of magnitude lower than in our GRD model. In addition, the abundance ratio of solid $\rm CO_2$ to $\rm H_2O$ drops to 5\%. Thus, only the model that includes both photodissociation-induced reactions and sequential reactions matches all the observational data considered in this work, with the role of sequential reactions particularly important for the abundances of major ice species. The introduction of excited 3-body reactions \citep[as defined in Section 2.5 in][]{JinGarrod20} to our model worsens the agreement with observational data due to the overproduction of methyl formate. The details of this mechanism are particularly uncertain. Thus, it is not included in our GRD and MRD models. 

\subsection{COMs vs. Methanol}
\begin{figure*}
\begin{center}
\includegraphics[width=0.95\linewidth]{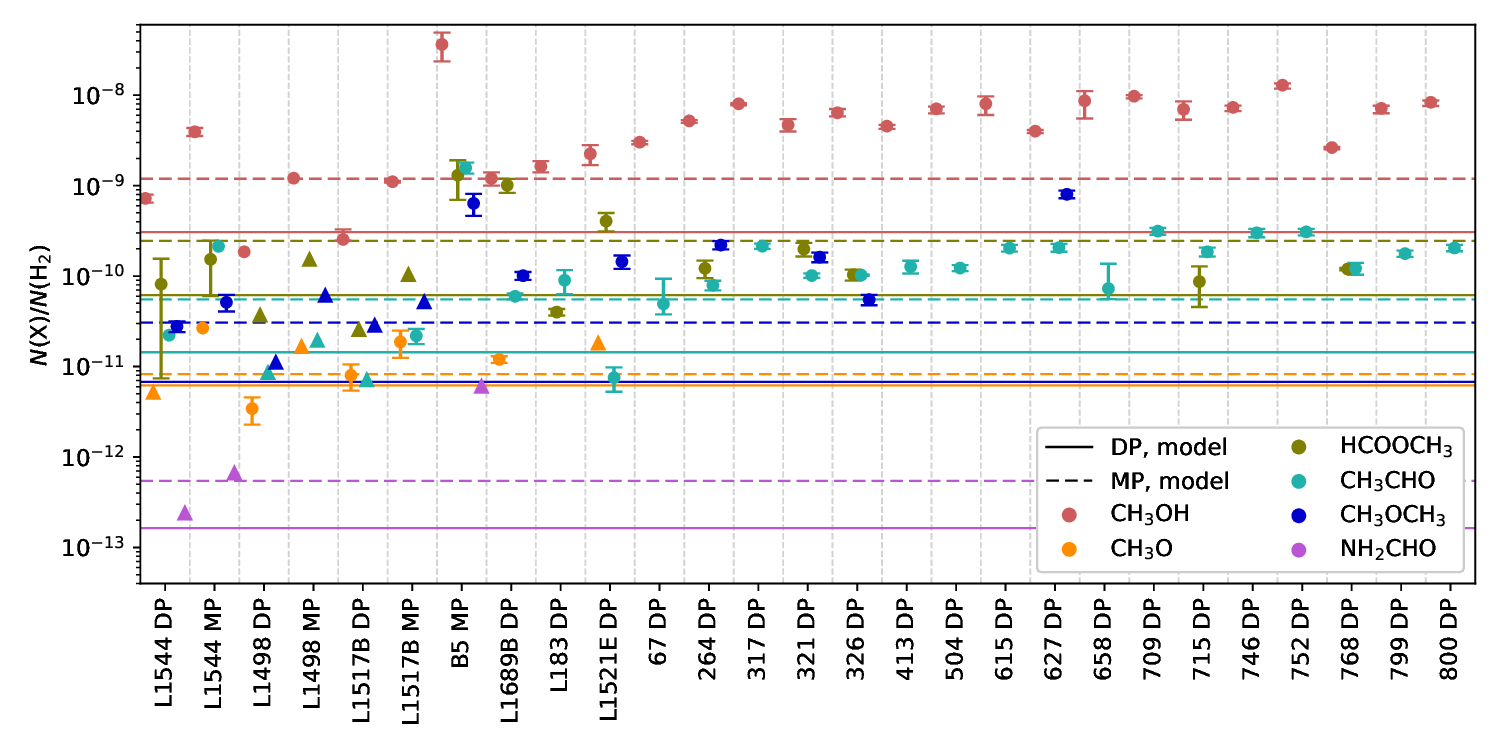}
\includegraphics[width=0.95\linewidth]{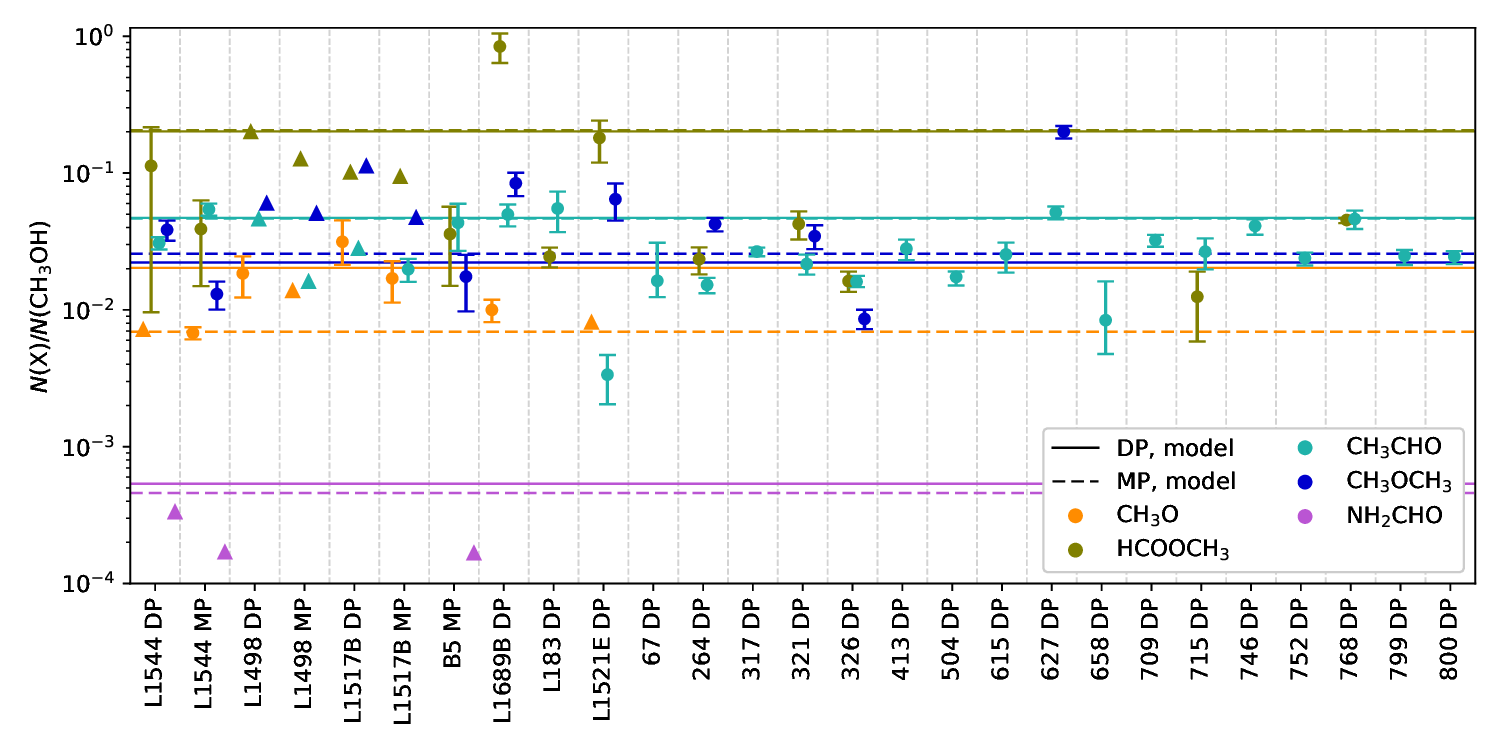}
\end{center}
\caption{Top: COMs observational abundances in cold cores in comparison with the abundances obtained as column densities ratios with our GRD model for the prestellar core L1544. Bottom: same for COMs-to-$\rm CH_3OH$ column densities ratios. (Note that the physical model of L1544 may not be representative of other sources in this comparison.) Circles denote observational values with error bars, triangles denote observational upper limits. DP is for the dust peak, MP is for the methanol peak. The modeled species abundances and ratios are given with horizontal lines of corresponding colors. The observational data is collected from \cite{Jimenez-Serra16} --- L1544; \cite{Chacon-Tanarro_ea2019_CH2DOH} --- L1544: methanol; \cite{Spezzano16} --- L1544: methanol (peak); \cite{Jimenez-Serra21} --- L1498; \cite{Megias_ea23} --- L1517B; \cite{Taquet_ea2017} --- B5; \cite{Wirstrom_ea2014} --- B5: $N({\rm H_2})$; \cite{Bacmann_ea12} --- L1689B; \cite{Bacmann_Faure2016} --- L1689B: methanol and methoxy radical, L1521E: methoxy radical; \cite{Lattanzi_ea2020} --- L183; \cite{Nagy_ea19} --- L1521E: methanol; \cite{Scibelli_ea21} --- L1521E: methyl formate, acetaldehyde, dimethyl ether; \cite{Scibelli_ea24} --- cores 67--800 in Perseus.} \label{fig:cold_cores}
\end{figure*}

We compared our GRD model gas phase COMs abundances convolved with the 26$''$ beam to the COMs abundances reported in observations toward cold cores (see Figure~\ref{fig:cold_cores}). We also show COMs-to-$\rm CH_3OH$ column density ratios. The abundances of COMs and methanol toward L1544, L1498, L1517B, L1521E (Taurus MC), L1689B (Ophiuchus MC), L183 are in agreement with the abundances predicted by our model for L1544, showing larger and smaller values within an order of magnitude of the model results (see caption of Fig.~\ref{fig:cold_cores} for the references). The abundances of CH$_3$OH, CH$_3$CHO, CH$_3$OCH$_3$ toward B5 and cores 67--800 (Perseus MC) are systematically higher than the modeled ones. This may be a scale effect, that is a combination of $\approx$~60$''$ beam and $\approx$~300~pc distance to Perseus (the beam covers $\approx$~18000~au) compared to $\approx$~30$''$ beam and $\approx$~130~pc distance to the other cores (the beam covers $\approx$~3900~au). This means that the abundances toward the Perseus cores should be compared with the model values for the methanol peak rather than dust peak, and then the observation results agree with the simulations. Another possibility is the higher density of the cloud surrounding the Perseus cores (which mostly reside in the region of clustered star-formation NGC~1333) compared to the relatively isolated L1544, L1498, L1517B, L1521E, L1689B, L183 cores. However, in cold cores and other evolutionary states (including comets), COMs abundance w.r.t. $\rm CH_3OH$  vary roughly within an order of magnitude \citep[see e.g.][]{Scibelli_ea24}. This strongly hints for similar chemical evolution between different stages of low-mass star formation.

\subsection{Comparison to \citet[][]{Vasyunin17} and the importance of tunneling through diffusion barriers}
The results obtained with our Model~MRD could not reproduce the observed abundances of COMs in the gas phase of L1544. This, to some extent, contradicts the previous results obtained with unmodified diffusive code presented in~\citet[][]{Vasyunin17}. The reactive desorption of several non-diffusively formed COMs from grains is negligible in both~\citet[][]{Vasyunin17} and the MRD model. On the other hand, the gas-phase COMs formation routes proposed in~\citet[][]{Vasyunin17} are still included in the MRD model. This discrepancy originates from the introduction of newly constrained values for reaction rates into the MRD model following recent laboratory and theoretical data. Those updates largely explain the fact that MRD model is not capable of reproducing the observed abundances of COMs in L1544.

In~the best-fit model presented in \citet[][]{Vasyunin17}, gas-phase chemistry of methanol plays a key role in the formation of both formaldehyde and dimethyl ether. Methanol has an abundance of $2.7 \cdot 10^{-8}$ at the location of the methanol peak. In the Model~MRD, gas-phase abundance of methanol is $2 \cdot 10^{-9}$. This renders the efficiency of gas-phase chemistry responsible for the formation of O-bearing COMs in~\citet[][]{Vasyunin17}. The lower abundance of gas-phase methanol in the MRD model in comparison to~\citet[][]{Vasyunin17} is due to the following differences between the models. Firstly, there is a difference in the widths of the activation barriers in the reactions $\rm H + \rm CO \rightarrow \rm HCO$ and $\rm H + \rm H_2CO \rightarrow CH_2OH/CH_3O$: 1.35~\AA, see Table~\ref{tab:CH3OH_chain} vs. 1.2~\AA~~in~\citet[][]{Vasyunin17}. This makes surface hydrogenation of CO to CH$_3$OH slower, reducing its grain-to-gas transfer rate via RD. Secondly, hydrogen abstraction loops introduced for the intermediate steps of the CO hydrogenation sequence in this work further reduce the efficiency of $\rm CH_3OH$ formation on surface and $\rm CH_3OH$ transfer to the gas phase. Importantly, switching on tunneling for diffusion of H and H$_2$ doesn't help to improve the agreement between the results of the MRD model and observed values. Another key discrepancy with the reaction rates used by~\citet[][]{Vasyunin17} is the implementation of the recent estimate for the $\rm CH_3 + \rm CH_3O \rightarrow \rm CH_3OCH_3 + h\nu$ reaction rate proposed by~\citet[][]{Tennis21}. This rate is an order of magnitude lower than the value used in~\citet[][]{Vasyunin17}. The factors listed above made the MRD model not capable of explaining gas-phase abundances of COMs in L1544 in contrast to the model presented in~\citet[][]{Vasyunin17}.

Note also that the abundance of solid methanol in~\citet[][]{Vasyunin17} reaches 40\% w.r.t. water ice, while in this work its abundance is about four times smaller. Overproduction of solid methanol in \citet[][]{Vasyunin17} can be partially explained by the lack of efficient mechanisms for the competitive reactions resulting in the formation of CO$_2$ ice from $\rm CO$ molecules. Indeed, the diffusive surface reactions CO~+~OH~$\rightarrow$~CO$_2$~+~H and CO~+~O~$\rightarrow$~CO$_2$ are not efficient at dust temperatures of $\le$10~K in \citet[][]{Vasyunin17} because their model lacks non-diffusive reaction mechanisms and has diffusion-to-binding energy ratio for molecules set to a moderately high value of 0.5. Implementation of non-diffusive reaction mechanisms is required to produce observed amounts of CO$_2$ ice under such conditions.

The better agreement of our GRD model results with the observations compared to the results by \cite{Vasyunin17} is also explained by different formation routes for COMs. In our simulations, COMs predominantly form on dust grains in non-diffusive radical-radical reactions \citep[$\rm CH_3CHO$ also forms in the chain suggested by][]{Fedoseev22}, arriving to the gas via efficient chemical desorption \citep{Garrod07}, enhanced by H-addition/abstraction loops. In \cite{Vasyunin17}, COMs were predominantly produced in gas-phase reaction chains involving $\rm CH_3OH$ as a precursor. COMs abundances were strongly dependent on methanol abundance, and their COM-to-methanol ratios were lower than ours.

Interestingly, partial grain-to-gas transfer is a key mechanism regardless of whether gas-phase or grain-surface chemistry is responsible for the formation of COMs in prestellar cores. In the first case, RD delivers precursors of COMs to the gas. In the second case, if COMs are formed on grains, they shall be somehow delivered to the gas. In cold, dark and dense environments of prestellar clouds, RD is a promising candidate to explain this delivery.

\subsection{The impact of different $E_{\rm diff}/E_{\rm des}$ ratios} \label{subsec:EbEd_grid}
The exact ratio between diffusion energy and desorption energy is still debatable. Experiments and models suggest many different values for the ratio of diffusion energy to desorption energy. \cite{Ruffle_Herbst2000} used $E_{\rm diff}/E_{\rm des} = 0.77$ in their simulations. In \citet[][]{Ruaud_ea2016} and \cite{Wakelam17}, all species including H and H$_2$ are assumed to diffuse by thermal hopping only, with $E_{\rm diff}/E_{\rm des}$~=~0.4.

More detailed models published so far, include different values of diffusion-to-desorption energy ratios for individual species. For example, \cite{Acharyya_2022} summarized the desorption and diffusion energies for CO from a number of experiments, and explored the impact of E$_b$/E$_D$ ratios varied in a range of 0.1~--~0.5 for CO, and in a range of 0.3~--~0.5 for other species, on the results of astrochemical modeling. They found that models with higher diffusion barriers provide a relatively better agreement with the observational data compared to models with lower diffusion barriers. \cite{Furuya_ea2022_EbEd_not_clear} found no clear relation between E$_b$ and E$_D$ an ASW ice. According to their study, the E$_b$/E$_D$ ratio may vary in a range of 0.2~--~0.7, depending on a species.
Moreover, binding and probably diffusion energies of species may also depend on surface coverage by the species~\cite[see e.g.][]{He_ea2016_binding}. Finally, on a microscopic level, each surface binding site is characterized by its own value of binding energy, and thus a distribution of binding and diffusion energies on the surface shall take place~\citep[see e.g.][]{Grassi_ea2020, Furuya24}. However, such a level of detail is normally not implemented in models based on rate equations: in this work, we consider only two values of E$_b$/E$_D$ for adsorbed species: one for atoms and another for molecules.

\begin{figure} [ht!]
  \includegraphics[width=1.0\linewidth]{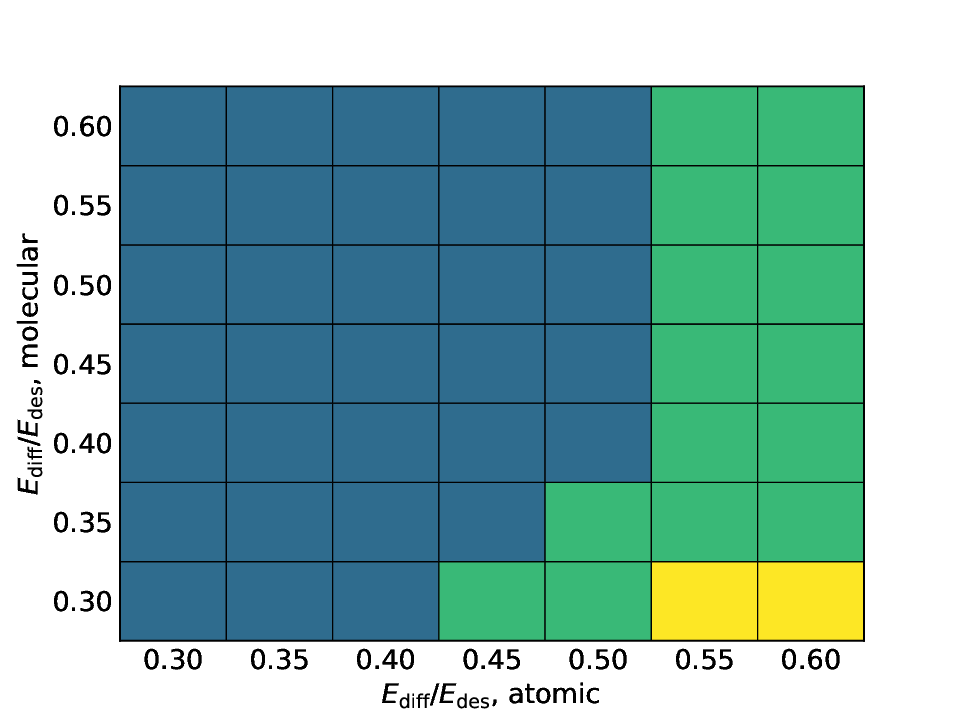}
\caption{The generalizing diagram of the success of different models in reproducing observed values of gas-phase abundances of complex organic species in the L1544 prestellar core. Blue: models do not reproduce the observational data; green: non-diffusive models reproduce the observations, yellow: both non-diffusive and diffusive models reproduce the observations.} \label{fig:table_ebed}
\end{figure}

In order to explore how well the simulations with various $E_{\rm diff}/E_{\rm des}$ ratios reproduce the observational data, we run two sets of models, using Models~GRD and MRD, with $E_{\rm diff}/E_{\rm des}$ values in an interval of 0.30--0.60 and a step of 0.05, and with disabled tunneling for diffusion. The $E_{\rm diff}/E_{\rm des}$ ratios for atomic and for molecular species have been varied separately. The simulations with MRD do not provide any combinations of atomic and molecular $E_{\rm diff}/E_{\rm des}$ ratios with which the model reproduces the observational data well. The results of the simulations with GRD are summarized in Figure~\ref{fig:table_ebed}. Each cell in Figure~\ref{fig:table_ebed} represents a particular pair of molecular and atomic $E_{\rm diff}/E_{\rm des}$ values. A cell is colored in green if the non-diffusive model with a given $E_{\rm diff}/E_{\rm des}$ reproduces the observed abundances of COMs in the gas phase. Blue colored cells indicate models that do not reproduce the observational data. Yellow color indicates the pairs of $E_{\rm diff}/E_{\rm des}$ values where both the diffusive (with non-diffusive reactions switched off) and non-diffusive models reproduce the observational gas-phase values well (see Appendix~\ref{sec:exploratory_models}). In the case of Figure~\ref{fig:table_ebed}, we consider that a model reproduces observational data well if a time moment $t$ exists at which the modeled CO depletion factor is within a factor of two compared to the observational value, and modeled abundances of all six studied gas-phase species are {\it in agreement} with observed values simultaneously at the locations of dust and methanol peaks.

As seen in~Figure~\ref{fig:table_ebed}, the atomic $E_{\rm diff}/E_{\rm des}$ values have in general more effect on species' abundances than the molecular $E_{\rm diff}/E_{\rm des}$. With the increase of atomic $E_{\rm diff}/E_{\rm des}$ ratios, grows the reproducibility of the observational data. For the atomic $E_{\rm diff}/E_{\rm des}$ values of 0.55--0.60, non-diffusive models reproduce the observational data well with any molecular $E_{\rm diff}/E_{\rm des}$ ratio. However, the relevance of such high $E_{\rm diff}/E_{\rm des}$ ratios is limited by the bulk composition. In our main model (where $E_{\rm diff}/E_{\rm des}$ is 0.5 for atomic species and 0.3 for molecular species), the abundance of unreacted H atoms entrapped in bulk ice is $10^{-7}$ at the dust peak, whereas increasing the atomic $E_{\rm diff}/E_{\rm des}$ only to 0.55 leads to the growth of H bulk abundance up to $2.4 \cdot 10^{-6}$. The situation is similar for diffusive models. Thus, we assumed the atomic $E_{\rm diff}/E_{\rm des}$ value 0.5 to be more acceptable than the larger ones.

With the atomic $E_{\rm diff}/E_{\rm des}$ value of 0.5, the ice composition (both main ice components and COMs) remains nearly stable when varying the molecular $E_{\rm diff}/E_{\rm des}$ value. At the central area of the core, abundance variations for the time moment of $10^5$ years do not exceed one tenth of an order of magnitude. However, in the models with molecular $E_{\rm diff}/E_{\rm des} \geq 0.4$, the modeled gas-phase abundances for the dust peak fit the observational values only in the narrow time interval of $\approx 2.0 \cdot 10^4 - 4.0 \cdot 10^4$ years, when gaseous CO abundance is still higher than $2.0 \cdot 10^{-5}$, and thus the observed CO depletion factor is not attained. Thus, we discard these models and employ $E_{\rm diff}/E_{\rm des} = 0.3$ for molecular species.

\subsection{Sticking coefficient}\label{subsec:sticking}
Several numerical and laboratory studies assess the temperature-dependent sticking coefficient for light species H and $\rm H_2$ \citep[e.g.][]{He_ea2016_sticking, Acharyya_2014, Veeraghattam_ea2014, Cazaux_ea2011, Matar_ea2010, Al-Halabi_Dishoeck2007, Masuda_Takashi1997, Hollenbach_Salpeter1970} and heavier molecules such as $\rm N_2$, $\rm O_2$ and CO \citep[e.g.][]{Bisschop_ea2006, Fuchs_ea2006, Oeberg_ea2005}.

The H-atom sticking coefficient is difficult to measure in laboratory conditions. The classical molecular dynamics simulations by \cite{Veeraghattam_ea2014} on an amorphous ice substrate predict the sticking probability of H atoms in the range of 0.7--1.0 for the temperature of the amorphous ice substrate of 10~K and the incident energy of 100~K (Figure~4 in their paper). The calculations by \cite{Buch_Zhang1991} claimed 0.85 for H atoms.

In our simulations, sticking probabilities are equal to unity for all species except H. For atomic hydrogen, sticking probability is calculated according to \cite{Hollenbach_McKee1979} and equals $\approx 0.8$ at 10~K dust temperature.

\subsection{Sputtering by cosmic rays}
In this work, we found that in order to match the observed values of abundances of complex organic molecules in the gas phase of L1544, one has to employ treatment of reactive desorption that yields desorption probability upon formation for COMs of $\sim$~0.1\%. That is, consistent with the RD treatment based on RRKM theory proposed in~\citet[][]{Garrod_ea06}. However, there are other types of non-thermal desorption that could be efficient for complex organic molecules. Recently, the experimental works by \cite{Dartois_ea18, Dartois_ea20, Dartois_ea21} provided the yields for sputtering by cosmic rays for some abundant ices, including water, carbon monoxide, carbon dioxide and methanol. Similarly to the approach suggested in \cite{Wakelam_ea21}, we implemented sputtering by cosmic rays into our modeling as a type of desorption. The sputtering rate has been calculated taking into account the fractions of CO and $\rm CO_2$ ices on the surface and in the bulk, and for the rest of the mantle the parameters for $\rm H_2O$ ice have been used.

In our simulations, the only significant effect of sputtering appeared to be an enhancement of abundances of gaseous COMs toward the prestellar core center, typically within one order of magnitude ($\rm CH_3OH$: from $7.8 \cdot 10^{-13}$ to $3.4 \cdot 10^{-12}$, $\rm HCOOCH_3$: from $5.0 \cdot 10^{-14}$ to $2.4 \cdot 10^{-13}$, $\rm CH_3CHO$: from $6.2 \cdot 10^{-14}$ to $1.9 \cdot 10^{-13}$, $\rm CH_3OCH_3$: from $2.8 \cdot 10^{-16}$ to $1.6 \cdot 10^{-15}$, $\rm NH_2CHO$: from $1.9 \cdot 10^{-18}$ to $3.6 \cdot 10^{-16}$). However, this does not affect the abundances calculated as column densities $N{\rm (X)}/N{\rm (H_2)}$, which we use for the comparison with the observations. This result is somewhat in agreement with the conclusion of \cite{Wakelam_ea21}, who suggest that at high densities, sputtering by cosmic rays dominates the desorption for molecules formed on dust grains, such as $\rm CH_3OH$ and $\rm CH_3OCH_3$. In our simulations, though, this effect becomes noticeable for gas densities higher than $\approx 3 \cdot 10^5$ cm$^{-3}$, while \cite{Wakelam_ea21} state $4 \cdot 10^4$ cm$^{-3}$ as a reference point. It should be noted that their and our chemical models significantly differ from each other; for example, their model includes the chemistry of Van Der Waals complexes \citep{Ruaud_ea15} and does not include non-diffusive chemistry. Ice COMs and main ice constituents abundances in our simulations remain the same as without sputtering.

\section{Conclusions} \label{sec:conclusions}
In this study, we performed chemical modeling of the formation and evolution of complex organic molecules and icy mantles of interstellar grains in the prestellar core L1544. For that, we utilized an updated version of the MONACO code, that now includes basic treatment for non-diffusive mechanism of surface chemical reactivity, surface reactions with H$_2$ molecules, H-atom induced abstraction routes, as well as some recent important updates on gas-phase and surface chemical processes. The most important results may be summarized as follows.
\begin{itemize}
\item The updated MONACO code, with the inclusion of the treatment for non-diffusive chemical reactivity in icy mantles of interstellar dust grains and and the reaction network updated with the most recent laboratory and theoretical results, provides results that demonstrate a very good agreement with the observational data on COMs in the prestellar core L1544. It also provides reasonable composition of icy mantles in the core, that is in agreement with the observations of L1544 and similar interstellar objects. For the first time, both the abundances of COMs in the gas phase and the location of methanol peak in L1544 are successfully reproduced with a model that includes non-diffusive chemical reactivity.

\item 
The mechanism of non-diffusive reactivity of radicals proposed in~\citet[][]{Fedoseev_ea15} and \citet[][]{Ioppolo_ea21} is implemented to the MONACO code following \citet[][]{JinGarrod20}. It appears to be efficient in producing complex organic molecules in icy mantles of cold ($T_{\rm dust}<10~{\rm K}$) interstellar grains with abundances 0.1--3\% with respect to water ice. Importantly, abundances of COMs in the ice are similar to the gas-phase abundances of those species observed in hot cores/corinos. This supports a scenario where COMs observed in the gas of hot cores are formed earlier than the warm-up transition phase from cold prestellar cores to hot cores/corinos occurs.

\item We found that parametrization of the efficiency of chemical desorption utilized in the model strongly affects the abundances of COMs in the gas phase, but only moderately --- the abundances of both simple and complex species --- in ices. Thus, the formation of COMs in ices, and delivery of COMs to the cold gas of prestellar cores may be considered as two separate problems. As shown in this study, the fraction of surface-formed COMs needed to be transferred to the gas upon formation to match observed abundances is $\sim$~0.1\%. In our model, such desorption rate for complex molecules is only achieved with reactive desorption with rates calculated following the RRKM theory. Neither cosmic ray-induced sputtering implemented following \citet[][]{Wakelam_ea21} nor the reactive desorption parametrization proposed in \citet[][]{Minissale_ea16b} have the similar efficiency for COMs.
\end{itemize}

\begin{acknowledgments}
The authors wish to thank the anonymous referees for valuable
comments, which helped us to improve the manuscript. This research has made use of NASA’s Astrophysics Data System. KB, GF and AV acknowledge the support of the Russian Science Foundation via the Project 23-12-00315 and Russian Ministry of Science and Higher Education via the project FEUZ-2025-0003 (Appendix~A).
GF is also benefited from the Xinjiang Tianchi Talent Program (2024).
I.J-.S acknowledges funding from grant PID2022-136814NB-I00 funded by the Spanish Ministry of Science, Innovation and Universities/State Agency of Research MICIU/AEI/10.13039/501100011033 and by “ERDF/EU”, and from the ERC grant OPENS (project number 101125858) funded by the European Union. Views and opinions expressed are however those of the author(s) only and do not necessarily reflect those of the European Union or the European Research Council Executive Agency. Neither the European Union nor the granting authority can be held responsible for them.
\end{acknowledgments}

%







\appendix

\section{Time dependence} \label{sec:time_dependence}

\begin{figure*}[h!]
  \includegraphics[width=0.5\linewidth]{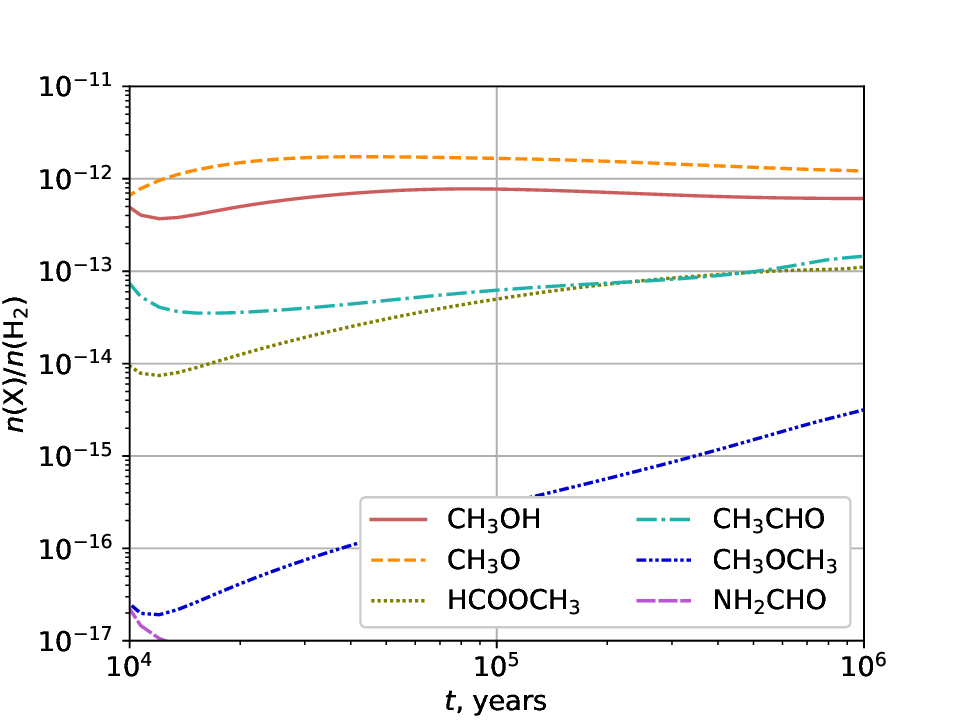}
  \includegraphics[width=0.5\linewidth]{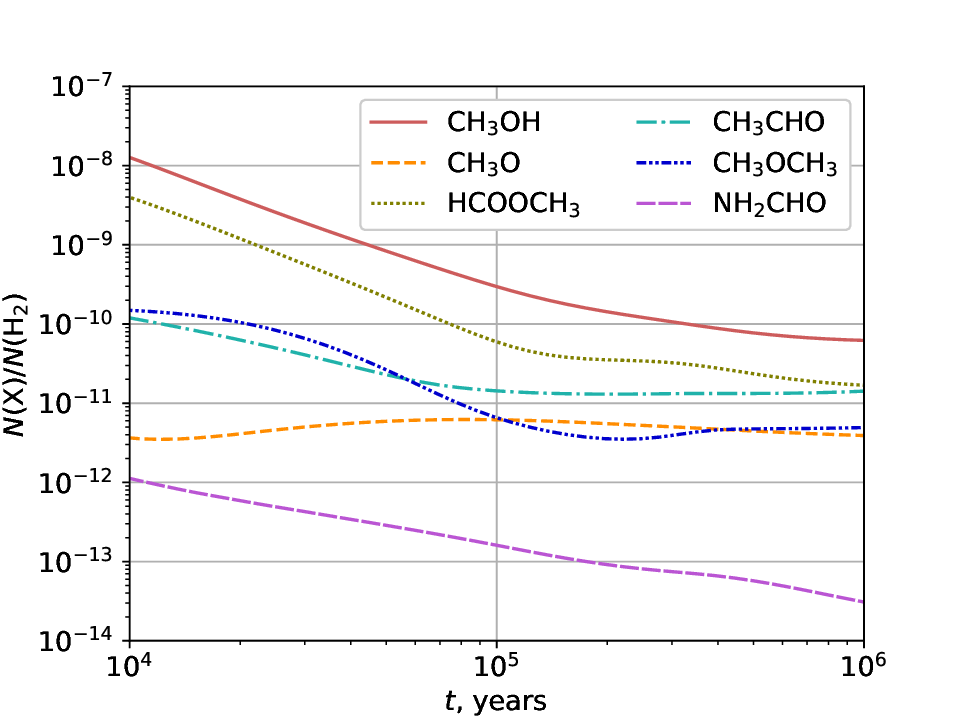}
  \includegraphics[width=0.5\linewidth]{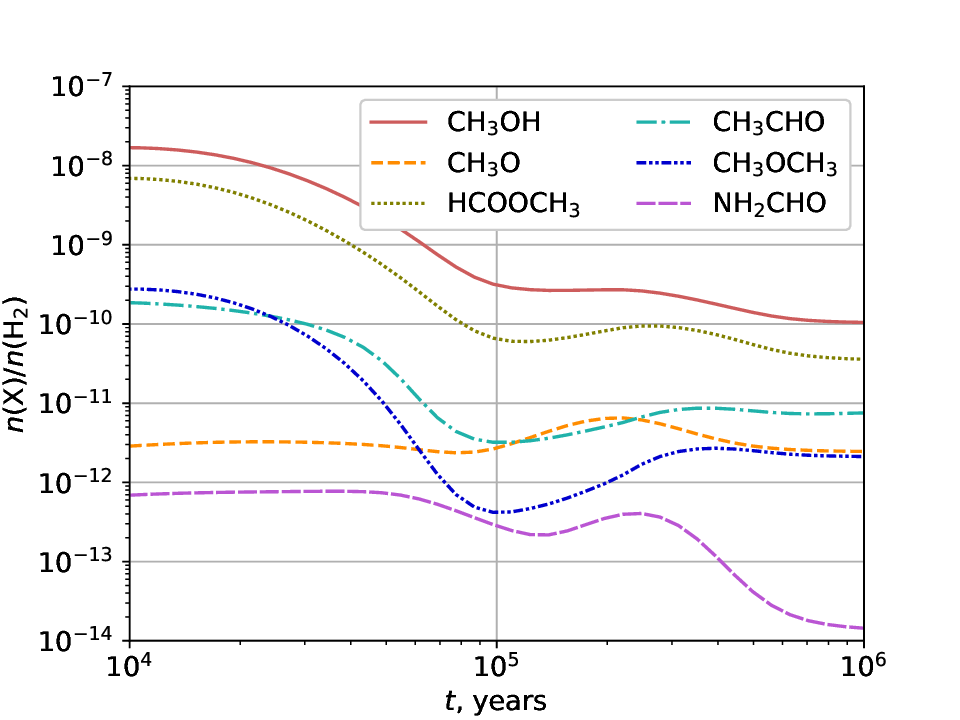}
  \includegraphics[width=0.5\linewidth]{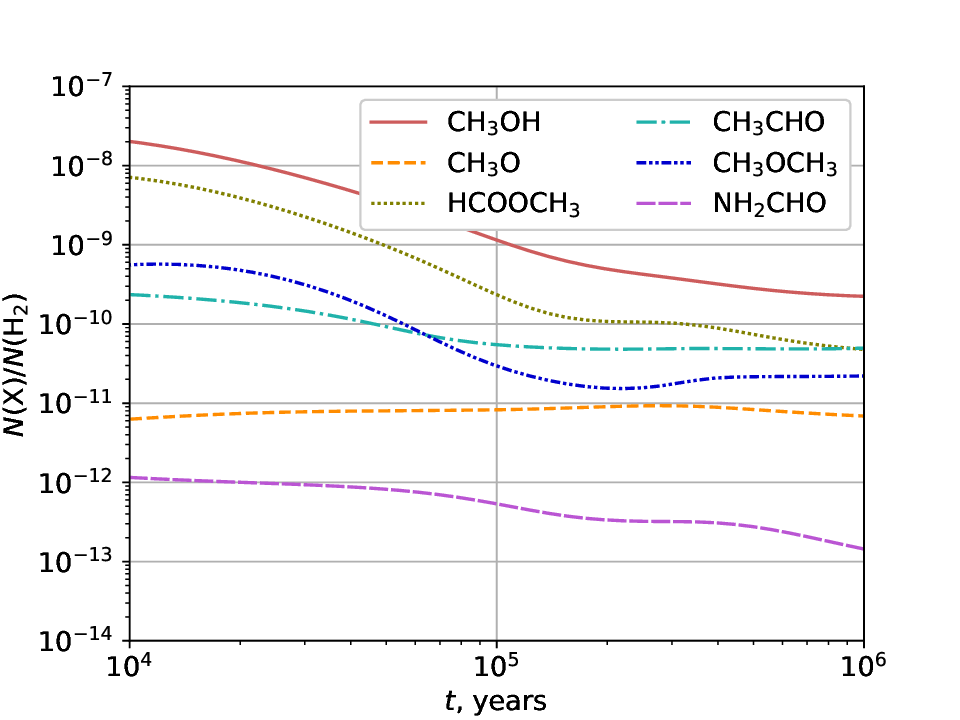}
\caption{Time profiles of COMs abundances obtained with our GRD model for the prestellar core L1544. Top: profiles of modeled abundances for the dust peak (top left) and abundances derived as modeled column density ratios for the dust peak (top right). Abundances derived as column densities are smoothed over the 26$''$ Gaussian beam. Bottom: same as in the top, but for the methanol peak position, 4000~au. (Note different abundance scale for the top left image.)} \label{fig:COM_time}
\end{figure*}

\section{Updates to the Chemical Network}
\label{subsec:network_updates}

The network of chemical reactions used in this study is mainly based on that presented in~\citet[][]{Jimenez-Serra21}. However, a number of important updates are introduced to this network following recent experimental studies and theoretical findings.

First, chemistry of CO hydrogenation on grain surface has been extended. Hydrogen addition reactions that lead to the gradual hydrogenation of CO into methanol (CH$_3$OH) through several intermediates (HCO, H$_2$CO, CH$_3$O/CH$_2$OH) were complemented by hydrogen abstraction reactions following \cite{JinGarrod20} (see Table~{\ref{tab:CH3OH_chain}}). We also included a reaction between metoxy radical and formaldehyde producing methanol ($\rm gCH_3O + gH_2CO \rightarrow gCH_3OH + gHCO$) with a barrier of 2670~K proposed in \cite{Alvarez-Barcia18}; recently the reaction was experimentally confirmed by~\citet[][]{Santos_ea2022}. 

\begin{deluxetable}{l|c|r}\label{tab:CH3OH_chain}
\tablecaption{Solid methanol formation chemistry.} 
\tablehead{
\colhead{Reaction} & \colhead{Bar., \AA} & \colhead{$E_{\rm act}$, K}
}
\startdata
 $\rm gH + gCO \rightarrow gHCO$                  & 1.35 & 2320 \\
 $\rm gH + gHCO \rightarrow gH_2CO$               & 1.00 &    0 \\
 $\rm gH + gH_2CO \rightarrow gCH_2OH$            & 1.35 & 4500 \\
 $\rm gH + gH_2CO \rightarrow gCH_3O$             & 1.35 & 2320 \\
 $\rm gH + gCH_2OH \rightarrow gCH_3OH$           & 1.00 &    0 \\
 $\rm gH + gCH_3O \rightarrow gCH_3OH$            & 1.00 &    0 \\
 $\rm gH + gHCO \rightarrow gCO + gH_2$           & 1.00 &    0 \\
 $\rm gH + gH_2CO \rightarrow gHCO + gH_2$        & 1.22 & 2960 \\
 $\rm gH + gCH_2OH \rightarrow gH_2CO + gH_2$     & 1.00 &    0 \\
 $\rm gH + gCH_3O \rightarrow gH_2CO + gH_2$      & 1.00 &    0 \\
 $\rm gH + gCH_3OH \rightarrow gCH_2OH + gH_2$    & 1.00 & 4380 \\
 $\rm gH + gCH_3OH \rightarrow gCH_3O + gH_2$     & 1.00 & 6640 \\
 $\rm gCH_3O + gH_2CO \rightarrow gCH_3OH + gHCO$ & --   & 2670$^a$ \\
 $\rm gCH_3O + gCH_3O \rightarrow gCH_3OH+gH_2CO$ & --   &    0 \\
 $\rm gC + gH_2O \rightarrow gH_2CO$              & --   &    0$^{b,c}$ \\
\enddata
\tablecomments{Parameters for hydrogen addition/abstraction reactions are taken from~\cite{JinGarrod20}. $^a$~\cite{Alvarez-Barcia18}; $^b$~\cite{Molpeceres21}; $^c$ also see Appendix~\ref{subsec:network_updates}.}
\end{deluxetable}

Recently, \cite{Molpeceres21} showed that formaldehyde (H$_2$CO), a species that can be converted to methanol via two subsequent additions of hydrogen atoms, can be formed in a reaction $\rm gC + gH_2O \rightarrow gH_2CO$. This reaction may occur efficiently before the ``catastrophic freeze-out'' of CO on grains \citep[][]{Caselli99}, thus facilitating chemistry of methanol and other COMs on earlier stages of formation of prestellar cores. Given the two-stage nature of our model (see Section~\ref{subsec:physmod}), we included this reaction in the chemical network. 

Second, formation routes for other COMs have been updated. As a route of acetaldehyde formation, we included the reaction chain proposed by \cite{Fedoseev22} (see Figure~3 in their paper). Table~{\ref{tab:AA_formation}} contains reactions from the chain leading to solid acetaldehyde. Recent theoretical results by \cite{Ferrero_ea2023} suggest that acetaldehyde ice can possibly form in the conditions similar to those of interstellar medium via successive hydrogenation of iced ketene $\rm CH_2CO$ (helped by H tunneling through the reaction barrier) and then the acetyl radical $\rm CH_3CO$. In the previous work by \cite{Vasyunin17}, acetaldehyde is mainly formed in the gas-phase reaction $\rm CH + CH_3OH \rightarrow CH_3CHO + H$. The rate constants for this reaction were taken from~\citet[][]{Johnson_ea00}. However, these rate constants are obtained at high temperatures (298--753~K) and high pressures (100--600 Torr helium), which are far from the conditions typical for prestellar cores. Our updated model demonstrates results similar to the observational ones without including this reaction, therefore, we removed it. We also switched off the reaction $\rm gN + gCH_2OH \rightarrow gNH_2CHO$; although it is present in KIDA~\citep[][]{Wakelam_ea12} with the note that it was listed in the OSU gas-grain code from Eric Herbst group in 2006, this seems to be an inefficient way of producing formamide in the solid-state because of the differences in chemical structure of $\rm CH_2OH$ and $\rm NH_2CHO$. For the reaction $\rm NH_2 + H_2CO \rightarrow NH_2CHO + H$, \cite{Vasyunin17} exploit the rates provided by~\citet[][]{Skouteris_ea17}. However, in a more recent work, \citet[][]{Douglas_ea22} suggested that below 110~K the branching ratio for this channel is effectively zero. Thus, under the conditions of the prestellar core characterized by low dust and gas temperatures, we switched this reaction off.

Important surface formation routes for methyl formate (HCOOCH$_3$), dimethyl ether (CH$_3$OCH$_3$) and acetaldehyde (CH$_3$CHO) in our non-diffusive model are radical-radical reactions ${\rm HCO}+{\rm CH_3O}\rightarrow~{\rm HCOOCH_3}$, ${\rm CH_3}+{\rm CH_3O}\rightarrow~{\rm CH_3OCH_3}$ and ${\rm CH_3}+{\rm HCO}~\rightarrow~{\rm CH_3CHO}$ \citep[][]{AllenRobinson77}, although the possibility of other products for the last reaction was also reported~\citep[][]{ Lamberts_ea2019}. Following \cite{JinGarrod20}, we also added H-addition/abstraction loops for those species. We included these loops in our network with the barriers for H-abstraction reactions taken from \cite{Garrod13} and \cite{JinGarrod20} (Table~{\ref{tab:COM_loops}}); see also \cite{Alvarez-Barcia18} for more details. H-addition reactions forming these species are barrierless. The loops may be important as iterative attempts to desorb a species that result in a net increase of RD probability.

\begin{deluxetable}{l|r}\label{tab:AA_formation}
\tablecaption{Acetaldehyde formation route \citep{Fedoseev22}} 
\tablehead{
\colhead{Reaction} & \colhead{$E_{\rm act}$, K}
}
\startdata
 $\rm gC + gCO \rightarrow gCCO$                 &    $0^{a}$\\
 $\rm gH + gCCO \rightarrow gHC_2O$    &    0\\
 $\rm gH + gHC_2O \rightarrow gCH_2CO$           &    0\\
 $\rm gH + gCH_2CO \rightarrow gCH_3CO$           & $975^{b}$\\
 $\rm gH + gCH_3CO \rightarrow gCH_3CHO$           &    0\\
\enddata
\tablecomments{$^{a}$~\cite{Papakondylis19}; $^{b}$~\cite{Umemoto84}.}
\end{deluxetable}

\begin{deluxetable}{l|r}\label{tab:COM_loops}
\tablecaption{H-abstraction reactions for COMs.} 
\tablehead{
\colhead{Reaction} & \colhead{$E_{\rm act}$, K}
}
\startdata
 $\rm gH + gCH_3CHO \rightarrow gH_2 + gCH_3CO$      & $2120^{a,b}$ \\
 $\rm gH + gCH_3OCH_3 \rightarrow gCH_3OCH_2 + gH_2$  & $4450^{c,d}$ \\
 $\rm gH + gHCOOCH_3 \rightarrow gH_2 + gCH_3OCO$  & $3970^{a,e}$ \\
\enddata
\tablecomments{$^{a}$~\cite{Garrod13}; $^{b}$~\cite{Warnatz84}; $^{c}$~\cite{JinGarrod20}; $^{d}$~\cite{Takahashi07}; $^{e}$~\cite{GoodFrancisco02}.}
\end{deluxetable}

\cite{Lamberts22} showed that along with reactions with H \citep{Qasim20}, reactions with $\rm H_2$ are also important for the hydrogenation of carbon atoms into methane on grain surfaces. We updated our reaction barriers according to their results (see Table~\ref{tab:CH4_formation}). As for the reaction $\rm gC + gH_2 \rightarrow gCH_2$, \cite{Lamberts22} point out that its activation energy strongly depends on the neighborhood of a binding site of the carbon atom. We have chosen the minimal activation energy provided by \cite{Lamberts22} equal to 30 kJ mol$^{-1}$ (3600~K).

\begin{deluxetable}{l|r}\label{tab:CH4_formation}
\tablecaption{Grain-surface methane formation routes \citep{Lamberts22}.} 
\tablehead{
\colhead{Reaction} & \colhead{$E_{\rm act}$, K}
}
\startdata
 $\rm gC + gH \rightarrow gCH$               &    0 \\
 $\rm gCH + gH \rightarrow gCH_2$            &    0 \\
 $\rm gCH_2 + gH \rightarrow gCH_3$          &    0 \\
 $\rm gCH_3 + gH \rightarrow gCH_4$          &    0 \\
 $\rm gC + gH_2 \rightarrow gCH_2$           &    3600 \\
 $\rm gCH + gH_2 \rightarrow gCH_3$          &    0 \\
 $\rm gCH_2 + gH_2 \rightarrow gCH_3 + H$    &    5900 \\
 $\rm gCH_3 + gH_2 \rightarrow gCH_4 + H$    &    5300 \\
\enddata
\end{deluxetable}

In previous models of COMs formation in cold clouds, the gas-phase reaction of radiative association between CH$_3$ and CH$_3$O leading to dimethyl ether (${\rm CH_3}+{\rm CH_3O}\rightarrow {\rm CH_3OCH_3}$) was shown to play a key role~\citep[][]{Balucani15, Vasyunin17}. Rates of reactions of radiative association are known poorly. In~\citet[][]{Balucani15} and \citet[][]{Vasyunin17} the rate of this reaction at 10~K was taken equal to 3$\cdot$10$^{-10}$~cm$^3$/s. Recently, \cite{Tennis21} studied the gas-phase formation path for dimethyl ether by the radiative association of $\rm CH_3$ and $\rm CH_3O$ radicals. They calculated the rate coefficient by two methods, canonical and phase-space, and provided the rate constants for modified Arrhenius rate expressions. We adopted the rate constant obtained by the phase-space method, as \cite{Tennis21} report it to be more precise. The Arrhenius coefficients for the rate are $\alpha = 1.37 \cdot 10^{-12}\ \rm cm^3 s^{-1}$, $\beta = -0.96$, $\gamma = 0.00$. This gives a rate constant of the reaction at 10~K equal to 3.6$\cdot$10$^{-11}$~cm$^3$/s, which is almost an order of magnitude lower than the values used previously.

Binding energies of species utilized in this study are the same as in~\citet[][]{Jimenez-Serra21} with exception for molecular hydrogen H$_2$ and atomic carbon. For molecular hydrogen, we adopted the value of binding energy equal to 380~K which is closer to the estimate provided in~\citet[][]{Minissale22}. The binding energy of atomic carbon was taken equal to 10000~K~\citep[][]{Wakelam17}. Such a high value reflects the fact that carbon atoms can be partially chemisorbed on amorphous H$_2$O ice. In this work, we do not use systematically the values of binding energies and pre-exponential factors of surface species presented in~\citet[][]{Minissale22}. In a narrow range of grain temperatures that exists in our model of L1544 (7--15~K), those new factors combined with accordingly adjusted binding energies introduce very little changes to chemistry calculated using values utilized in~\citet[][]{Jimenez-Serra21}.  

\section{Diffusive model} \label{sec:exploratory_models}
\subsection{Agreement with observations}
\begin{figure*} [ht!]
  \includegraphics[width=0.5\linewidth]{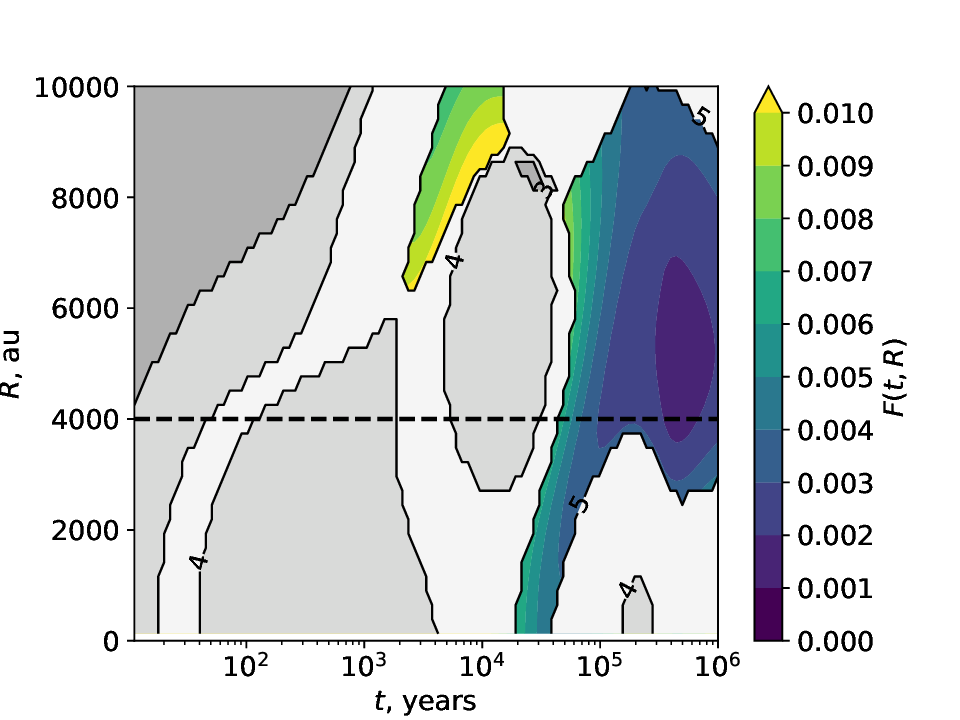}
  \includegraphics[width=0.5\linewidth]{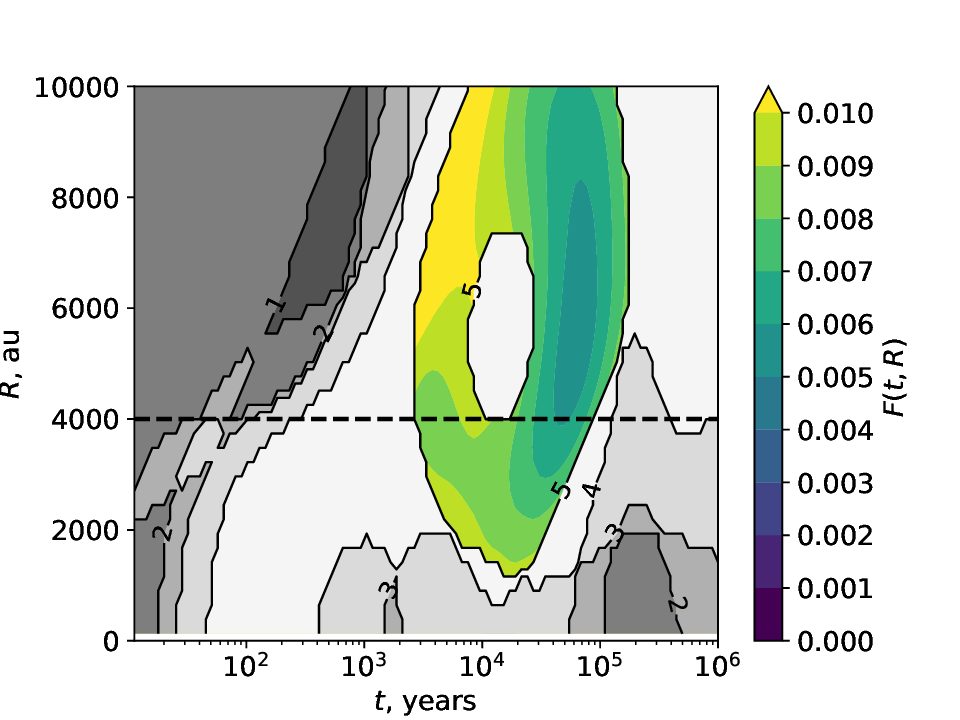}
\caption{Agreement maps for the dust peak (left) and for the methanol peak (right), the diffusive model. The details are the same as in Figure~\ref{fig:agreement_maps}.} \label{fig:agreement_maps_diff}
\end{figure*}

To find out the role of non-diffusive grain chemistry for COM formation in comparison with other processes, we also run a model with the same set of parameters and GRD chemical desorption, but the non-diffusive reactions switched off. This model is in the following referred to as the diffusive model. Like in the case of our GRD model, we created agreement maps to compare our modeling data with the observational data (Figure~\ref{fig:agreement_maps_diff}). At the central area of the core, the modeled abundances fit the observational data only in a narrow time interval from $2 \cdot 10^4$ to $4 \cdot 10^4$ years, which is not suitable for us because CO is not depleted yet. The direct comparison with Figure~\ref{fig:agreement_maps} shows that the non-diffusive model demonstrates a wider range of ``agreement'' over the chosen parameter space. Therefore, we provide the results for the diffusive model at the time point of $10^5$ years to perform a reasonable comparison with our GRD model.

\begin{figure*} [p!]
  \includegraphics[width=0.5\linewidth]{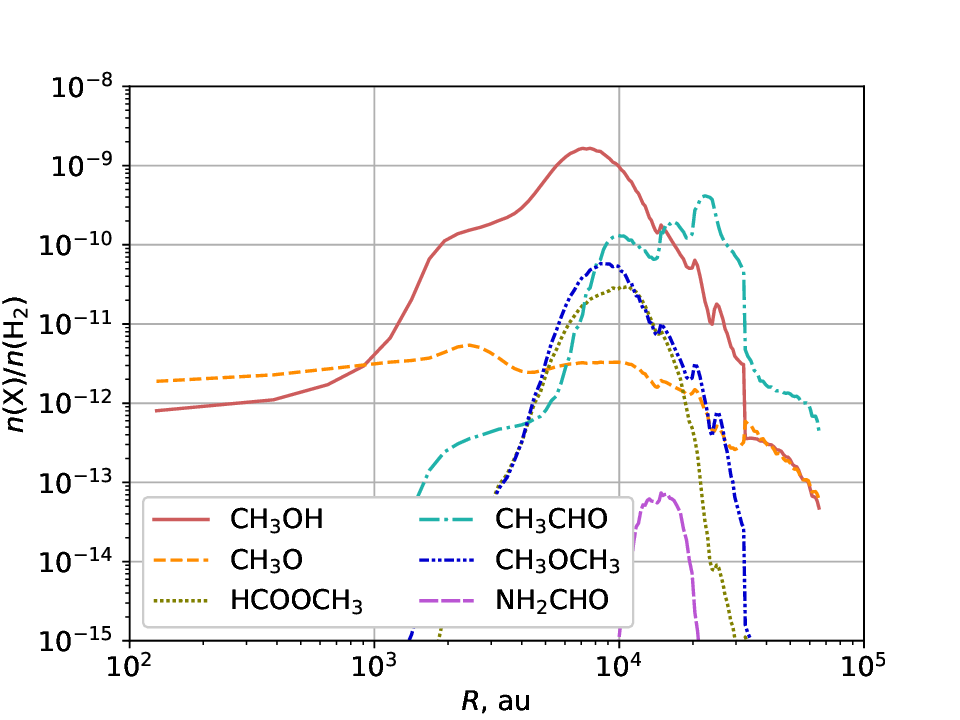}
  \includegraphics[width=0.5\linewidth]{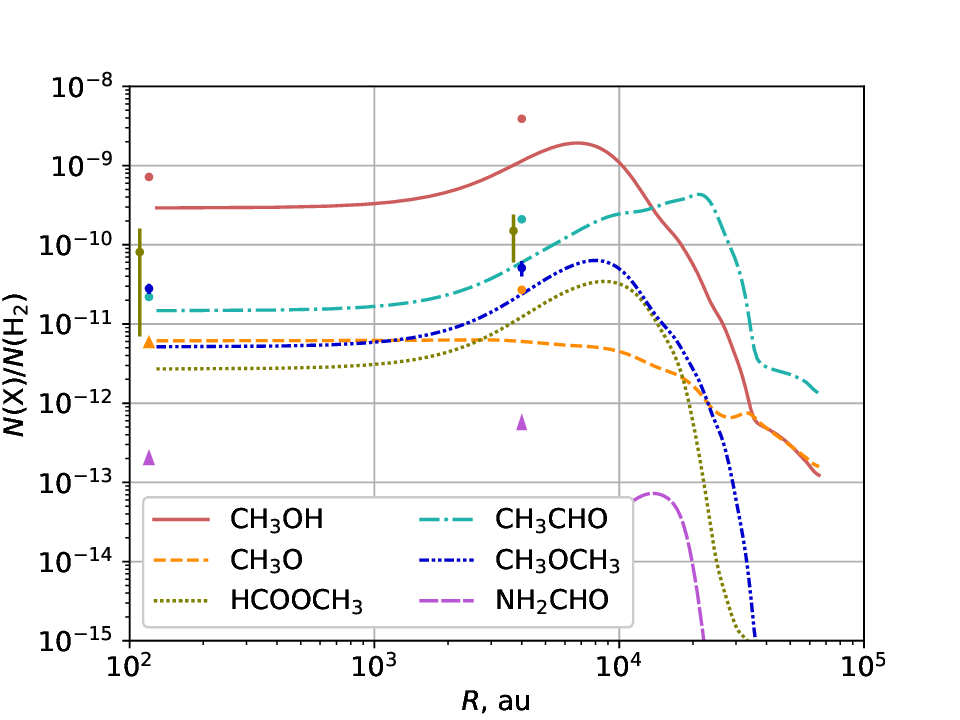}
  \includegraphics[width=0.5\linewidth]{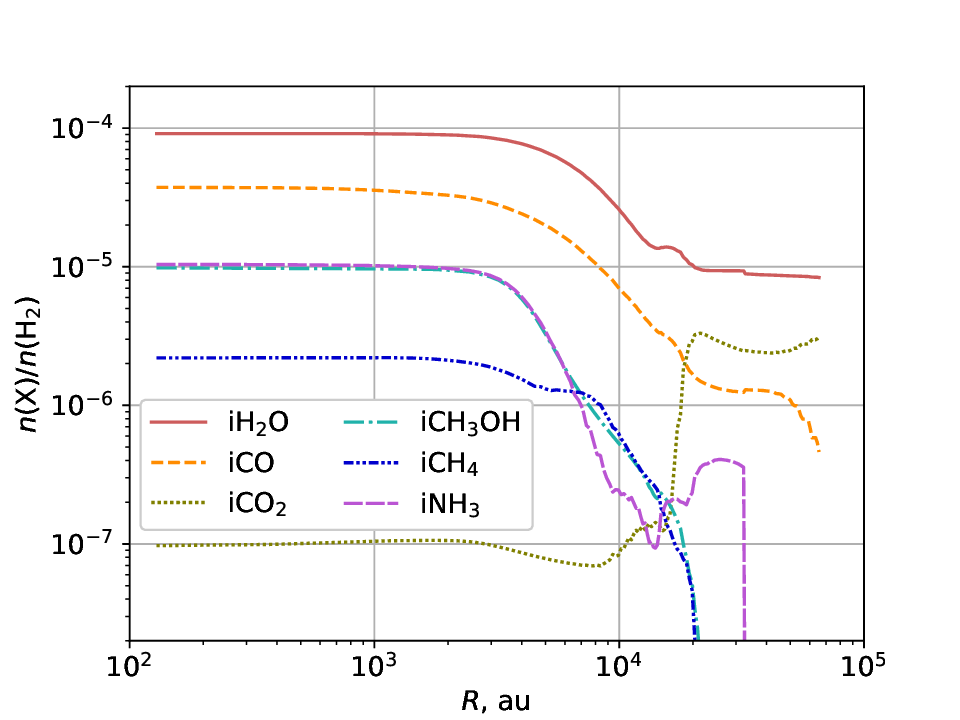}
  \includegraphics[width=0.5\linewidth]{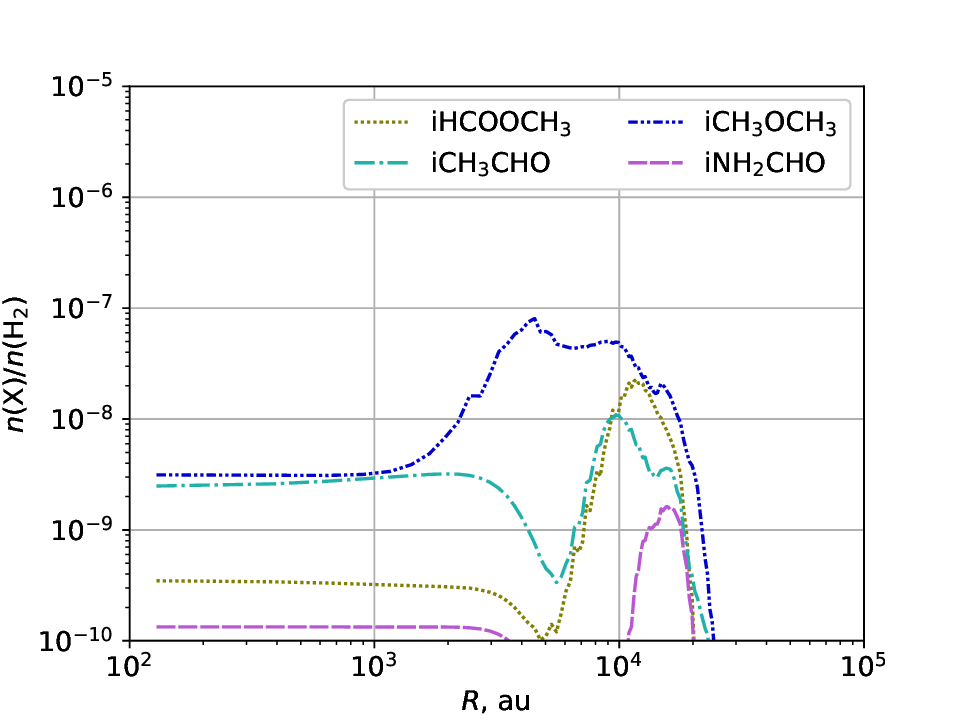} \\
  \includegraphics[width=0.5\linewidth]{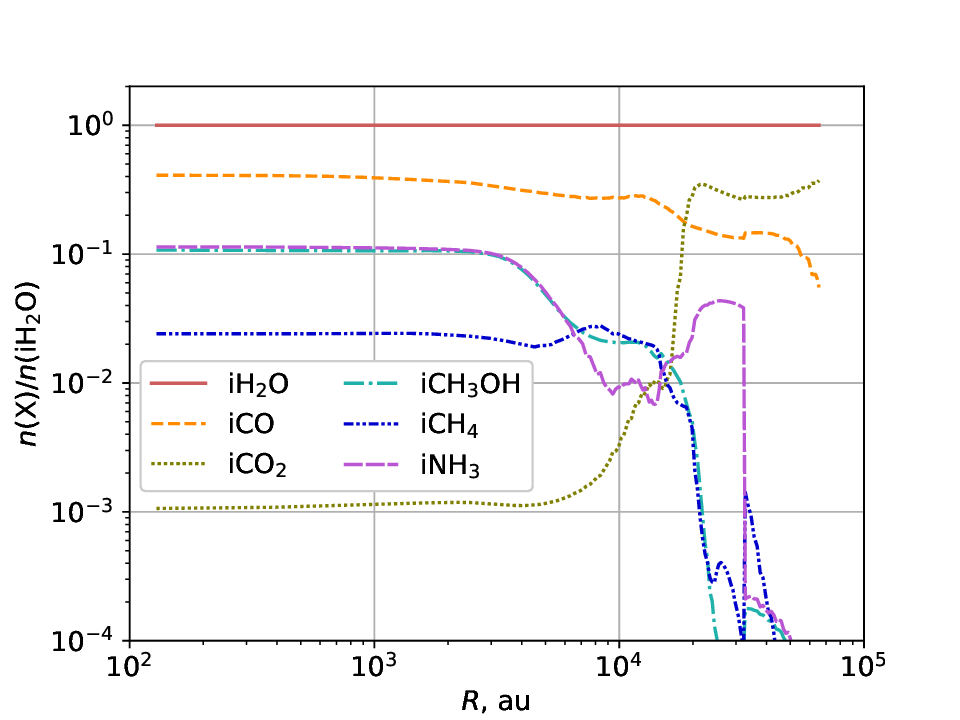}
  \includegraphics[width=0.5\linewidth]{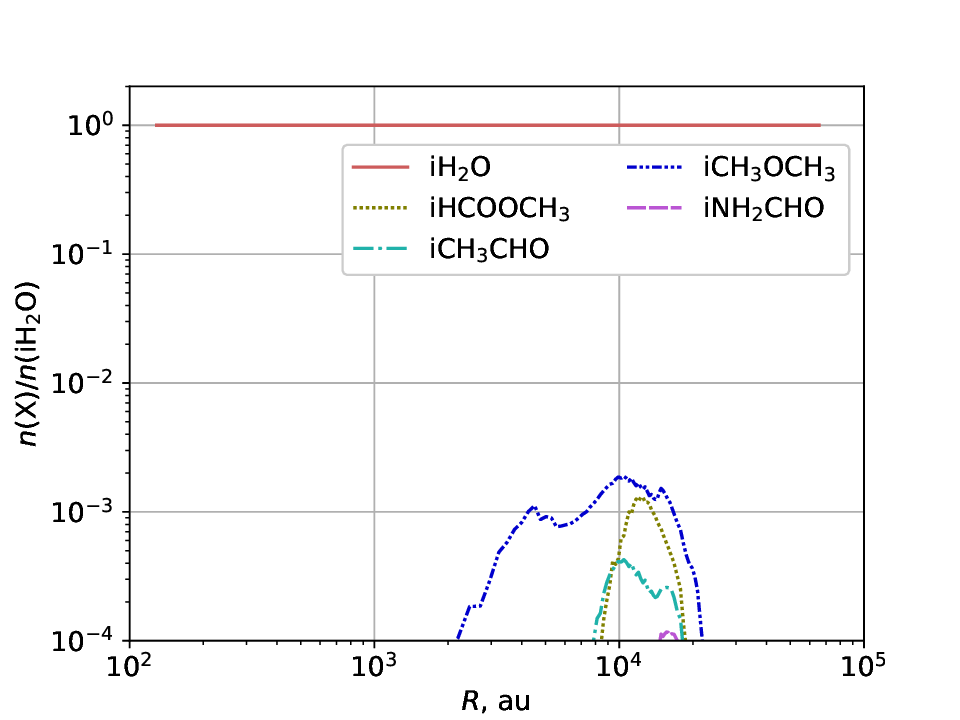}
\caption{Radial profiles obtained with our diffusive model at 10$^{5}$ years. Top: profiles of modeled abundances (top left) and abundances derived as column density ratios (top right) of complex organic species. Abundances derived as column densities are smoothed over the 26$''$ Gaussian beam. Middle: profiles of abundances of major ice constituents (middle left) and selected complex organic molecules in the ice (middle right) w.r.t. H$_2$. Bottom: same as in the middle, but w.r.t. H$_2$O. Colored dots in the top right panel denote observational values by \cite{Chacon-Tanarro_ea2019_CH2DOH} for $\rm CH_3OH$ and by \cite{Jimenez-Serra16} for other species, vertical lines are for error bars, arrows are for observational upper limits.} \label{fig:COM_bestfit_diff}
\end{figure*}

In the case of our diffusive model, almost all species of our interest demonstrate the modeling abundances {\it in agreement} (in the sense of the definition given in Section~\ref{subsec:agreement}) with the observational results for gaseous COMs (Figure~\ref{fig:COM_bestfit_diff}, top panel). The exception is $\rm HCOOCH_3$, whose modeled abundance is more than an order of magnitude lower than the observational one. However, when taking into account observational errors, this difference does not look crucial. It is also worth noting that $\rm CH_3O$ modeled abundance does not show a peak in our diffusive model. Nevertheless, the agreement with the observational results is still reasonable. The abundances of the main ice components are also reasonable except $\rm CO_2$ (which is produced via the non-diffusive processes in our GRD model), and ice COMs practically disappear in the central parts of the core (Figure~\ref{fig:COM_bestfit_diff}, middle and bottom panels).

\subsection{Chemistry at the methanol peak for the diffusive model}\label{subsec:chemistry_diff_model}

We describe the chemistry of COMs at the position of the methanol peak in our diffusive model. All values are given for the time moment of $\approx 10^5$~years.

$CH_3OH$. Like in our GRD model, the H-addition reactions are the major paths to the adsorbed species constituting the methanol formation chain (Table~\ref{tab:CH3OH_chain}). The exception is $\rm CH_2OH$ ice, produced by the H-abstraction reaction from $\rm CH_3OH$.

$HCOOCH_3$. In our diffusive model, the only efficient channel of methyl formate production is the gas-phase reaction $\rm CH_3OCH_2 + O \rightarrow HCOOCH_3 + H$ \citep{Balucani15}. The predominant source of $\rm CH_3OCH_2$ radical is RD in the reaction $\rm gCH_2 + gCH_3O \rightarrow gCH_3OCH_2$ \citep{JinGarrod20}. In contrast to the position of the dust peak, this reaction has a non-negligible rate at the position of the observed methanol peak. As in our GRD model, switching off the reaction $\rm CH_3OH_2^+ + HCOOH \rightarrow HC(OH)OCH_3^+ + H_2O$ in our diffusive model does not affect $\rm HCOOCH_3$ abundance. In the absence of non-diffusive processes, RD following the surface reactions $\rm gH + gCH_3OCO \rightarrow gHCOOCH_3$ and $\rm gHCO + gCH_3O \rightarrow gHCOOCH_3$ provides less than 1\% of gaseous methyl formate abundance.

$CH_3CHO$. The reaction $\rm gH + gCH_3CO \rightarrow gCH_3CHO$ accounts for almost all dust-to-gas $\rm CH_3CHO$ transfer rate and for 84\% of gaseous acetaldehyde production. Gas-phase processes accounting for the rest of acetaldehyde formation rate are the dissociative recombination of protonated acetaldehyde $\rm (CH_3CHO)H^+$ and the reaction $\rm O + C_2H_5 \rightarrow CH_3CHO + H$. As for $\rm (CH_3CHO)H^+$, 55\% of its formation rate is due to the reactions $\rm H_3O^+ + C_2H_2 \rightarrow (CH_3CHO)H^+$ and $\rm H_2CO^+ + CH_4 \rightarrow (CH_3CHO)H^+ + H$, and the rest is due to the loops including the reactions of acetaldehyde with $\rm H_3^+$, $\rm HCO^+$ or $\rm H_3O^+$.

Surface acetaldehyde is the end species in the chain of hydrogenation reactions $\rm gHC_2O \rightarrow gCH_2CO \rightarrow gCH_3CO \rightarrow gCH_3CHO$. Thus, similarly to our GRD model, here we find a part of the acetaldehyde ice production path described in \cite{Fedoseev22} and \cite{Ferrero_ea2023}. Importantly, at the position of the methanol peak, $\rm gHC_2O$ predominantly accretes from gas -- the efficiency of the reaction $\rm gH + gCCO \rightarrow gHC_2O$ is very low because of low CCO abundance. Toward the edge of the core, at a distance of $\approx 20000$~au, the reaction $\rm gC + gCO \rightarrow gCCO$ proceeds efficiently even in a diffusive mode, which enhances surface acetaldehyde production and therefore its dust-to-gas desorption rate. At large radii, gas-phase acetaldehyde abundance becomes even greater than methanol gas-phase abundance.

$CH_3OCH_3$. The H-addition surface reaction $\rm gH + gCH_3OCH_2 \rightarrow gCH_3OCH_3$ is the predominant source of gaseous $\rm CH_3OCH_3$ via RD and also the major source of dimethyl ether ice. The radical $\rm gCH_3OCH_2$ is produced via the reaction $\rm gCH_2 + gCH_3O \rightarrow gCH_3OCH_2$.

Gas-phase reactions practically do not contribute to the production of $\rm CH_3OCH_3$. The only noticeable process is the dissociative recombination of protonated dimethyl ether $\rm (CH_3)_2OH^+$, which is predominantly produced in loops via the reactions of dimethyl ether with $\rm H_3^+$, $\rm HCO^+$ and $\rm H_3O^+$. Only 8\% of $\rm (CH_3)_2OH^+$ is a product of the reaction $\rm CH_3^+ + CH_3OH \rightarrow (CH_3)_2OH^+$.
 
$NH_2CHO$. Due to the lack of non-diffusive mechanisms, the reaction $\rm gNH_2 + gHCO \rightarrow gNH_2CHO$ responsible for formamide formation on grain surface appears to be slow. Since we switch the gas-phase reaction of $\rm NH_2$ and $\rm H_2CO$ off under the conditions of the prestellar core~\citep[][]{Douglas_ea22}, gaseous formamide has no efficient production paths and demonstrates an abundance of $\sim 10^{-14}$ at the position of the methanol peak.

\subsection{Issues of the diffusive model and possible indicators for non-diffusive mechanisms efficiency}

The only efficient process supplying gaseous methyl formate and dimethyl ether in our diffusive model is the reaction $\rm gCH_2 + gCH_3O \rightarrow gCH_3OCH_2$. At the methanol peak with the dust temperature $\sim 10$~K, it proceeds efficiently even in the diffusive mode. When switching it off, the $\rm HCOOCH_3$ and $\rm CH_3OCH_3$ calculated gas-phase abundances drop by about two orders of magnitude at the methanol peak position. In our reactions network, $\rm CH_2$ has a desorption energy of 1050~K, however, higher values exist. KIDA astrochemical database provides 1400~K as the desorption energy for $\rm CH_2$ \citep{Wakelam17}. When we incorporate this value in our network, the gas-phase abundances of methyl formate and dimethyl ether drop by some tenths of an order of magnitude.

Creating the diagram for the diffusive model as the one in Figure~\ref{fig:table_ebed} shows that the model provides abundances for all the studied gaseous COMs {\it in agreement} with the observations only in a very narrow range of $E_{\rm diff}/E_{\rm des}$~=~0.55--0.60 for atomic species and $E_{\rm diff}/E_{\rm des}$~=~0.30 for molecular species. At these parameters, ice composition does not look reasonable, with too high abundances of atomic H and free radicals.

Going back to the diffusive model with our default parameters ($E_{\rm diff}/E_{\rm des} = 0.50$ for atomic species and 0.30 for molecular ones), we find the abundance of $\rm CO_2$ ice dropping to $10^{-7}$ in the absence of non-diffusive processes. Other main ice components do not demonstrate significant variations in abundance between the non-diffusive and diffusive models. In our GRD model, $\rm CO_2$ ice is a product of non-diffusive processes, however, there exist ways to obtain relevant $\rm CO_2$ ice abundance other than implementing non-diffusive reactions --- for example, changing physical conditions during the translucent cloud phase. Being formed during the ``pre-core'' phase, $\rm CO_2$ ice may survive in the central area of the core in the absence of strong photolysis or radiolysis. Interestingly, \citet{Clement_ea23} in their simulations produce significant amounts of $\rm CO_2$ ice at 10~K without invoking non-diffusive chemistry. Their main $\rm gCO_2$ formation route is $\rm gCO + gO$, with the older binding energy value of 800~K assumed for atomic oxygen instead of 1660~K.

\begin{figure*} [ht!]
  \includegraphics[width=0.5\linewidth]{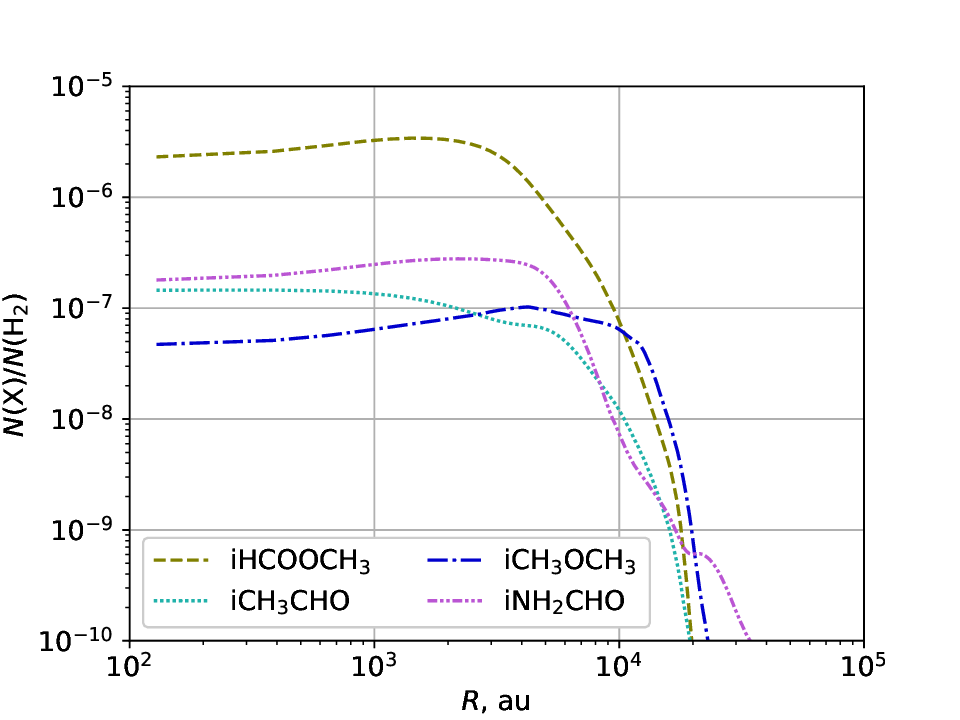}
  \includegraphics[width=0.5\linewidth]{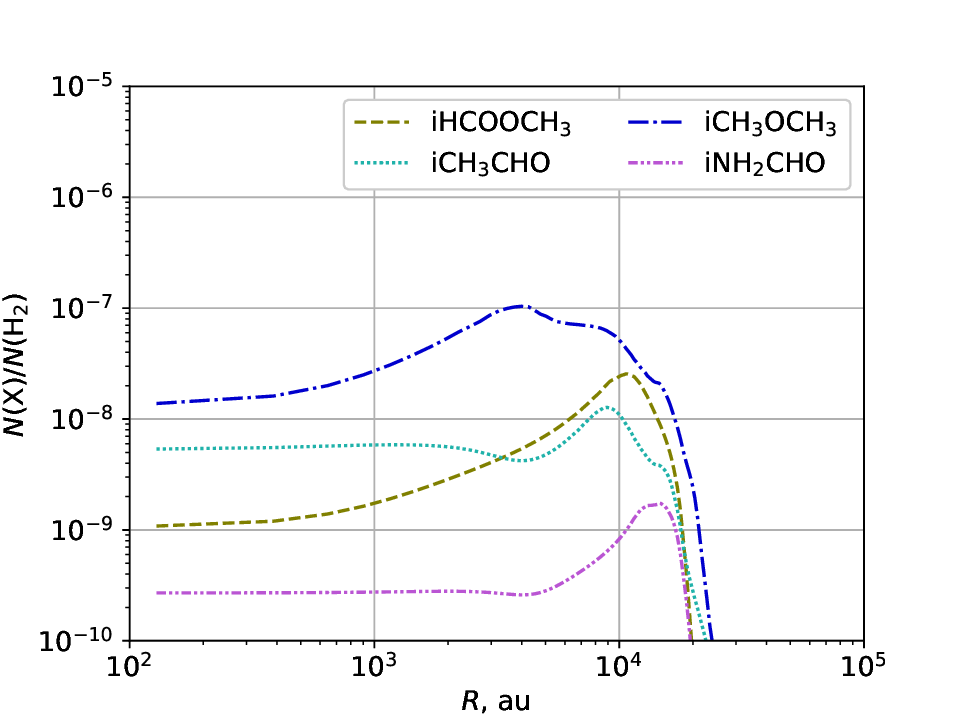}
\caption{Comparison of ice COM abundances derived as column densities ratios (not smoothed over a beam in this case) for the GRD model (left) and the diffusive model (right).} \label{fig:iceCOM_NXtoNH2_compare}
\end{figure*}

Since both diffusive and GRD models provide reasonable gaseous COM abundances (except for methyl formate in the diffusive model), it is good to have some indicator species in addition to $\rm CO_2$ whose observational abundances may help to learn if non-diffusive processes do have a significant effect on ice chemistry. The comparison of ice COMs abundances derived from column densities (Figure~\ref{fig:iceCOM_NXtoNH2_compare}) shows that at the central area of the core, $\rm CH_3OCH_3$ ice abundance differs no more than an order of magnitude between the diffusive and GRD models, but other ice COMs demonstrate much larger variations in abundance: more than one order of magnitude for $\rm CH_3CHO$, more than two orders of magnitude for $\rm NH_2CHO$ and more than three orders of magnitude for $\rm HCOOCH_3$. Future ice composition observations with the James Webb Space Telescope (JWST) may help to find out which type of processes dominates ice chemistry.

\end{document}